\documentclass[sigconf,dvipsnames]{acmart}

\usepackage{booktabs} 
\usepackage{subcaption}
\usepackage{gensymb}
\usepackage{multirow}
\usepackage{xspace}
\usepackage{algorithm}
\usepackage[noend]{algpseudocode}
\usepackage{algorithmicx}
\floatname{algorithm}{Template}
\usepackage{siunitx}
\usepackage{fixltx2e}

\setcopyright{none}
\renewcommand\footnotetextcopyrightpermission[1]{}
\usepackage{ifthen}
\usepackage{calc}
\usepackage{pifont}
\usepackage{color}
\usepackage{fancyhdr}
\usepackage{tikz}
\usepackage{anyfontsize}
\usepackage{xfrac}

\newcounter{hours}
\newcounter{minutes}

\newcommand{\vdd}[0]{{$V_{array}$}\xspace}
\newcommand{\vddhalf}[0]{$V_{array}/2$\xspace}

\newcommand{\myitem}[1]{\emph{(#1)}\xspace}

\newcommand{\varr}[0]{{{$V_{array}$}}\xspace}
\newcommand{\vperi}[0]{{{$V_{peri}$}}\xspace}
\newcommand{\vmin}[0]{$V_{min}$\xspace}

\newcommand{\vddbl}[0]{\emph{full voltage}\xspace}
\newcommand{\vddblhalf}[0]{\emph{half voltage}\xspace}
\newcommand{\vddzero}[0]{\emph{zero voltage}\xspace}
\newcommand{\patt}[1]{\texttt{0x#1}\xspace}

\newcommand{\squishlist} {
    \begin{list}{$\bullet$} {
        \setlength{\itemsep}{-2pt}
            \setlength{\parsep}{2pt}
            \setlength{\topsep}{0pt}
            \setlength{\partopsep}{0pt}
            \setlength{\leftmargin}{1.0em}
            \setlength{\labelwidth}{1em}
            \setlength{\labelsep}{0.5em}
    }
}
\newcommand{\squishend} {
    \end{list}
}

\newboolean{timeofmake}
\setboolean{timeofmake}{false}

\newcommand{\ignore}[1]{}

\newcommand{\fix}[1]{#1}
\newcommand{\fixII}[1]{#1}
\newcommand{\fixIII}[1]{#1}
\newcommand{\fixIV}[1]{#1}
\newcommand{\fixV}[1]{#1}
\newcommand{\fixVI}[1]{#1}
\newcommand{\changes}[1]{#1}
\newcommand{\dhlfix}[1]{#1}
\newcommand{\response}[1]{#1}

\newcommand{\voltron}[0]{{Voltron}\xspace}
\newcommand{\memdvfs}[0]{{MemDVFS}\xspace}

\newcommand{\tras}[0]{{{tRAS}}\xspace}
\newcommand{\trp}[0]{{tRP}\xspace}

\newcommand{\trcd}[0]{{{tRCD}}\xspace}
\newcommand{\tcl}[0]{{tCL}\xspace}
\newcommand{\tcwl}[0]{{tCWL}\xspace}

\newcommand{\trcdmin}[0]{$tRCD_{min}$\xspace}
\newcommand{\trpmin}[0]{$tRP_{min}$\xspace}

\newcommand{\act}[0]{\texttt{\small{ACTIVATE}}\xspace}

\newcommand{\crd}[0]{\texttt{\small{READ}}\xspace}
\newcommand{\cwr}[0]{\texttt{\small{WRITE}}\xspace}
\newcommand{\pre}[0]{\texttt{\small{PRECHARGE}}\xspace}

\newcommand*\circled[1]{\tikz[baseline=(char.base)]{
            \node[shape=circle,draw,inner sep=0.8pt,fill=white,text=black] (char) {#1};}}

\newcommand{\figputWS}[3]{
\begin{figure*}[t]
\begin{minipage}{\linewidth}
\begin{center}
\includegraphics[scale=#2]{plots/#1}
\end{center}
\vspace{-0.15in}
\caption{#3 \label{fig:#1}}
\end{minipage}
\end{figure*}
}

\newcommand{\figputHW}[2]{
\begin{figure}[h]
\begin{minipage}{\linewidth}
\footnotesize 
\begin{center}
\includegraphics[width=1.0\linewidth]{plots/#1}
\end{center}
\vspace{-0.2in}
\caption{#2 \label{fig:#1}}
\end{minipage}
\end{figure}
}

\newcommand{\figputHWL}[3]{
\begin{figure}[h]
\begin{minipage}{\linewidth}
\footnotesize 
\begin{center}
\includegraphics[width=1.0\linewidth]{plots/#1}
\end{center}
\vspace{-0.1in}
\caption{#2 \label{fig:#3}}
\end{minipage}
\end{figure}
}

\newcommand{\figputHS}[3]{
\begin{figure}[h]
\begin{minipage}{\linewidth}
\begin{center}
\includegraphics[scale=#2]{plots/#1}
\end{center}
\vspace{-0.1in}
\caption{#3 \label{fig:#1}}
\end{minipage}
\end{figure}
}

\newcommand{\figref}[1]{Figure~\ref{fig:#1}}
\newcommand{\tabref}[1]{Table~\ref{tab:#1}}

\newcommand{\secref}[1]{Section~\ref{sec:#1}}
\newcommand{\ssecref}[1]{Section~\ref{ssec:#1}}

\newcommand{\paratitle}[1]{\textbf{#1.}\xspace}

\newcommand{\module}[3]{{{\textit #1}$_{\mathrm{#2}}^{\mathrm{#3}}$}\xspace}

\algnewcommand\algorithmicforeach{\textbf{for each}}
\algdef{S}[FOR]{ForEach}[1]{\algorithmicforeach\ #1\ \algorithmicdo}

\newfloat{afloat}{t!}{log}
\newcommand\afloatname{Algorithm}
\floatname{afloat}{\afloatname}

\begin{document}
\title{Understanding Reduced-Voltage Operation in Modern DRAM
Chips:\\Characterization, Analysis, and Mechanisms}


\author{
    Kevin K. Chang$^\dagger$ \quad Abdullah Giray
    Ya\u{g}l{\i}k\c{c}{\i}$^\dagger$ \quad Saugata Ghose$^\dagger$ \quad Aditya
    Agrawal$^\P$ \quad Niladrish Chatterjee$^\P$ \quad \\
        Abhijith Kashyap$^\dagger$ \quad
        Donghyuk Lee$^\P$ \quad
        Mike O'Connor$^{\P,\ddagger}$ \quad
        Hasan Hassan$^\S$ \quad
        Onur Mutlu$^{\S,\dagger}$
}

\author{
  \large
        $^\dagger$Carnegie Mellon University \qquad
        $^\P$NVIDIA \qquad
        $^\ddagger$The University of Texas at Austin \qquad
        $^\S$ETH Z{\"u}rich \qquad
}

\renewcommand{\shortauthors}{K. K. Chang et al.}

\begin{abstract}

The energy consumption of DRAM is a critical concern in modern computing
systems. Improvements in manufacturing process technology have allowed DRAM
vendors to lower the DRAM supply voltage conservatively, which reduces some of
the DRAM energy consumption. We would like to reduce the DRAM supply voltage
more aggressively, to further reduce energy. Aggressive supply
voltage reduction requires a thorough understanding of the effect voltage
scaling has on DRAM access latency and DRAM reliability.

In this paper, we take a comprehensive approach to understanding and exploiting
the latency and reliability characteristics of modern DRAM when the supply
voltage is lowered below the \fixII{nominal voltage level} specified by
\fixIV{DRAM standards}. Using an FPGA-based testing platform, we perform an
experimental study of 124 real DDR3L (low-voltage) DRAM chips manufactured
recently by three major DRAM vendors. We find that reducing the supply voltage
below a certain point introduces bit errors in the data, and we comprehensively
characterize the behavior of these errors. We discover that these errors can be
avoided by increasing the latency of three major DRAM operations (activation,
restoration, and precharge). \fixII{We perform} detailed DRAM circuit
simulations to validate and explain our experimental findings. We also
characterize the various relationships between reduced supply voltage and error
locations, stored data patterns, DRAM temperature, and data retention.

Based on our observations, we propose a new DRAM energy reduction mechanism,
called \emph{Voltron}. The key idea of Voltron is to use a performance model to
determine \fix{by} how much we can reduce the supply voltage without introducing errors
and without exceeding a user-specified threshold for performance loss. Our
evaluations show that Voltron reduces the average DRAM and system energy
consumption by 10.5\% and 7.3\%, respectively, while limiting the average system
performance loss to only 1.8\%, for a variety of \fix{memory-intensive} quad-core
workloads. We also show that Voltron significantly outperforms prior dynamic
voltage and frequency scaling mechanisms for DRAM.

\end{abstract}

%
%


%
%


\maketitle


\section{Introduction}

In a wide range of modern computing systems, spanning from warehouse-scale data
centers to mobile platforms, energy consumption is a first-order
concern~\cite{hoelzle-book2009,ibm-power7,deng-asplos2011,
  david-icac2011,malladi-isca2012,yoon-isca2012,choi-memcon2013,superfri,mutlu-imw2013}.
In these systems, the energy consumed by the DRAM-based main memory system
constitutes a significant fraction of the total energy.  For example,
\fix{experimental} studies of production systems have shown that DRAM consumes
40\% of the total energy in servers~\cite{hoelzle-book2009,ibmpower7-hpca}
\fixII{and 40\%} of the total power in graphics cards~\cite{paul-isca2015}.

The energy consumed by DRAM is correlated with the \emph{supply voltage} used
within the DRAM chips. The supply voltage is distributed to the two major
components within DRAM: the \emph{DRAM array} and the \emph{peripheral
  circuitry}~\cite{vogelsang-micro2010,lee-thesis2016}. The DRAM array consists
of thousands of capacitor-based DRAM cells, which store data as charge within
the capacitor. \changes{Accessing data stored in the \emph{DRAM array} requires
  a DRAM chip to perform a series of fundamental \fix{operations:} activation,
  restoration, and precharge.\footnote{We explain the detail of each of these
    operations in \secref{background}.} A memory controller \fix{orchestrates}
  each of the DRAM operations \fixII{while obeying the DRAM} timing parameters.
  On the other hand, the \emph{peripheral circuitry} consists of control logic
  and I/O drivers that connect the DRAM array to the memory channel, which is
  responsible for transferring commands and data between the memory controller
  and the DRAM chip. Since the DRAM supply voltage is distributed to both the
  DRAM array and the peripheral circuitry, \fixII{changing} the supply voltage
  would affect the energy consumption of both components in the entire DRAM
  chip.}


To reduce the energy consumed by DRAM, vendors have developed low-voltage
variants of DDR (Double Data Rate) memory, such as LPDDR4 (Low-Power
DDR4)~\cite{jedec-lpddr4} and DDR3L (DDR3 Low-voltage)~\cite{jedec-ddr3l}. For
example, in DDR3L, the internal architecture remains the same as DDR3 DRAM, but
vendors lower the nominal supply voltage to both the DRAM array and the
peripheral circuitry \fix{via} improvements in manufacturing process technology.
In this work, we would like to reduce DRAM energy by \emph{further reducing DRAM
  supply voltage}. Vendors choose a conservatively high supply voltage, to
provide a \emph{guardband} that allows DRAM chips with the worst-case process
variation to operate without errors \fixIV{under the worst-case operating
  conditions}~\cite{david-icac2011}.
The exact amount of supply voltage guardband varies across chips, and lowering
the voltage below the guardband can result in erroneous or even undefined
behavior. Therefore, we need to understand how DRAM chips behave during
reduced-voltage operation. To our knowledge, no previously published work
examines the effect of using a wide range of different supply voltage values on
the reliability, latency, and retention characteristics of DRAM chips.

\changes{\textbf{Our goal} in this work is to \myitem{i}~characterize and
understand the relationship between supply voltage reduction and various
characteristics of DRAM, including DRAM reliability, latency, and data
retention; and \myitem{ii}~use the insights derived from this characterization
and understanding to design a new mechanism that can aggressively lower the
supply voltage to reduce DRAM energy consumption while keeping performance loss
under a bound.} To this end, we build an FPGA-based testing platform that allows
us to tune the DRAM supply voltage~\cite{hassan-hpca2017}. Using this testing
platform, we perform experiments on 124~real DDR3L DRAM chips~\cite{jedec-ddr3l}
\fixII{from \fix{three} major vendors}, contained within 31~dual in-line memory
modules (DIMMs). Our comprehensive experimental characterization provides four
major observations on how DRAM \fixII{latency, reliability, and data retention
  time are} affected by reduced supply voltage.

First, we observe that we can reliably access data when DRAM supply voltage is
lowered below the nominal voltage, {\em until a certain voltage value}, \vmin,
which is the minimum voltage level at which no bit errors occur. Furthermore, we
find that we can reduce the voltage below \vmin to attain further energy
savings, but that errors start occurring in some of the data read from memory.
As we drop the voltage further below \vmin, the number of erroneous bits of data
increases exponentially \fixII{with the voltage drop}.

Second, we observe that while reducing the voltage below \vmin introduces bit
errors in the data, we can prevent these errors if we \emph{increase} the latency of
the three fundamental DRAM operations (activation, restoration, and precharge).
When the supply voltage is reduced, the capacitor charge takes a longer time to
change, thereby causing these DRAM operations to become slower to complete.
Errors are introduced into the data when the memory controller does \emph{not}
account for this slowdown in the DRAM operations.  We find that if the memory
controller allocates extra time for these operations to finish when the supply
voltage is below \vmin, errors no longer occur. We validate, analyze, and
explain this behavior using \fixII{detailed} circuit-level simulations.

Third, we observe that when only a small number of errors occur due to reduced
supply voltage, these errors tend to \emph{cluster} physically in certain
\emph{regions} of a DRAM chip, as opposed to being randomly distributed
throughout the chip.  This observation implies that when we reduce the supply
voltage to the DRAM array, we need to increase the fundamental operation
latencies for \emph{only} the regions where errors can occur.



Fourth, we observe that reducing the supply voltage does \emph{not} impact the
data retention guarantees of DRAM.  Commodity DRAM chips guarantee that all
cells can safely retain data for 64ms, after which the cells are
\emph{refreshed} to replenish charge that leaks out of the capacitors. Even when
we reduce the supply voltage, the rate at which charge leaks from the capacitors
is so slow that no data is lost during the 64ms \fixIV{refresh interval} at
20$\celsius$ and 70$\celsius$ \fix{ambient temperature}.

Based on our experimental observations, we propose a new low-cost DRAM energy
\fix{reduction} mechanism called \emph{Voltron}. \changes{The key idea of
  Voltron is to use a performance model to determine \fixIV{by} how much we can
  reduce the DRAM array voltage at runtime without introducing errors and
  without exceeding a user-specified threshold for \fixII{acceptable}
  performance loss.} Voltron consists of two components: \fix{\emph{array
    voltage scaling} and \emph{performance-aware voltage control}}.

\emph{Array voltage scaling}
leverages minimal hardware modifications within DRAM to reduce the
voltage of \emph{only} the DRAM array, without affecting the voltage
of the peripheral circuitry. \changes{If Voltron were to reduce the voltage of the
peripheral circuitry, we would have to reduce the operating frequency of DRAM.
This is because the \emph{maximum} operating frequency of DRAM is a
function of the peripheral circuitry voltage~\cite{david-icac2011}.
A reduction in the operating frequency reduces the memory data throughput, which
can significantly harm the performance of applications that require high memory
\fix{bandwidth}, \fixIV{as we demonstrate in this paper.}}

\emph{Performance-aware voltage control} uses \fixII{performance counters}
within the processor to build a piecewise linear model of how the performance of
an application decreases as the DRAM array supply voltage is lowered (due to
longer operation latency to prevent errors), and uses the model to select a
supply voltage that \fixII{keeps performance above} a
\fixIV{user/system-specified} performance target.


Our evaluations of Voltron show that it \fix{significantly reduces} both DRAM and system energy
consumption, at the expense of very modest performance degradation. For example,
at \fixII{an average} performance loss of only 1.8\% over seven \fix{memory-intensive}
quad-core workloads from SPEC2006, Voltron reduces DRAM energy consumption by an
average of 10.5\%, which translates to an overall system energy consumption of
7.3\%. We also show that \voltron effectively saves DRAM and system energy on
even \fix{non-memory-intensive} applications, with very little performance impact.


This work makes the following major contributions:
\squishlist

\item We perform the first detailed experimental characterization \fix{of how
the reliability and latency of \fixIII{modern DRAM chips are} affected when the
supply voltage is lowered below the nominal voltage level.  We comprehensively
test and analyze} 124 real DRAM chips from three major DRAM vendors. \fixII{Our
characterization reveals four new major observations, which can be useful for
developing new mechanisms \fixIV{that improve or better trade off between}
DRAM energy/power, latency, and/or reliability.}


\item We experimentally demonstrate that reducing the supply voltage below a
  certain point introduces bit errors in the data read from DRAM. We show that we
  can avoid these bit errors by increasing the DRAM access latency when the
  supply voltage is reduced.



\item We propose Voltron, a mechanism that \fixII{\myitem{i}~reduces the supply
voltage to only the DRAM array without affecting the peripheral circuitry, and
\myitem{ii}~uses a performance model to select a voltage that does not degrade
performance beyond a chosen threshold.  We show that Voltron is effective at
improving system energy consumption, with only a small impact on performance.}

\item \fix{We open-source our FPGA-based experimental characterization \fixIII{infrastructure} and
DRAM circuit simulation infrastructure, \fixIV{used in this paper}, for \fixIII{evaluating} reduced-voltage operation~\cite{volt-github}.}

\ignore{reduces DRAM and system energy consumption
without introducing errors and without incurring significant application performance
loss. Voltron \fixII{\myitem{i}} reduces the supply voltage to only the DRAM array without
affecting the peripheral circuitry, to avoid hurting memory data throughput, and
\fixII{\myitem{ii}} uses a performance model to select a voltage that does not degrade performance
by more than a threshold.}

\ignore{Onur says: "We show that Voltron is effective at improving system energy
consumption, with only a small impact on performance." This is a bit redundant
with the previous sentence.}

  \ignore{ Voltron reduces DRAM and system energy consumption by an average of
  10.8\% and 7.4\%, respectively, with only a 3.5\% reduction in system
  performance over seven \fix{memory-intensive} quad-core workloads.}


\squishend


\section{Background and Motivation}
\label{sec:background}

In this section, we first provide necessary DRAM background and terminology. We then
discuss related work on reducing the voltage and/or frequency of DRAM, to motivate
the need for our study.

\subsection{DRAM Organization}
\label{ssec:dram_org}

\begin{figure}[h]
  \captionsetup[subfigure]{justification=centering}

  \subcaptionbox{DRAM System\label{fig:dram-organization}}[0.2\linewidth][l] {
    \vspace{0.6cm}
    \includegraphics[scale=0.65]{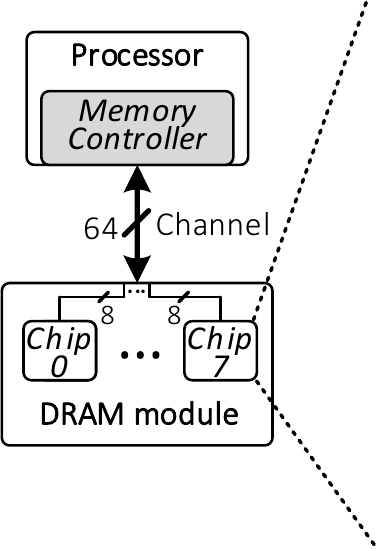}
}
  \subcaptionbox{DRAM Chip\label{fig:chip-organization}}[0.78\linewidth][r]
  {
    \includegraphics[scale=0.7]{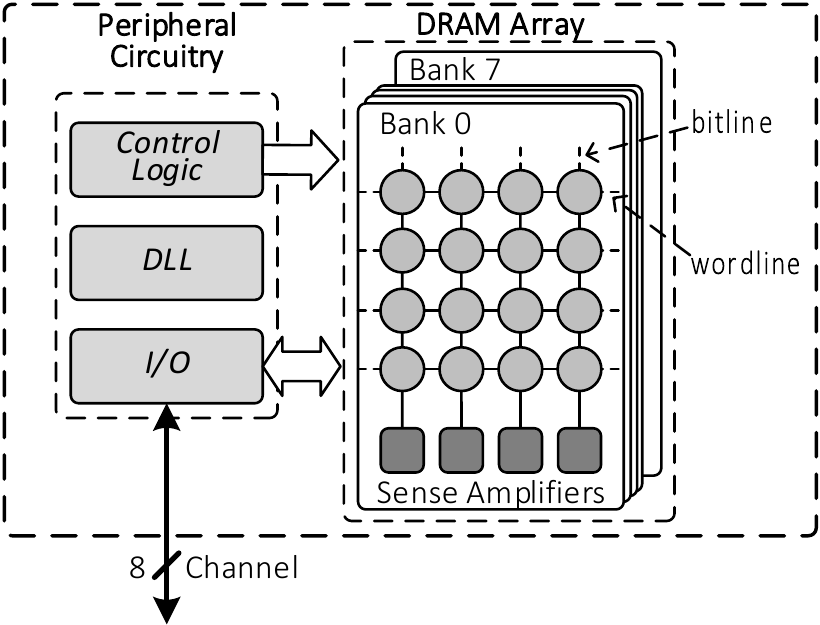}
  }%
  \caption{DRAM system and chip organization.}
  \label{fig:background}
\end{figure}

\figref{dram-organization} shows a high-level overview of \fixII{a modern}
memory system organization. A processor (CPU) is connected to a DRAM module via
a \emph{memory channel}, which is a bus used to transfer data and commands
between the processor and DRAM. A DRAM module is also called a \emph{dual
  in-line memory module} (\fixII{DIMM}) and it consists of multiple \emph{DRAM
  chips}, which are controlled together\fixIV{.\footnote{\fixIV{In this paper, we
study DIMMs that contain a single \emph{rank} (i.e., a group of chips in a
single DIMM that operate in lockstep).}}} \changes{Within each DRAM chip,
  illustrated in \figref{chip-organization}, we categorize the internal
  components into two broad categories: \myitem{i} the \emph{DRAM array}, which
  consists of multiple banks of DRAM cells organized into rows and columns, and
  \myitem{ii} \emph{peripheral circuitry}, which consists of the circuits that
  sit outside of the DRAM array. For a more detailed view of the components in a
  DRAM chip, we refer the reader to prior
  work\fixII{s}\fixIV{~\cite{vogelsang-micro2010, kim-isca2012, lee-hpca2013,
      liu-isca2012, lee-hpca2015, hassan-hpca2016, chang2016low,
      chang-thesis2017, lee-thesis2016, kim-thesis2015, lee-sigmetrics2017,
      chang-sigmetrics2016,lee-pact2015,seshadri-micro2013,seshadri-arxiv2016,lee-arxiv2016,chang-hpca2014,seshadri-thesis2016,seshadri-micro2015}}.}

\ignore{\fixIV{The DRAM array in each DRAM chip} is divided into multiple banks
  (typically eight \fixIV{in DDR3 DRAM~\cite{jedec-ddr3l, jedec-ddr3}}).
  \fixIV{Because the chips are controlled together, the arrays from each
    chip behave as if they are a single larger array.  Thus, each bank spans across
    all of the chips in the module. }
}

A DRAM array is divided into multiple banks (typically eight \fixIV{in DDR3
DRAM~\cite{jedec-ddr3l, jedec-ddr3}}) that can process DRAM commands
independently from each other to increase parallelism. A bank \fixIV{contains} a
2-dimensional array of DRAM cells. Each cell uses a capacitor to store \fixIV{a
  single bit} of data. Each array of cells is connected to a row of sense
amplifiers via vertical wires, called \emph{bitlines}. \fixII{This} row of sense
amplifiers \fixII{is called the} \emph{row buffer}. The row buffer senses the
data stored in one row of DRAM cells and serves as a temporary buffer for the
data. A typical row \fixIV{in a DRAM module (i.e., across all of the DRAM chips
  in the module)} is 8KB wide, comprising 128 \fixIV{64-byte} cache lines.


The peripheral circuitry has three major components. First, the I/O component is
used to receive commands or transfer data between the DRAM \fixIV{chip} and the
processor via the memory channel. Second, a typical DRAM chip uses a delay-lock
loop (DLL) to synchronize its data signal with the external clock to coordinate
data transfers on the memory channel.
Third, the control logic decodes DRAM commands sent across the memory channel
and selects the row \fixIV{and column} of cells to read data from or write data into.

\subsection{Accessing Data in DRAM}
\label{ssec:access_data}

To access data stored in DRAM, \fixII{the} \emph{memory controller} (shown in
\figref{dram-organization}) issues DRAM commands across the memory channel
to the DRAM chips. Reading a cache line from DRAM requires three essential
commands, as shown in \figref{background_timeline}: \act, \crd, and \pre.
Each command requires some time to complete, and the DRAM
\fixII{standard}~\cite{jedec-ddr3l} define\fixII{s} the latency of the commands
\fixII{with} a set of \emph{timing parameters}. The memory controller can be
programmed to obey different sets of timing parameters through the
BIOS~\cite{intel-xmp, lee-hpca2015, amd-bkdg, amd-opteron}.

\figputHS{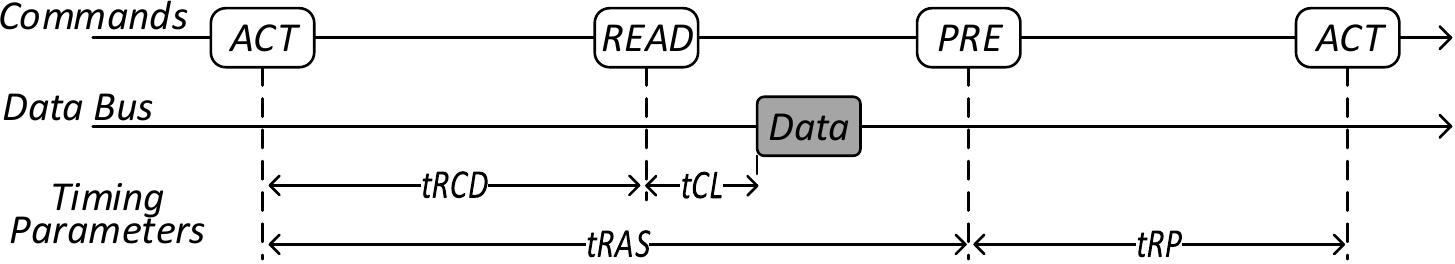}{0.57}{DRAM commands and timing parameters
when reading one cache line of data.}

\paratitle{Activate Command} To open the target row of data in the bank
that contains the desired cache line, the memory controller first
\fixII{issues} an \act command \fixII{to the target DRAM bank}. During
activation, the electrical charge stored in the target row starts to
propagate to the row buffer. The charge propagation triggers the row buffer
to latch the data stored in the row after some amount of time. The latency
of an \act command, or the \fixII{\emph{activation latency}}, is defined as
the minimum amount of time \fixII{that is required to pass} from the issue
time of an \act until the issue time of a column command (i.e., \crd or
\cwr). The timing parameter for the activation latency is called \trcd, as
shown in \figref{background_timeline}, and is typically set to 13ns in
DDR3L~\cite{micronDDR3L_2Gb}.

Since an activation drains charge from the target row's cells to latch
\fixII{the cells'} data into the row buffer, the \fixII{cells'} charge needs to
be restored to prevent data loss. The row buffer performs \emph{charge
restoration} simultaneously with activation. Once the cells' charge is
fully restored, the row can be closed \fixII{(and thus the DRAM array be
prepared for the next access)} by \fixII{issuing} a \pre command \fixII{to the
DRAM bank}. The DRAM standard specifies the restoration latency as the
minimum amount of time the controller must wait after \act before issuing
\pre. The timing parameter for restoration is called \tras, as shown in
\figref{background_timeline}, and is typically set to 35ns in
DDR3L~\cite{micronDDR3L_2Gb}.

\paratitle{Read Command} Once the row data is latched in the row buffer after
the \act command, the memory controller \fixII{issues} a \crd command. The row
buffer contains multiple cache lines of data (8KB), and the \crd command enables
\fixIV{all $n$ DRAM chips in the} DRAM module to select the desired cache line
(64B) from the row buffer. \fixIV{Each DRAM chip on the module then drives
  \fixV{($\sfrac{1}{n}$)\textsuperscript{th}} of the cache line from the row buffer to
  the I/O component within the peripheral circuitry.}
The peripheral circuitry \fixII{of each chip} then sends \fixIV{its
  \fixV{($\sfrac{1}{n}$)\textsuperscript{th}} of the cache line} across the memory
channel to the memory controller. \response{Note that the column access time to
  read and write the cache line is defined by the timing parameters \tcl and
  \tcwl, respectively, \fixII{as shown in \figref{background_timeline}}. Unlike
  the activation latency (\trcd),  \tcl and \tcwl are
  \fixIV{\emph{DRAM-internal}} timings that are determined by a clock inside
  DRAM~\cite{micronDDR3L_2Gb}. Therefore, our \fix{FPGA-based} experimental
  \fixIII{infrastructure} (described in \secref{fpga}) \emph{cannot} evaluate
  the effect of changing \tcl and \tcwl.}

\paratitle{Precharge Command} After reading the data from the row buffer,
the memory controller may contain a request that needs to access data from
a \emph{different} row within the same bank.  To prepare \fixII{the bank to service
this request}, the memory controller \fixII{issues} a \pre command \fixII{to
the bank}, which closes the currently-activated row and resets the bank
\fixII{in preparation} for the next \act command. Because closing the
activated row and resetting the bank takes some time, the standard
specifies the precharge latency as the minimum amount of time the
controller must wait \fixII{for} after \fixII{issuing} \pre before \fixII{it
issues an} \act. The timing parameter for precharge is called \trp, as
shown in \figref{background_timeline}, and is typically set to 13ns in
DDR3L~\cite{micronDDR3L_2Gb}.


\changes{
    \subsection{Effect of DRAM Voltage and Frequency on Power
    \fixII{Consumption}}
\label{sec:dram_power}

DRAM power \fixII{is divided} into dynamic and static power. \fixII{Dynamic}
power \fixII{is} the power consumed by executing the access commands: \act,
\pre, and \crd/\cwr. \fixIV{Each} \act and \pre consumes \fixIV{power in the
  DRAM array and the peripheral circuitry due to the activity in the DRAM
  array and control logic.} \fixII{Each} \crd/\cwr consumes power in the DRAM
array by accessing data in the row buffer, and in the peripheral circuitry by
driving data on the channel. On the other hand, static power \fixII{is} the
power that is consumed \emph{regardless} of the DRAM accesses, and it is mainly
due to transistor leakage. DRAM power is governed by both the supply voltage and
\fixII{operating clock} frequency: $Power \propto Voltage^2 \times
Frequency$~\cite{david-icac2011}. As shown in \fixIV{this} equation, power consumption
scales quadratically with supply voltage, and linearly with frequency.

DRAM supply voltage is distributed to both the DRAM array and the
peripheral circuitry through \fixII{respective} power pins on the
DRAM chip, \fixII{dedicated separately to the DRAM array and the peripheral
circuitry}. We call the voltage supplied to the DRAM array, \varr, and the
voltage supplied to the peripheral circuitry, \vperi. Each DRAM standard
requires a specific nominal supply voltage value, which depends on many
factors, such as the architectural design and process technology. In this
work, we focus on the widely used DDR3L DRAM design that requires a nominal
supply voltage of 1.35V~\cite{jedec-ddr3l}. To remain operational when the
\fixII{supply} voltage is unstable, DRAM can tolerate a small amount of
\fixIV{deviation from the nominal supply}
voltage. In particular, DDR3L DRAM \fixII{is specified to} operate
with a supply voltage ranging from 1.283V to 1.45V~\cite{micronDDR3L_2Gb}.

The DRAM \fixII{channel} frequency value of a DDR DRAM chip is typically
specified \fixIV{using the} \emph{channel data rate}, measured in mega-transfers per second
(MT/s). The size of each data transfer is dependent on the width of the data
bus, which ranges from 4 to 16~bits for a DDR3L chip~\cite{micronDDR3L_2Gb}.
Since a modern DDR channel transfers data on both the positive and \fixII{the}
negative clock edge\fixII{s (hence the term \emph{double data rate}, or DDR)},
the channel frequency \fixII{is} \emph{half of the data rate}. For example, a
DDR data rate of 1600 MT/s means that the frequency \fixII{is} 800 MHz. To run
the channel at a specified data rate, the peripheral circuitry requires a
certain minimum voltage (\vperi) for stable operation. As a result, the supply
voltage scales directly \fixIV{(i.e., linearly)} with DRAM frequency, and it
determines the maximum operating
frequency~\cite{deng-asplos2011,david-icac2011}.

\ignore{
Modern DDR DRAM chips are \emph{synchronous} state machines, driven by an
external clock signal from the memory controller. This allows the memory
controller to have timing control on when to send commands to or receive data
from DRAM. The DRAM frequency value of a DDR DRAM chip is typically specified in
channel data rate, measured in mega-transfers per second (MT/s). The size of
each data transfer is dependent on the width of the data bus, which ranges from
4 to 16 bits for a DDR3L chip~\cite{micronDDR3L_2Gb}. Since a modern DDR channel
transfers data on both the positive and negative clock edge, the channel
frequency operates at half of the data rate. For example, a DDR data rate of
1600 MT/s means that the channel operates at 800 MHz.\footnote{\reftiny Note that
  several prior works (imprecisely) use MT/s and MHz interchangeably to discuss
  channel frequency.} A DRAM chip typically supports several frequency values
(e.g., 1066, 1333, 1600 MT/s) that can be chosen by the processor only during the
system boot time. The DRAM frequency not only affects the data throughput, but
also the DRAM power. Except for the dynamic power of \act and \pre, which are
\emph{asynchronous operations}, the remaining DRAM power scales \emph{linearly} with
frequency: $Power \propto Frequency$.
}
}

\subsection{Memory Voltage and Frequency Scaling}
\label{ssec:dvfs}

\ignore{The memory system has become a major energy consumer in modern computing
systems, consuming 40\% of the total system energy in
servers~\cite{hoelzle-book2009,ibmpower7-hpca}, and 40\% of the total system
power in GPUs~\cite{paul-isca2015}).}

One \fixII{proposed} approach to reducing memory energy consumption is to scale
the voltage and/or the frequency of DRAM based on the \fixIV{observed memory
  channel} utilization. We briefly describe two different ways of scaling
frequency and/or voltage below.



\paratitle{Frequency Scaling} To \fix{enable the power reduction} \fixII{that
comes with reduced} \fixIV{DRAM} frequency, prior works propose to apply
\emph{dynamic frequency scaling} (DFS) by adjusting the DRAM \fixIV{channel}
frequency based on \fixIV{the} memory bandwidth demand \fixIV{from the DRAM
  channel}~\cite{deng-asplos2011,deng-islped2012,deng-micro2012, paul-isca2015,
  begum-iiswc2015,sundriyal2016}. A major \fixIV{consequence of} lowering the
frequency is the \fixII{likely} performance loss that occurs, as it takes a
longer time to transfer data across the DRAM channel \fixIV{while operating at a
  lower frequency}. The clocking logic within the peripheral circuitry requires
a \emph{fixed number of DRAM cycles} to transfer the data, \fixII{since} DRAM
sends data on each edge of the clock cycle. For a 64-bit memory channel with a
64B~cache line size, the transfer typically takes four DRAM
cycles~\cite{jedec-ddr3}. Since \fixII{lowering} the frequency increases the
time required for each cycle, the total amount of time spent on data transfer,
in nanoseconds, increases accordingly. As a result, \fix{not only does memory
  latency increase, but} \fixIII{also} memory \fix{data} throughput decreases,
\fixII{making} DFS \fixII{undesirable to use} when the running workload's memory
bandwidth \fixII{demand} \fixIV{or memory latency sensitivity} is high. The
extra transfer \fix{latency from DRAM} can also cause longer queuing times for
requests waiting at the memory controller\fixIII{~\cite{lee-tech2010,
    ipek-isca08,
    kim-rtas2014,subramanian-iccd2014,subramanian-tpds2016,kim-rts2016}},
further exacerbating the performance loss \fixII{and potentially delaying
  latency-critical applications}~\cite{deng-asplos2011,david-icac2011}.


\paratitle{Voltage and Frequency Scaling} While decreasing the channel
frequency \fixII{reduces} the peripheral circuitry power and static power, it
does \fix{\emph{not}} affect the dynamic power consumed by the operations performed on
the DRAM array (i.e., activation, restoration, precharge). This is because
\fixII{DRAM array operations} are asynchronous\fix{, i.e., independent of}
the channel frequency~\cite{micron-tr}. As a result, these operations
require a fixed \fixIII{time (in nanoseconds)} to complete.  For example, the
activation latency in a DDR3L DRAM module \fixII{is} 13ns, regardless of the DRAM
frequency~\cite{micronDDR3L_2Gb}.  If the channel frequency is \fixII{doubled
from 1066~MT/s to 2133~MT/s}, the memory controller \fixII{doubles} the
number of cycles for the \act timing parameter (i.e., \trcd) (from 7~cycles
to 14~cycles), to maintain the 13ns latency.


\changes{ In order to reduce the dynamic power consumption of the DRAM
array as well, prior work \fixII{proposes} \emph{dynamic voltage and
frequency scaling} (DVFS) for DRAM, which reduces the supply voltage along
with the \fixII{channel} frequency~\cite{david-icac2011}. This mechanism
selects a DRAM frequency based on the current memory bandwidth utilization
and finds the \emph{minimum operating voltage} (\vmin) for that frequency.
\vmin is defined to be the lowest voltage that still provides ``stable
operation'' for DRAM (i.e., no errors occur within the data). There are two
significant limitations for this proposed DRAM DVFS mechanism. The first
limitation is due to a lack of understanding of how voltage scaling affects
the DRAM behavior. No prior work \fixIV{provides} experimental characterization
or analysis of \fixII{the effect of} reducing \fixIV{the DRAM supply} voltage \fixII{on latency,
reliability, and data retention} in real DRAM chips. As the DRAM behavior \fix{under reduced-voltage operation}
is unknown \fixII{to \fix{satisfactorily} maintain the latency and reliability of DRAM}, the
\fixII{proposed DVFS} mechanism~\cite{david-icac2011} can reduce \fixII{supply}
voltage only \emph{very conservatively}. The second limitation is that this
prior work
\fixII{reduces the
supply} voltage only when it \fixII{reduces the channel frequency, since
a lower channel frequency requires a lower supply voltage for stable operation.}
As a result, \fixII{DRAM} DVFS results in the same
performance issues experienced by \fixII{the DRAM} DFS mechanisms. In
\ssecref{pavc_eval}, we evaluate \fixII{the main} prior
work~\cite{david-icac2011} on memory DVFS to quantitatively demonstrate its
\fixII{benefits and limitations}.}


\changes{ \subsection{Our Goal} The goal of this work is to \myitem{i}
\fixII{experimentally} characterize and analyze \emph{real modern DRAM chips} operating at different
supply voltage levels, \fixII{in order to develop a solid and thorough
understanding of how \fix{reduced-voltage} operation affects latency,
reliability, and data retention in DRAM;} and \myitem{ii} develop a mechanism that can reduce DRAM
energy consumption \fix{by reducing DRAM voltage,} without having to sacrifice
memory \fixII{data} throughput,
\fixII{based on the insights obtained from \fixIV{comprehensive} experimental}
characterization. Understanding how DRAM characteristics change at different
voltage levels is imperative not only for enabling memory DVFS in real systems,
but also for developing other low-power \fixII{and low-energy} DRAM designs that can effectively reduce
the DRAM voltage. We \fixII{experimentally analyze} the \fixII{effect} of
reducing \fixII{supply} voltage \fixII{of modern} DRAM chips
in \secref{dram_exp}, and \fixII{introduce} our \fixII{proposed new} mechanism \fixII{for
reducing DRAM energy} in \secref{varray}. }


\section{Experimental Methodology}
\label{sec:fpga}

To study the behavior of real DRAM chips under \fixII{reduced} voltage, we build
an FPGA-based infrastructure based on SoftMC~\cite{hassan-hpca2017}, which
allows us to have precise control over the DRAM modules. This method was used in
many previous works\fixIII{~\cite{jung-memsys2016, jung-patmos2016,
    kim-isca2014, chang-sigmetrics2016, khan-sigmetrics2014,
    khan-dsn2016,hassan-hpca2017,lee-sigmetrics2017,mathew-rapido2017,
    lee-hpca2015, qureshi-dsn2015, khan2016case, lee-thesis2016,
    chang-thesis2017, kim-thesis2015, liu-isca2013}} as an effective way to
explore different DRAM characteristics (e.g., latency, reliability, \fixVI{and}
\fixIV{data retention time}) that have not been known or exposed to the public
by DRAM manufacturers. Our testing platform consists of a Xilinx ML605 FPGA
board and a host PC that communicates with the FPGA via a PCIe bus
(\figref{fpga}). We adjust the supply voltage to the DRAM by using a USB
interface adapter~\cite{ti-usb} that enables us to tune the power rail connected
to the DRAM module directly. The power rail is connected to all the power pins
of every chip on the module (as shown in Appendix~\ref{sec:pin_layout}).

\figputHS{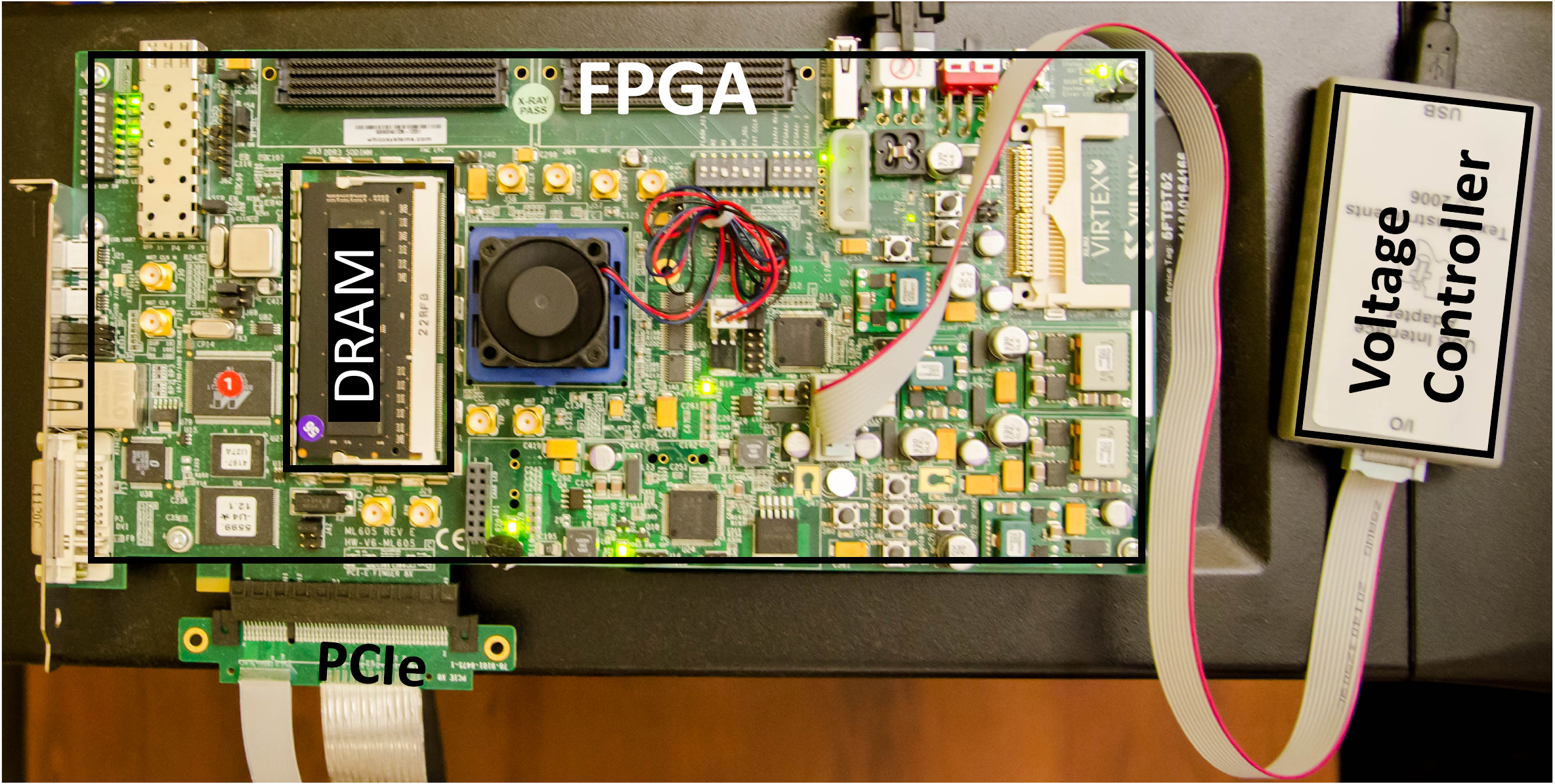}{0.35}{FPGA-based DRAM testing platform.}

\paratitle{Characterized DRAM Modules} In total, we tested 31 DRAM DIMMs,
comprising \fix{of} 124 DDR3L (low-voltage) chips, from the three \fixII{major}
DRAM chip vendors that hold more than 90\% of the DRAM market
share~\cite{bauer-mckinsey2016}.  Each chip has a 4Gb density. Thus,
\fixII{each} of our DIMMs \fixII{has} a 2GB capacity. The DIMMs support up to a
1600 MT/s channel frequency. Due to \fixII{our} FPGA's maximum operating
frequency limitations, all of our tests are conducted at 800 MT/s.  Note that
the experiments we perform \fixII{do \emph{not} require us to adjust} the
channel frequency. \tabref{dimm_list} describes the relevant information about
\fixIV{the} tested DIMMs. \fixIV{Appendix~\ref{sec:dimm_info} provides
  detailed information on each DIMM}. Unless otherwise specified, we test our
DIMMs at an ambient temperature of 20$\pm$1\celsius. We examine the effects of
high \fixIV{ambient} temperature \fixIV{(i.e., 70$\pm1\celsius$)} in
Section~\ref{ssec:temperature}.


\begin{table}[h]
  \centering
    \setlength{\tabcolsep}{.35em}
    \begin{tabular}{cccc}
        \toprule
        \multirow{2}{*}{Vendor} & Total Number & Timing (ns) &
        Assembly  \\
        & of Chips & (\trcd/\trp/\tras) & Year  \\
        \midrule
        A (10 DIMMs) & 40 & 13.75/13.75/35 & 2015-16\\
        B (12 DIMMs) & 48 & 13.75/13.75/35 & 2014-15 \\
        C (9 DIMMs) & 36 & 13.75/13.75/35 & 2015 \\
        \bottomrule
    \end{tabular}
  \caption{\fixIII{Main} properties of the tested DIMMs.}
  \label{tab:dimm_list}
\end{table}


\paratitle{DRAM Tests} \label{ssec:dramtest} At a high level, we develop a test
(Test~\ref{test}) that writes/reads data to\fixII{/from} \emph{every} row in the
\emph{entire} DIMM, \fixIV{for a given supply voltage}. The test takes in
several different input parameters: activation latency (\trcd), precharge
latency (\trp), and data pattern. The goal of the test is to examine if any
errors occur under \fixIV{the given supply voltage with} the different input
parameters.

\floatname{algorithm}{Test}

\begin{algorithm}[h]

\algnewcommand\algorithmicto{\textbf{to}}
\algrenewtext{For}[3]{\algorithmicfor\ #1 $\gets$ #2 \algorithmicto\ #3 \algorithmicdo}

\algrenewcommand\algorithmicfunction{}
\algrenewcommand\algorithmicdo{}
\algrenewcommand\algorithmicindent{1.2em}
\algrenewcommand\alglinenumber[1]{\footnotesize\texttt{#1}}
\small

\begin{algorithmic}[1]
\Function{VoltageTest}{$\mathit{DIMM}, \mathit{tRCD}, \mathit{tRP},
\mathit{data}, \overline{\mathit{data}}$}

\For{bank}{1}{$\mathit{DIMM.Bank}_{\mathit{MAX}}$}
\For{row}{1}{$\mathit{bank.Row}_{\mathit{MAX}}$} \Comment{Walk through every
  row within the current bank}
\State {\tt WriteOneRow(}$\mathit{bank}, \mathit{row}, \mathit{data}${\tt)}
\Comment{Write the data pattern into the current row}
\State {\tt WriteOneRow(}$\mathit{bank}, \mathit{row+1},
\overline{\mathit{data}}${\tt)} \Comment{Write the inverted data pattern into
  the next row}
\State {\tt ReadOneRow(}\trcd, \trp, $\mathit{bank}, \mathit{row} ${\tt)}
\Comment{Read the current row}
\State {\tt ReadOneRow(}\trcd, \trp, $\mathit{bank}, \mathit{row+1} ${\tt)}
\Comment{Read the next row}
\State {\tt RecordErrors()} \Comment{Count errors in both rows}

\EndFor
\EndFor

\EndFunction
\end{algorithmic}

    \caption{Test DIMM with specified tRCD/tRP and data \fix{pattern}.}
\label{test}
\end{algorithm}


In the test, we iteratively test two consecutive rows at a time. The two rows
hold data that are the inverse of each other (i.e., $\mathit{data}$ and
$\overline{\mathit{data}}$). Reducing \trp lowers the amount of time the
precharge unit has to reset the bitline voltage from either \vddbl (bit value 1)
or \vddzero (bit value 0) to \vddblhalf. If \trp were reduced too much, the
bitlines would float at some other intermediate voltage value between \vddblhalf
and \emph{full/zero voltage}. As a result, the next activation can potentially
start before the bitlines are fully precharged. If we were to use the same data
pattern in both rows, the sense amplifier would require \emph{less} time to
sense the value during the next activation, as the bitline is already
\emph{biased} \fixII{toward} those values. By using the \emph{inverse} of
\fixII{the data pattern in} the row \fixII{that is} precharged for the next row
\fixII{that is} activated, we ensure that the partially-precharged state of the
bitlines does \emph{not} unfairly favor the \fix{access to the} next
row~\cite{chang-sigmetrics2016}.
In total, we use three different groups of data patterns for our test: (0x00,
0xff), (0xaa, 0x33), and (0xcc, 0x55). Each specifies the $\mathit{data}$ and
$\overline{\mathit{data}}$, placed in consecutive rows in the same bank.

\section{Characterization of DRAM Under Reduced Voltage}
\label{sec:dram_exp}

In this section, we present our major observations from our detailed
experimental characterization of 31 commodity DIMMs \changes{(124 chips)} from
three vendors, when \changes{the DIMMs} operate under \fix{reduced} supply
voltage (i.e., below the nominal voltage level specified by \fixIV{the DRAM
  standard}).
First, we analyze the reliability of DRAM chips as we reduce the \changes{supply}
voltage without changing the DRAM access latency (\ssecref{volt_sensitivity}).
Our experiments \changes{are designed} to identify if lowering \changes{the
  supply} voltage induces bit errors (i.e., \emph{bit flips}) in data.
\fix{Second}, we present our findings on the effect of increasing the activation
and precharge latencies for DRAM operating under \fixIV{reduced} supply voltage
(\ssecref{low_volt_latency}). The purpose of this experiment is to understand
the trade-off between access latencies (which \fix{impact} performance) and the
supply voltage of DRAM \fixIII{(which impacts energy consumption)}. We use detailed circuit-level DRAM simulations to
validate \changes{and explain} our observations on the relationship between
access latency and supply voltage. \fix{Third}, we examine the spatial locality
of errors \changes{induced due to reduced-voltage operation} (\ssecref{spatial})
and the distribution of errors in the data sent across the memory channel
(\ssecref{ecc}).  \fix{Fourth}, we study the effect of temperature on
\changes{reduced}-voltage operation (\ssecref{temperature}). \fix{Fifth}, we
study \fix{the effect of reduced voltage on the} data retention times within
DRAM (\ssecref{refresh}). We present a summary of our findings in
\ssecref{char_summary}.

\subsection{DRAM Reliability as Supply Voltage Decreases}
\label{ssec:volt_sensitivity}

We first study the reliability of DRAM chips under low voltage, which was not
studied by prior works on DRAM voltage scaling
\fix{(e.g.,~\cite{david-icac2011}).} For these experiments, we use the minimum
activation and precharge latencies that we experimentally determine to be
reliable (i.e., they do not induce any errors) under the nominal voltage of
1.35V \changes{at 20$\pm$1$\celsius$ temperature}. As shown in \fix{prior
  works~\cite{chang-sigmetrics2016, lee-hpca2015, chandrasekar-date2014,
    lee-sigmetrics2017, hassan-hpca2017, chang-thesis2017, lee-thesis2016,
liu-isca2012, agrawal-hpca2014, qureshi-dsn2015,
venkatesan-hpca2006,bhati-isca2015,lin-iccd2012,ohsawa-islped1998,
patel-isca2017, khan-sigmetrics2014, khan-dsn2016, khan2016case}},
\changes{DRAM} manufacturers adopt a pessimistic standard latency that
\fix{incorporates} a large margin as a \fix{safeguard to ensure that each chip
  \fixIII{deployed in the field} operates correctly under a wide range of
  conditions.  Examples of these conditions include process variation, which
  causes some chips or some cells within a chip to be slower than others, or
  high operating temperatures, which can affect the time required to perform
  various operations within DRAM.}
Since our goal is to understand how the inherent DRAM latencies vary with
voltage, we \fixIII{conduct} our experiments \emph{without} such an excessive
margin. We identify that the reliable \trcd and \trp latencies are
\changes{both} 10ns (instead of \fix{the 13.75ns latency} specified by the
\fix{DRAM} standard) at \fix{20$\celsius$}, which agree with the values reported
by prior work on DRAM latency characterization\fix{~\cite{chang-sigmetrics2016,
    lee-hpca2015, chang-thesis2017, lee-thesis2016}}.

Using the \fix{\emph{reliable minimum latency values} (i.e., 10ns for all of the
DIMMs),} we run Test~\ref{test}, which accesses every bit within a DIMM at the
granularity of a 64B cache line. In total, there are 32~million cache lines in a \fix{2GB}
DIMM. We vary the supply voltage from the nominal voltage of 1.35V down to
1.20V, using a step size of 0.05V (50mV). Then, we change to a smaller step \fix{size} of
0.025V (25mV), until we reach the lowest voltage \fix{at which} the DIMM can
\fix{operate reliably (i.e., without any errors) while employing the reliable minimum latency values.}
\fix{(We examine methods to further reduce the \fixIII{supply} voltage in
  \ssecref{low_volt_latency}.)} For each voltage step, we run 30~rounds of
Test~\ref{test} for each DIMM.  \figref{dimm_errors_all} shows the fraction of
cache lines that \changes{experience} at least 1~bit of error (i.e., 1~bit flip)
in each DIMM \changes{(represented by each curve)}, \fixIV{categorized based
  on vendor.}

\figputHW{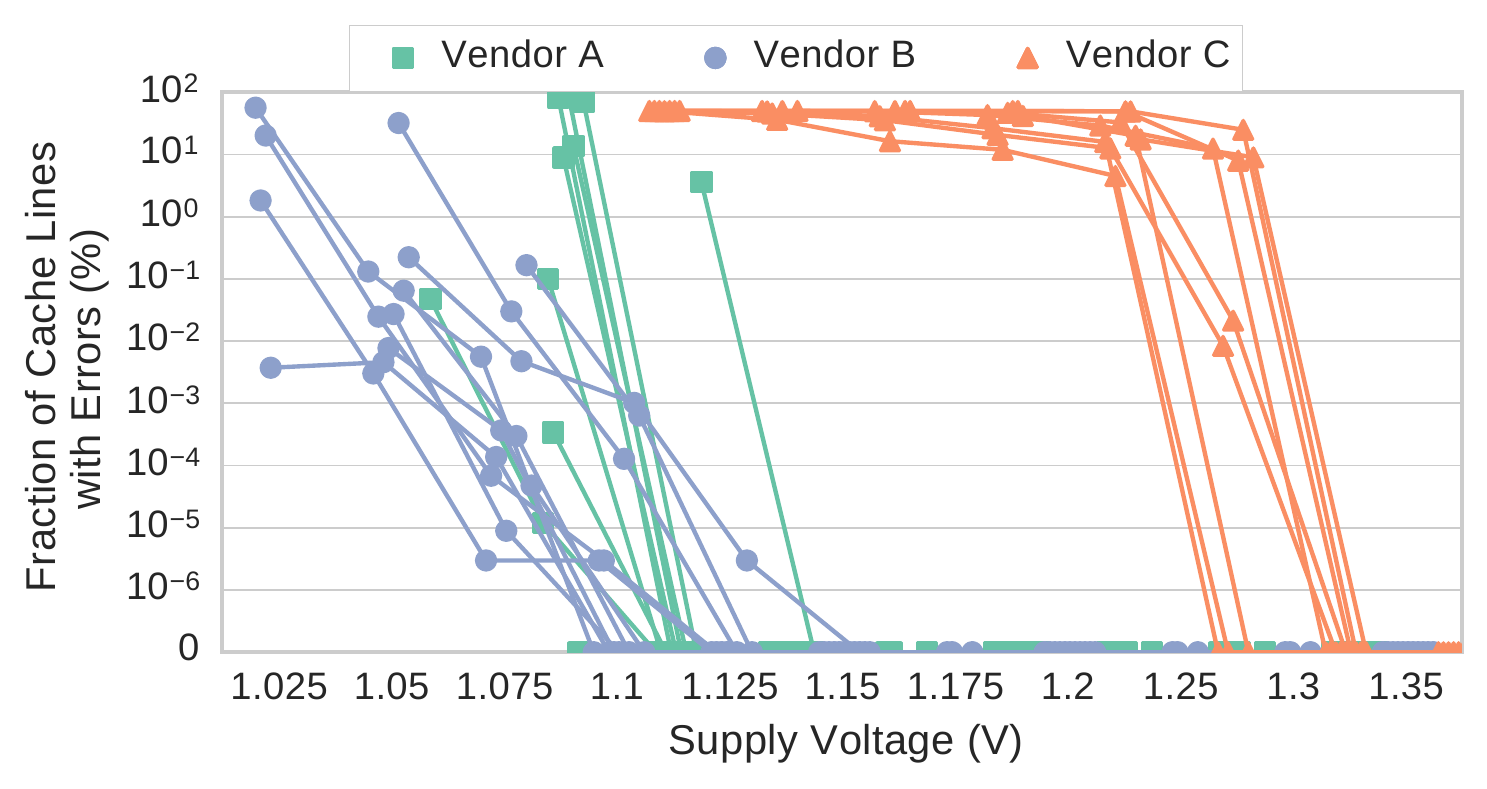}{The fraction of erroneous cache lines in each DIMM as
we reduce the supply voltage, with a fixed latency.}

We make three observations. First, when each DIMM runs below a certain voltage
level, errors start occurring. We refer to the \emph{minimum voltage level} of
each DIMM that allows error-free \changes{operation} as \vmin. For example, most
DIMMs from Vendor~C have $V_{min}=1.30V$.
\changes{Below \vmin, we observe errors because the fundamental DRAM array
operations (i.e., activation, restoration, precharge) \emph{cannot} fully
complete within the time interval specified by the latency parameters (e.g.,
\trcd, \tras) at low voltage.}
%
Second, not all cache lines exhibit errors for all supply voltage values below
\vmin. Instead, the number of erroneous cache lines \changes{for each DIMM}
increases as we reduce the voltage further below \vmin. Specifically, Vendor A's
DIMMs \fix{experience a} near-exponential increase \fix{in} errors as \fix{the} supply voltage reduces
below \vmin. This is mainly due to the \emph{manufacturing process and
  architectural variation}, which introduces strength and size variation across
the different DRAM cells within a chip\fixIII{~\cite{kim-edl2009, li11,
    chang-sigmetrics2016, lee-sigmetrics2017, lee-hpca2015,
    chandrasekar-date2014, lee-thesis2016, kim-thesis2015, chang-thesis2017}}. \changes{Third, variation in \vmin exists not only
  across DIMMs from different vendors, but also across DIMMs from the same
  vendor.} However, the variation across DIMMs \fix{from} the same vendor is
much smaller compared to cross-vendor variation, since the fabrication process
and circuit designs can differ drastically across vendors. These results
demonstrate that reducing voltage beyond \vmin, without altering the access
latency, has a negative impact on \fix{DRAM} reliability.


We also conduct an analysis of \fixIII{storing different \emph{data patterns}} on the error rate
\fixIII{during reduced-voltage operation} (see Appendix~\ref{sec:datapatt}).  In summary, our results show that the data
pattern does \emph{not} \changes{have a} consistent effect on the rate of errors
induced by reduced-voltage operation. For most supply voltage values, the data
pattern does \emph{not} have a statistically significant effect on the error
rate.

\paratitle{Source of Errors} To understand why errors occur in data as the
supply voltage \fixIV{reduces} below \vmin, we \changes{perform} circuit-level SPICE
simulations~\cite{nagel-spice,massobrio1993semiconductor}, which reveal more
detail on how the cell arrays operate. \fix{We develop a} SPICE model of the DRAM array \fix{that} uses a
sense amplifier design from prior work~\cite{baker-dram} with the \SI{45}{\nano\meter}
transistor model from the Predictive Technology Model (PTM)~\cite{ptm,
  zhao-isqed2006}. \fix{Appendix~\ref{spice_model} provides a detailed description of our SPICE
simulation model, which we have open-sourced~\cite{volt-github}.}

We vary the supply voltage of the DRAM array (\vdd) \fix{in our SPICE simulations} from
1.35V to 0.90V.
\figref{spice_act+pre_v3} shows the bitline voltage during activation and
precharge for different \vdd values. Times 0ns and 50ns correspond to when the
DRAM receives the \act and the \pre commands, respectively. An \act causes the
bitline voltage to increase from \vddhalf to \vdd in order to sense the stored
data value of ``1''. A \pre resets the bitline voltage back to \vddhalf in order
\fix{to enable the issuing of a later}
\act to another row within the same bank. In the figure, we mark the
points where the bitline reaches \fix{the} \circled{1} \emph{ready-to-access} voltage,
which we assume to be 75\% of \vdd; \circled{2} \emph{ready-to-precharge}
voltage, which we assume to be 98\% of \vdd; and \circled{3}
\emph{ready-to-activate} voltage, which we assume to be within 2\% of \vddhalf.
These points \fix{represent the minimum \trcd, \tras, and \trp values, respectively,}
required for reliable DRAM operation.
For readers who wish to understand \fix{the bitline voltage behavior in more detail,}
\fix{we refer them to recent
works~\cite{lee-hpca2015,lee-hpca2013,hassan-hpca2016, lee-sigmetrics2017, lee-thesis2016} that provide
extensive background on how the bitline voltage changes during the three DRAM
operations.}

\figputHW{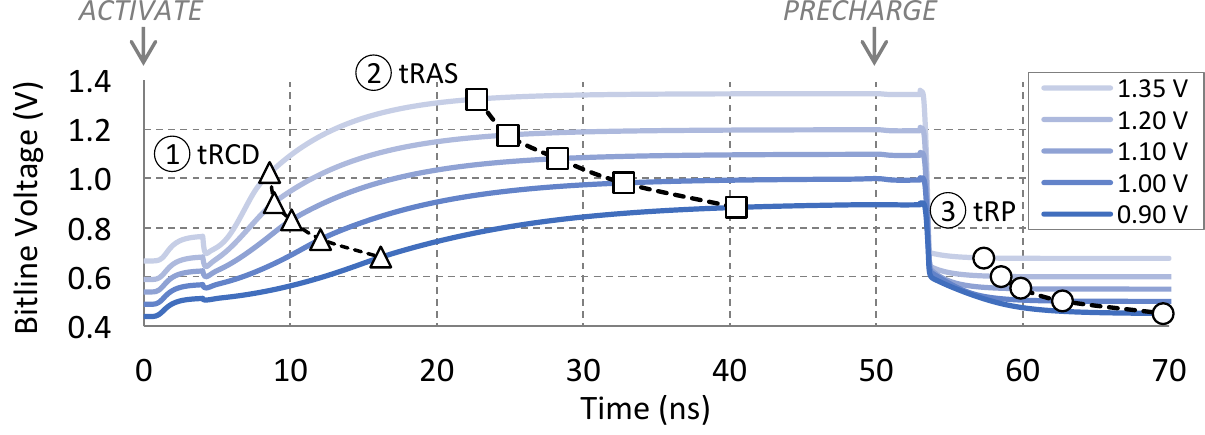}{Effect of reduced array supply
voltage on activation, restoration, and precharge, from SPICE simulations.}

We make two observations from our SPICE simulations. First, we observe that the
bitline voltage during activation increases at a different rate depending on the
\changes{supply} voltage \fix{of} the DRAM array (\vdd). \changes{Thus},
\fix{the} \changes{supply voltage affects} the latency of the three DRAM
operations (activation, restoration, and precharge). When the nominal voltage
level (1.35V) is used \fix{for \vdd}, the time (\trcd) it takes for the sense
amplifier to drive the bitline to the \emph{ready-to-access voltage
  \changes{level}} (75\% of \vdd) is much shorter than the time \fix{to do so at
  a} lower \vdd. As \vdd decreases, the sense amplifier needs more time to latch
in the data, increasing the activation latency. \ignore{Similarly, the
  \changes{restoration latency} (\tras)\changes{, i.e., the} time needed to
  restore the cell to 98\% \changes{of} \vdd and the \changes{precharge latency}
  (\trp), \changes{i.e, the time} for the bitline to reset back to \vddhalf
  increase as \vdd decreases.} Similarly, the \changes{restoration latency}
(\tras) and the \changes{precharge latency} (\trp) increase as \vdd decreases.

Second, the latencies of the three fundamental DRAM array operations (i.e.,
activation, restoration, precharge) \changes{do} \emph{not} correlate with the
channel (or clock) frequency \fixIV{(not shown in \figref{spice_act+pre_v3})}. This is
because these operations are clock-independent \changes{asynchronous} operations
that are a function of the cell capacitance, bitline capacitance, and
\vdd~\cite{keeth-dram-tutorial}.\fix{\footnote{\fix{In
      Appendix~\ref{spice_model}, we show a detailed circuit schematic of a DRAM
      array that operates asynchronously\fixIII{, which forms the basis of our
        SPICE circuit simulation model~\cite{volt-github}}.}}} As a result, the
channel frequency is \emph{independent} of the three fundamental \changes{DRAM}
operations.

Therefore, we hypothesize that DRAM errors occur at lower supply voltages
because the \fix{three} DRAM \fix{array} operations have insufficient latency to
fully complete \fix{at lower voltage levels}. In the next section, we
\changes{experimentally} investigate the effect of increasing latency values as
we vary the supply voltage \changes{on real DRAM chips}.

\subsection{Longer Access Latency Mitigates Voltage-Induced Errors}
\label{ssec:low_volt_latency}

%

To confirm our hypothesis from \ssecref{volt_sensitivity} that a lower supply
voltage requires a longer access latency, we test our DIMMs at supply voltages
below the nominal voltage (1.35V) while incrementally increasing the activation
and precharge latencies \changes{to be} as high as 20ns (2x higher than the
tested latency in \ssecref{volt_sensitivity}). At each supply voltage value, we
call the minimum required activation and precharge latencies that do \emph{not}
exhibit any errors \trcdmin and \trpmin, respectively.

\begin{figure*}[ht!]
    \centering
    \captionsetup[subfigure]{justification=centering}
    {
        \includegraphics[width=\linewidth]{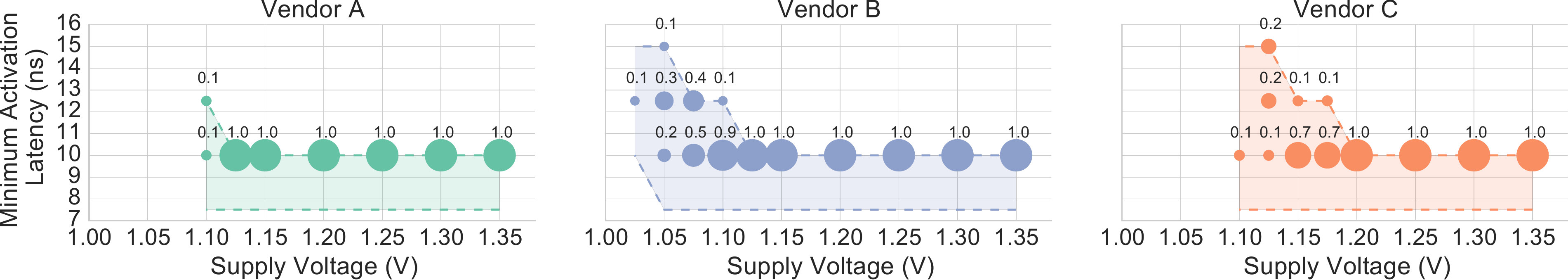}
    }

    {
        \includegraphics[width=\linewidth]{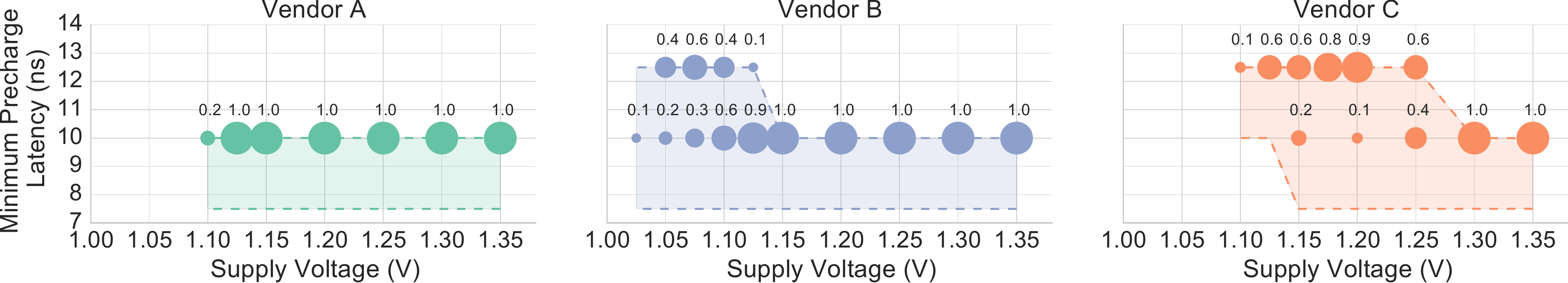}
        \vspace{-0.1in}
    }
    \vspace{-0.15in}
    \caption{Distribution of minimum reliable latency values as the supply
      voltage is decreased for 31 DIMMs. The number above each point indicates
      the fraction of DIMMs that \fixIV{work reliably at the specified voltage
      and latency}. Top row: \trcdmin; Bottom row: \trpmin.}
    \label{fig:fpga_latency_range}
\end{figure*}

\figref{fpga_latency_range} shows the distribution of \trcdmin (top row) and
\trpmin (bottom row) measured for all DIMMs across three vendors as we vary the
supply voltage. Each circle represents a \trcdmin or \trpmin value. A circle's
size indicates the DIMM population size, with bigger circles representing more
DIMMs. The number above each circle indicates the fraction of DIMMs that work
reliably at the specified voltage and latency. Also, we
shade the range of potential \trcdmin and \trpmin values. Since our
infrastructure can adjust the latencies at a granularity of
2.5ns, a \trcdmin or \trpmin value of 10ns is only an approximation of the
minimum value, as the precise \trcdmin or \trpmin falls between 7.5ns and 10ns.
We make three major observations.

First, when the supply voltage falls below \vmin\footnote{\fix{In
    \ssecref{volt_sensitivity}, we define \vmin as the minimum voltage level of
    each DIMM that allows error-free operation. \tabref{modules} in Appendix~\ref{sec:dimm_info}
shows the \vmin value we found for each DIMM.}},
the tested DIMMs show
that an increase of at least 2.5ns is needed for \trcdmin and \trpmin to read
data without errors. For example, some DIMMs require at least a 2.5ns increase
\fix{of} \trcdmin or \trpmin to read data without errors at 1.100V, 1.125V, and 1.25V
from \fix{Vendors}~A, B, and C, respectively. Since our testing platform can only
identify the minimum latency at a granularity of 2.5ns\fixIII{~\cite{hassan-hpca2017}}, we use circuit-level
simulations \fix{to obtain} a more precise latency measurement of \trcdmin and \trpmin
(\fix{which we describe} in the latter part of this section).

Second, DIMMs from different vendors \changes{exhibit} very different behavior
on how much \trcdmin and \trpmin need to increase for reliable operation as
supply voltage falls below \vmin. Compared to other vendors, many more of
Vendor~C's DIMMs require higher \trcdmin and \trpmin to operate at a lower \vdd.
This is particularly the case for the precharge latency, \fixIV{\trpmin}. For
instance, 60\% of Vendor~C's DIMMs require a \trpmin of 12.5ns to read data
without errors at 1.25V, whereas this increase is not necessary at all for
\fix{DIMMs from Vendor~A, which \emph{all} operate reliably at \fixIII{1.15V}.}
This reveals that different vendors may have different circuit architectures or
manufacturing process technologies, \fixIII{which lead to variations in} the
\fix{additional} latency required to compensate for a reduced \vdd in DIMMs.

%

Third, \changes{at very low supply voltages}, not all of the DIMMs have valid
\trcdmin and \trpmin values less than or equal to 20ns that \changes{enable
  error-free operation of the DIMM}. We see that the circle size gets smaller as
the \fixIV{supply} voltage reduces, indicating that the number of DIMMs
\fix{that can operate reliably (even at higher latency) reduces.} For example,
Vendor~A's DIMMs can no longer operate \changes{reliably (\fix{i.e.,}
  error-free)} when the voltage is below 1.1V.
We tested a small subset of DIMMs with latencies of more than 50ns and found
that these very high latencies still do \emph{not} prevent errors from
occurring. We hypothesize that this is because of signal integrity issues on the
channel, causing bits to flip during data transfer at very low supply voltages.

\fix{We correlate our characterization results with} our SPICE simulation results
from \ssecref{volt_sensitivity}, demonstrating that there is a direct
relationship between supply voltage and access latency. This \fix{new
  observation on the trade-off between supply voltage and access latency} is not
discussed or \fix{demonstrated} in prior work \changes{on DRAM voltage
  scaling}~\cite{david-icac2011}, where the access latency (in nanoseconds)
remains \emph{fixed} when performing memory DVFS. \changes{In conclusion, we
  demonstrate both experimentally and in circuit simulations that increasing the
  access latency (i.e., \trcd and \trp) allows us to lower the supply voltage
  while still reliably accessing data without errors.}


\paratitle{Deriving More Precise Access Latency Values}
One limitation of our experiments is that we cannot \emph{precisely} measure the
\emph{exact} \trcdmin and \trpmin values, due to the 2.5ns \fix{minimum latency} granularity of our
experimental framework~\cite{hassan-hpca2017}. Furthermore, supply voltage
is a continuous value, and it would take a prohibitively long time to study
the supply voltage \changes{experimentally} at a finer granularity.
We address these limitations by enriching our experimental results
with circuit-level DRAM SPICE simulations that model a DRAM array
(\fix{see} Appendix~\ref{spice_model} \fixIII{for details of our circuit simulation model}).

The SPICE simulation results highly depend on the specified transistor
parameters (e.g., transistor width). To fit our SPICE results with our
experimental results (for the supply voltage values that we studied
experimentally), we manually adjust the transistor parameters until the
simulated results fit within our \emph{measured} range of latencies.
\figref{spice+fpga_annotate} shows the latencies reported for activation and
precharge \fixIV{operations} using our final SPICE model\fix{,} based on the measured experimental data
for Vendor~B.

\figputHW{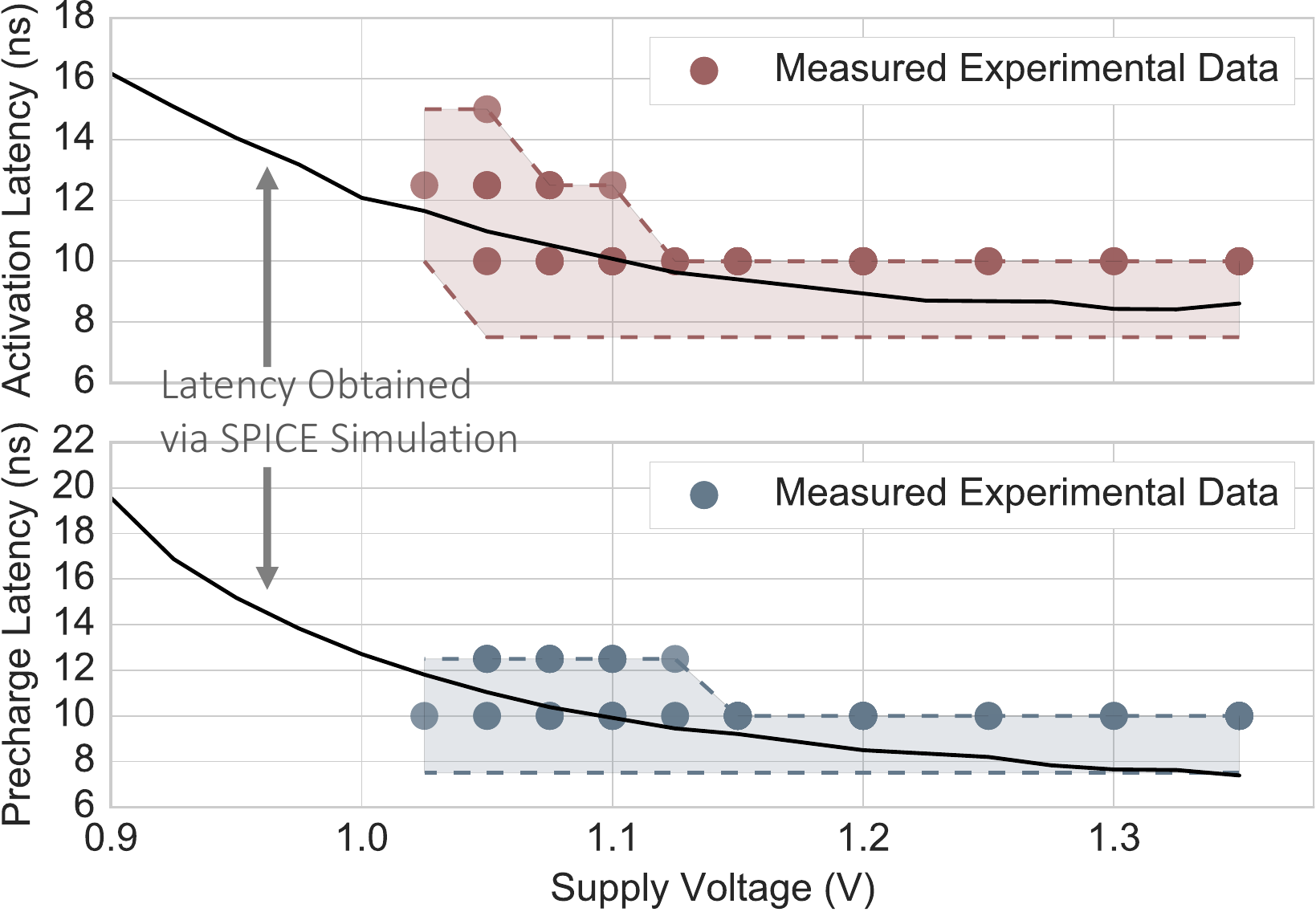}{SPICE simulation results compared with
  \fixIV{experimental} measurements
from 12 DRAM DIMMs for Vendor~B.}

We make \fix{two} major observations. First, \fix{we see} that the SPICE
\fix{simulation} results fit within the \fixIII{range of latencies measured
  during our experimental characterization}, confirming that our \fix{simulated}
circuit behaves close to the real DIMMs. \fix{As a result, our circuit model
  allows us to derive a more precise minimum latency for reliable operation than
  our experimental data.\fixIII{\footnote{\fixIII{The circuit model can further
        serve as a common framework for studying other characteristics of
        DRAM.}}}} Second, DRAM arrays can operate at a wide range of voltage
\fix{values} without \fix{experiencing} errors. This aligns with our hypothesis
that errors at very low supply voltages (e.g., 1V) occur during data transfers
\fix{across the channel rather than during DRAM array \fixIII{operations.}}
Therefore, our SPICE simulations not only validate \fixIV{our observation} that
a lower supply voltage requires longer access latency, but also provide us with
\fixIII{a} more precise \fix{reliable minimum operating latency
  \fixIII{estimate} for a given supply voltage.}




\subsection{Spatial Locality of Errors}
\label{ssec:spatial}

While reducing the supply voltage induces errors when the DRAM latency is not
long enough, we also show that not all DRAM locations experience errors at all
supply voltage levels. To understand the locality of the errors induced by a low
supply voltage, we show the probability of each DRAM row in a DIMM
\fix{experiencing} at least one bit of error across all experiments.
\response{We present results for two \fixIV{representative} DIMMs from two
  different vendors\fix{, as} the observations from these two DIMMs are similar
  to those we make \fixIV{for} the other tested DIMMs. Our results collected
  from \fix{each of the 31} DIMMs are publicly available~\cite{volt-github}.}

\begin{figure}[!h]
    \centering
    \subcaptionbox{DIMM B$_6$ of vendor~B at 1.05V.}[\linewidth][c]
    {
        \includegraphics[scale=1]{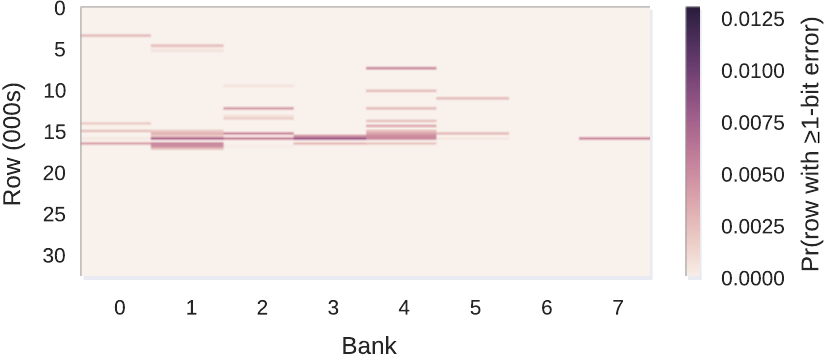}
    }
    \vspace{0.15in}

    \subcaptionbox{DIMM C$_2$ of vendor~C at 1.20V.}[\linewidth][c]
    {
        \includegraphics[scale=1]{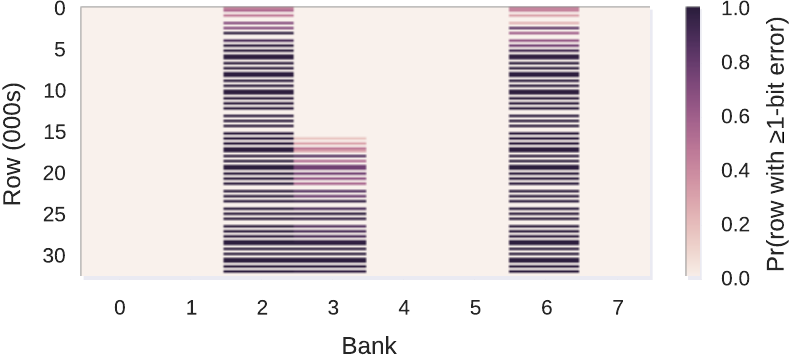}
    }
    \vspace{-0.1in}
    \caption{The probability of error occurrence for two representative
      DIMMs, \fixIV{categorized into different rows and banks}, due to reduced voltage.}

    \label{fig:locality_C}
\end{figure}

\figref{locality_C} shows the probability of each row \fix{experiencing} at
least a one-bit error due to reduced voltage in the two representative DIMMs.
For each DIMM, we choose the supply voltage when errors start appearing (i.e.,
the voltage \fix{level} one step below \vmin), and we do \emph{not} increase the
DRAM access latency \fixIV{(i.e., 10ns for both \trcd and \trp)}. The x-axis
and y-axis indicate the bank number and row number (in thousands), respectively.
Our tested DIMMs are divided into eight banks, and each bank consists of
32K~rows of cells.\footnote{Additional results showing the error locations at
  different voltage \fix{levels} are in Appendix~\ref{spatial}.}

Our main observation is that errors tend to cluster at certain locations. For
our representative DIMMs, we see that errors tend to cluster at certain rows
across multiple banks for Vendor~B. On the contrary, Vendor~C's DIMMs exhibit
errors in certain banks but not in other banks. We hypothesize that the error
concentration can be a result of \myitem{i} manufacturing process variation,
resulting in less robust components at certain locations, as observed in
Vendor~B\fix{'s DIMMs}; or \myitem{ii} architectural design variations in the power delivery
network. However, it is hard to verify our hypotheses without knowing the
specifics of the DRAM circuit design, which is proprietary information that
varies across different DRAM models within and across vendors.

Another implication of the spatial concentration of errors under low voltage is
that \emph{only those regions with errors require a higher access latency to
  read or write data correctly}, whereas error-free regions can be accessed
reliably with the standard latency. \fix{In \ssecref{eval_el}, \fixIII{we
    discuss} and evaluate a technique that exploits this \fixIV{spatial}
  locality \fixV{of errors} to improve system performance.}


\subsection{Density of Errors}
\label{ssec:ecc}

In this section, we investigate the density (i.e., the number) of error bits
that occur within each \emph{data beat} (i.e., the unit of data transfer,
\fixIV{which is 64 bits,} through the data bus) read back from DRAM.
Conventional error-correcting codes (ECC) used in DRAM detect and correct errors
at the granularity of a data beat. For example, SECDED
ECC\fixIII{~\cite{sridharan-asplos2015, luo-dsn2014}} can \fix{correct a
  single-bit error and detect two-bit errors} within a data beat.
\figref{err_density} shows the distribution of data beats that contain no
errors, \fix{a single-bit} error, two-bit errors, \fixVI{or} \fix{more than two
  bits of} errors, under different supply voltages for all DIMMs. These
distributions are collected from \fix{30 rounds of experiments that were tested
  on each of the 31 DIMMs per voltage} level, using 10ns of activation and
precharge latency. \fixIV{A round of experiment refers to a single run of
  Test~\ref{test}, as described in \secref{fpga}, on a specified DIMM.}

\figputHW{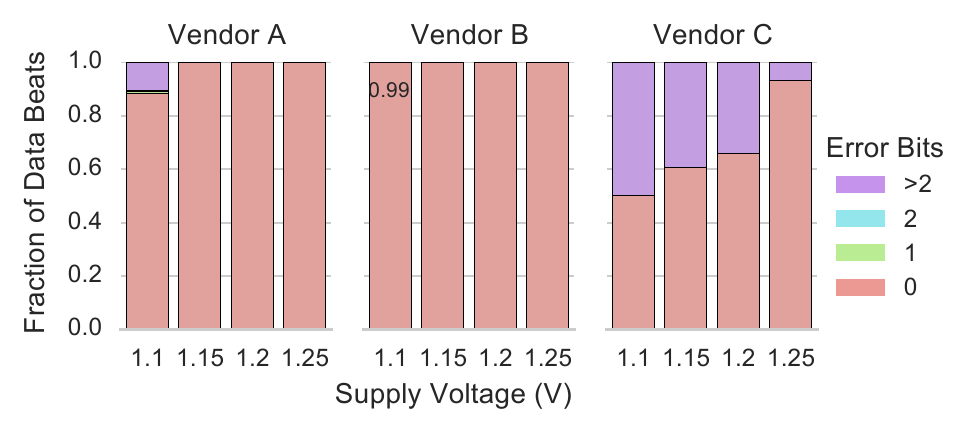}{Distribution of bit \dhlfix{errors} in data beats.}

The results show that lowering the supply voltage increases the fraction of
beats that contain \fix{more than two bits of} errors. There are very few beats
that contain only one or two error bits. This implies that the most
commonly-used \fix{ECC scheme, SECDED, is} unlikely to alleviate errors induced
by a low supply voltage. \dhlfix{\fix{Another ECC mechanism,
    Chipkill\fixIII{~\cite{sridharan-asplos2015, luo-dsn2014}},} protects
  multiple bit failures within a DRAM chip. However, it \fixIII{cannot} correct
  errors in \emph{multiple} DRAM chips.} Instead, \fixIV{we believe that}
increasing the access latency, as shown in \ssecref{low_volt_latency}, is a more
effective way of eliminating errors under low supply voltages.


\subsection{Effect of Temperature}
\label{ssec:temperature}

Temperature is an important external factor that can affect the behavior of
DRAM\fixIII{~\cite{schroeder-sigmetrics2009,el-sayed-sigmetrics2012,
    liu-isca2013, khan-sigmetrics2014, lee-hpca2015, liu-isca2012, kim-isca2014,
    lee-thesis2016}}. Prior works have studied the impact of temperature on
reliability\fixIII{~\cite{schroeder-sigmetrics2009, el-sayed-sigmetrics2012,
    kim-isca2014,kim-thesis2015}},
latency\fixIII{~\cite{lee-hpca2015,chang-sigmetrics2016, chang-thesis2017,
    lee-thesis2016}}, and retention
time~\cite{khan-sigmetrics2014,liu-isca2013,liu-isca2012,qureshi-dsn2015} at the
nominal supply voltage. However, no prior work has studied \dhlfix{the effect of
  temperature on the latency \fix{at which DRAM operates reliably},} as the
supply voltage changes.

To reduce the test time, we test 13~representative DIMMs under a high ambient
temperature of 70$\celsius$ using a closed-loop temperature
controller~\cite{hassan-hpca2017}. \figref{temperature} shows \fix{the} \trcdmin
and \trpmin values of tested DIMMs, categorized by \fix{vendor, at} 20$\celsius$
and 70$\celsius$. \fix{The error bars indicate the minimum and maximum latency
  values across all \fixIII{DIMMs we} tested that are from the same vendor. We
  increase the horizontal spacing between the low and high temperature data
  points at each voltage level to improve readability.}

\figputHW{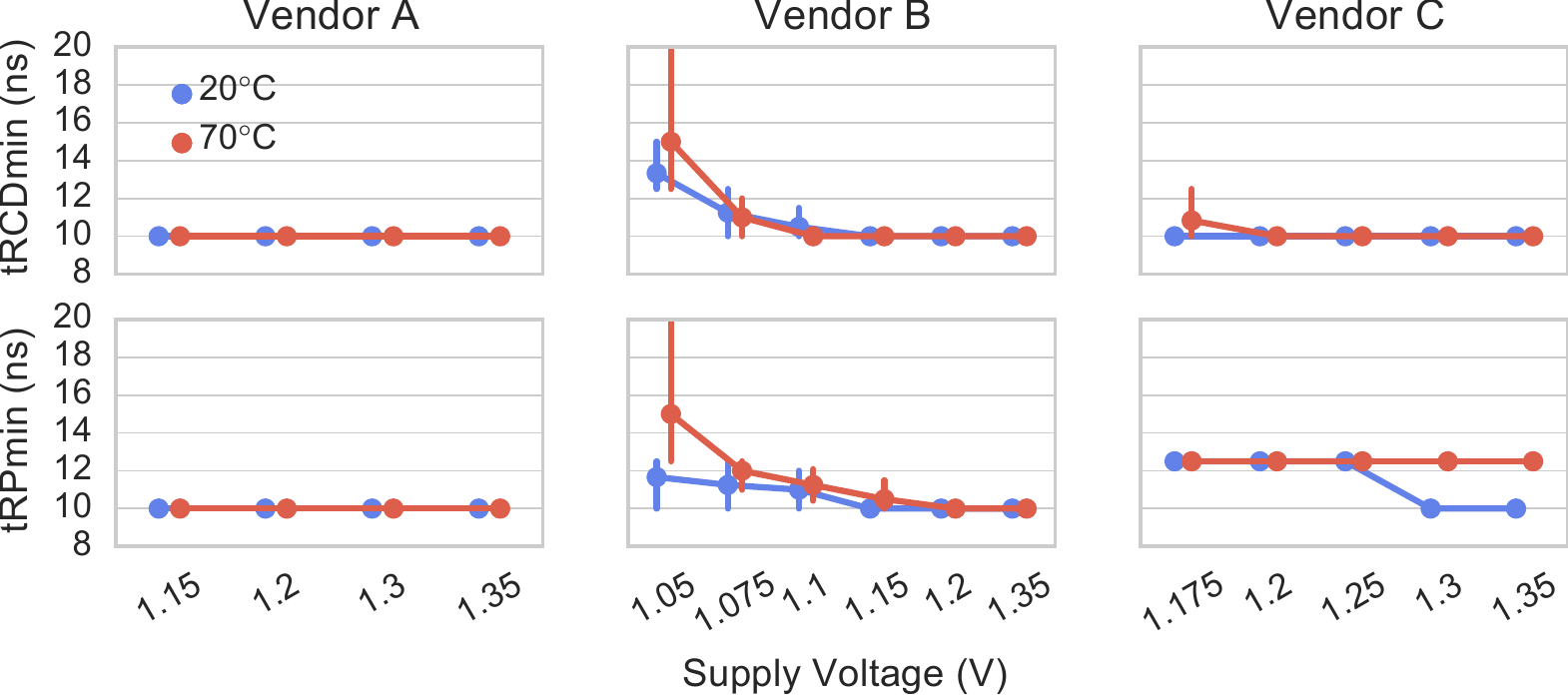}{Effect of high \fix{ambient} temperature (70$\celsius$) on minimum
  reliable operation latency at reduced voltage.}

We make two observations. First, temperature impacts vendors differently. On
Vendor~A's DIMMs, temperature does not have an observable impact on the
\dhlfix{reliable operation} latencies. \fix{Since our platform can test
  latencies with  a step size of only 2.5ns, it is possible that the effect of high
  temperature on the reliable minimum operating latency for Vendor~A's DIMMs
  may be within 2.5ns.}
On the other hand, the temperature effect on latency is measurable on DIMMs from
Vendors~B and C. DIMMs from Vendor~B are not strongly affected by temperature
when the supply voltage is above 1.15V. \fix{The precharge latency for
Vendor~C's DIMMs is affected by high temperature} at supply voltages of 1.35V and
1.30V, \fix{leading to an increase in the minimum} latency from 10ns to 12.5ns.
When the voltage is below 1.25V, the impact \fix{of high temperature on
precharge latency is} not observable, as the precharge
latency already needs to be raised by 2.5ns, to 12.5ns, at 20$\celsius$. Second,
the precharge latency is more sensitive to temperature than the activation
latency. Across all of our tested DIMMs, \trp increases \fix{with high
  temperature} under a greater number of supply voltage \fix{levels}, whereas
\trcd is less likely to be perturbed by temperature.

Since temperature can affect latency behavior under different voltage levels,
techniques that compensate for temperature changes can be used to dynamically
adjust the activation and precharge latencies, as proposed by prior
work~\cite{lee-hpca2015, lee-thesis2016}.


\subsection{Impact on Refresh Rate}
\label{ssec:refresh}

Recall from \ssecref{dram_org} that a DRAM cell uses a capacitor to store data.
The charge in the capacitor leaks over time. To prevent data loss, DRAM
periodically performs an operation called \emph{refresh} to restore the charge
stored in the cells. The frequency of refresh is determined by the amount of
time a cell can retain enough charge without losing information, commonly
referred to as a cell's \emph{retention time}. \fix{For DDR3 DIMMs, the}
\fixIV{worst-case} retention time \fixIV{assumed}
for a DRAM cell is 64ms (or 32ms \fixIII{at temperatures above
  85$\celsius$~\cite{liu-isca2013, liu-isca2012}}). \fixIV{Hence, each cell is
  refreshed every 64ms, which is the \fixV{DRAM-standard} refresh interval.}


\fix{When we reduce the supply voltage of the DRAM array, we expect the
  retention time \fixIV{of a cell}}
to \emph{decrease}, as less charge \fix{is} stored in each cell. This could
\fixIV{potentially}
require a shorter refresh interval (i.e., more frequent refreshes). To
investigate the impact of low supply voltage on retention time, our experiment
writes all 1s to every cell, and reads out the data after a given amount of
retention time, with refresh disabled. We test a total of seven different
retention times (in ms): 64 (the standard time), 128, 256, 512, 1024, 1536, and
2048. We conduct the experiment for \fix{ten}~rounds on every DIMM from all three
vendors. \figref{retention_line_new} shows the average number of \emph{weak}
cells (i.e., cells that experience bit flips due to too much leakage \fixIV{at a
given retention time}) across all
tested DIMMs, for each retention time, under both 20$\celsius$ and 70$\celsius$.
We evaluate three voltage levels, 1.35V, 1.2V, and 1.15V, that allow
\fix{data to be read} reliably with \fix{a} sufficiently long latency. The error bars indicate the 95\%
confidence interval. \fix{We increase the horizontal spacing between the curves
  at each voltage level to improve readability.}


\ignore{
\figref{retention_line} shows two heatmaps indicating the average number of
\emph{weak} cells (i.e., cells that experience bit flips due to too much
leakage) across all 17 DIMMs, for each tested retention time, under both
20$\celsius$ and 70$\celsius$. The x-axis shows four different tested voltage
levels, and the y-axis shows the tested retention time.
A heatmap entry with zero weak cells means that every cell across all the tested
DIMMs fully retains data throughout the given retention time.
}

\figputHW{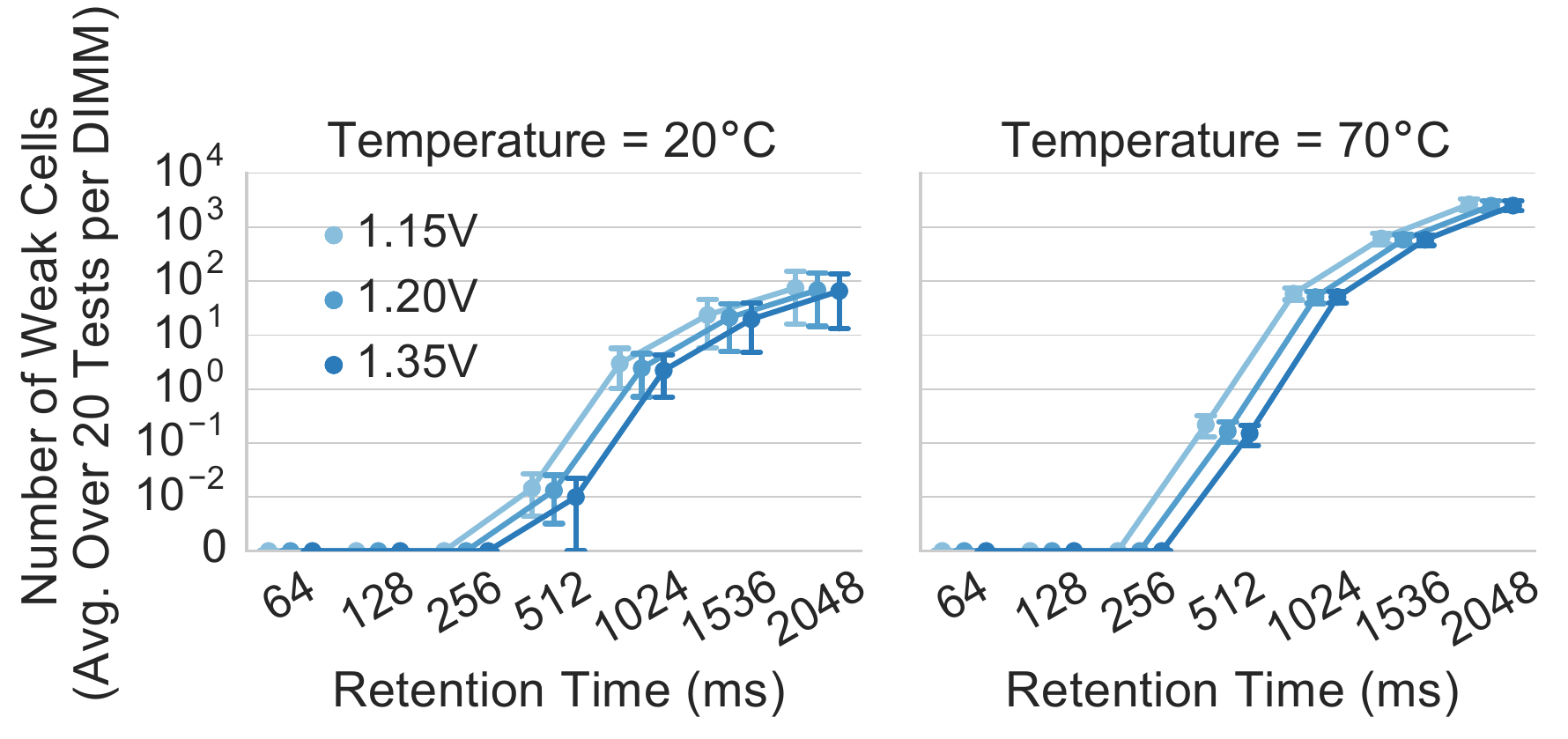}{The number of weak cells that experience errors
  under different retention times as supply voltage varies.}

\changes{Our results show that every DIMM can retain data for at least 256ms
\fix{before requiring a refresh operation}, which is 4x higher than the standard
\fixIV{worst-case} specification. These results align with prior works, which
also experimentally demonstrate that commodity DRAM cells have much higher
retention times than the standard specification of
64ms\fixIII{~\cite{kim-edl2009,liu-isca2013,khan-sigmetrics2014,lee-hpca2015,hassan-hpca2017,
    lee-thesis2016, patel-isca2017, qureshi-dsn2015}}. Even though higher
retention times (i.e., longer times without refresh) reveal more weak cells, the
number of weak cells is still very small, e.g., tens of weak cells out of
billions of cells, on average across all DIMMs \fixIV{at} under 20$\celsius$. Again, this
corresponds closely \fixIII{to} observations from prior \fix{works} showing that
there are relatively few weak cells with low retention time in DRAM
chips, \fixIV{especially at lower temperatures}\fixIII{~\cite{kim-edl2009,liu-isca2013,khan-sigmetrics2014,lee-hpca2015,hassan-hpca2017,qureshi-dsn2015,
    lee-thesis2016, patel-isca2017}}. }

We observe that the effect of \fixIV{the supply voltage} on retention times is
\fixV{\emph{not}} statistically significant. For example, at a 2048ms retention time,
the average \emph{number} of weak cells \fixV{in a DRAM module} increases by only 9~cells (out of a
population of billions of cells) when the supply voltage drops from \fix{1.35V
  (66~weak cells) to 1.15V (75~weak cells)} at $20\celsius$. \fixIV{For the same
  2048ms retention time at $70\celsius$, the average number of weak cells
  increases by only 131~cells when the supply voltage reduces from 1.35V
  (2510~weak cells) to 1.15V (2641~weak cells).}

  \ignore{ the average number of weak cells does not increase, but decreases by
  only 55 cells when the supply voltage drops from 1.35V (2519~weak cells) to
  1.15V (2464~weak cells). Although the average number of weak cells is lower at
  1.15V than 1.35V, the difference is only 2\% of the average number of weak
  cells at 1.35V. This small decrease of weak cells is likely due to a
  well-known phenomenon, called Variable Retention Time
  (VRT)~\cite{khan-sigmetrics2014, khan-dsn2016, khan2016case,
    patel-isca2017,liu-isca2013,qureshi-dsn2015,restle-iedm1992,yaney-iedm1987},
  which causes the retention times of some cells to shift randomly over time.}

\ignore{ For instance, out of Y rounds of tests performed on DIMM X, the number
of weak cells varies from round to round at 2048ms. The average number of weak
cells across all rounds is X with a maximum and minimum of x and z,
respectively. This shows that the retention times of cells change over time, as
experimentally demonstrated in prior works~\cite{liu-isca2013, kim-edl2009}. }


\fix{When we lower the supply voltage, we do not observe \emph{any} weak cells
until a retention time of 512ms, which is 8x the standard refresh interval of
64ms. Therefore, we conclude that using a reduced supply voltage does not
require any changes to the standard refresh interval \fixIII{at 20$\celsius$ and
  70$\celsius$ ambient temperature}.}


\subsection{Summary}
\label{ssec:char_summary}

\changes{We have presented extensive characterization results and analyses on
DRAM chip \fix{latency, reliability, and data retention time} behavior under
various supply voltage levels. We summarize our findings in six key points.
First, DRAM reliability worsens \fixIV{(i.e., more errors start appearing)} as
we reduce the supply voltage below \vmin. Second, we discover that
voltage-induced errors occur mainly because, at low supply voltages, the DRAM
access latency is no longer sufficient to allow the fundamental DRAM operations
to complete. Third, via both experiments on real DRAM chips and SPICE
simulations, we show that increasing the latency of activation, restoration, and
precharge \fix{operations in DRAM} can mitigate errors under low supply voltage
levels until a certain voltage level. Fourth, we show that voltage-induced
errors \fixIV{exhibit} strong spatial locality \fixIV{in a DRAM chip},
clustering at certain locations \fixIV{(i.e., \fixV{certain} banks and rows)}. Fifth,
temperature affects the reliable access latency at low supply voltage levels
\fix{and the effect is very vendor-dependent}. Sixth, we find that reducing the
supply voltage does \emph{not} require increasing the standard DRAM refresh
\fix{rate} for reliable \fixIII{operation below 70$\celsius$}.}


\section{Voltron: Reducing DRAM Energy Without Sacrificing Memory Throughput}
\label{sec:varray}

\changes{Based on the extensive understanding we developed on reduced-voltage
operation of real DRAM chips in \secref{dram_exp}, we propose a new mechanism
called \emph{\voltron}, which reduces DRAM energy without sacrificing memory
throughput. \voltron exploits the fundamental observation that reducing the
supply voltage to DRAM requires increasing the latency of the three DRAM
operations in order to prevent errors. Using this observation, the key idea of
Voltron is to use a performance model to determine \fix{by how much to reduce
  the DRAM supply voltage}, without introducing errors and without exceeding a
user-specified threshold for performance loss. \voltron consists of two main
components: \myitem{i} \emph{array voltage scaling}, a hardware mechanism that
leverages our \fix{experimental observations} to scale \emph{only} the voltage supplied
to the DRAM array; and \myitem{ii} \emph{performance-aware voltage control},
\fixIII{a software} mechanism\footnote{\fixIV{Note that this mechanism can also
    be implemented in hardware, or as a cooperative hardware/software
    mechanism.}} that automatically chooses the minimum DRAM array voltage that
meets a \fixIV{user-specified} performance target.}


\subsection{Array Voltage Scaling}
\label{ssec:avs}

\changes{ As we discussed in \secref{dram_power}, the DRAM supply voltage to the
peripheral circuitry determines the maximum operating frequency. If we reduce
the supply voltage directly, the frequency needs to be lowered as well.
As more applications become more sensitive to memory bandwidth, reducing DRAM
frequency can result in a substantial performance loss due to lower memory
throughput. In particular, we find that reducing the DRAM frequency from 1600
MT/s to 1066 MT/s significantly degrades performance of our evaluated
\fix{memory-intensive} applications by 16.1\%. Therefore, the design challenge
of \voltron is to reduce \fixIII{the} DRAM supply voltage \emph{without}
\fixIII{changing the} DRAM frequency.

To address this challenge, the key idea of Voltron's first component,
\emph{array voltage scaling}, is to reduce the voltage supplied to the
\emph{DRAM array} (\varr) \emph{without changing the voltage supplied to the
  peripheral circuitry}, thereby allowing the DRAM channel to maintain a high
frequency while reducing the power consumption of the DRAM array. To prevent
errors from occurring during reduced-voltage operation, \voltron increases the
latency of the three DRAM operations (activation, restoration, and precharge) in
every DRAM bank based on our observations in \secref{dram_exp}.


By reducing \varr, we effectively reduce \myitem{i} the dynamic DRAM power on
activate, precharge, and refresh \fixIV{operations}; and \myitem{ii} the portion
of the static power that comes from the DRAM array. These power components
decrease \emph{quadratically} with the square of \fix{the} array voltage
reduction in a modern DRAM chip~\cite{keeth-dram-tutorial, baker-dram}. The
trade-off is that reducing \varr requires \fixIV{increasing} the latency of the
three DRAM operations, \fixIV{for reliable operation}, \fix{thereby} leading to
some system performance degradation, which we quantify in our evaluation
(\secref{eval}). }



\subsection{Performance-Aware Voltage Control}
\label{ssec:pavc}


Array voltage scaling provides system users with the ability to decrease \varr
to reduce DRAM power. Employing a lower \varr provides greater power savings,
but at the cost of longer DRAM access latency, which leads to larger performance
degradation. This trade-off varies widely across different applications, as each
application has a different tolerance to the increased memory latency. This
raises the question of how to pick a ``suitable'' array voltage level for
different applications as a system user or designer. For this work, we say that
an array voltage level is suitable if it does not degrade system performance by
more than a user-specified threshold. Our goal is to provide a simple \fixIII{technique}
that can automatically select \fix{a} suitable \varr \fix{value} for
different applications. \changes{To this end, we propose
  \emph{performance-aware voltage control}, a \fix{power-performance} management
  policy that selects a minimum \varr that satisfies a desired performance
  constraint. The key observation is that an application's performance loss (due
  to increased memory latency) scales linearly with the application's memory
  \fix{demand \fixIII{(e.g., memory intensity)}.}
  Based on this \fixIV{empirical} observation \fixIV{we make}, we build a
  \emph{performance loss predictor} that leverages a linear model to predict an
  application's performance loss based on its characteristics at runtime. Using
  the performance loss predictor, Voltron finds a \varr that \fix{can keep the}
  predicted performance within a user-specified target at runtime. }

%

\begin{figure}[!h]
    \centering
    \captionsetup[subfigure]{justification=centering}
    \subcaptionbox{\fixV{Performance loss vs.\ last-level cache
        MPKI.}\label{fig:mpki}}[\linewidth]
    {
        \includegraphics[width=\linewidth]{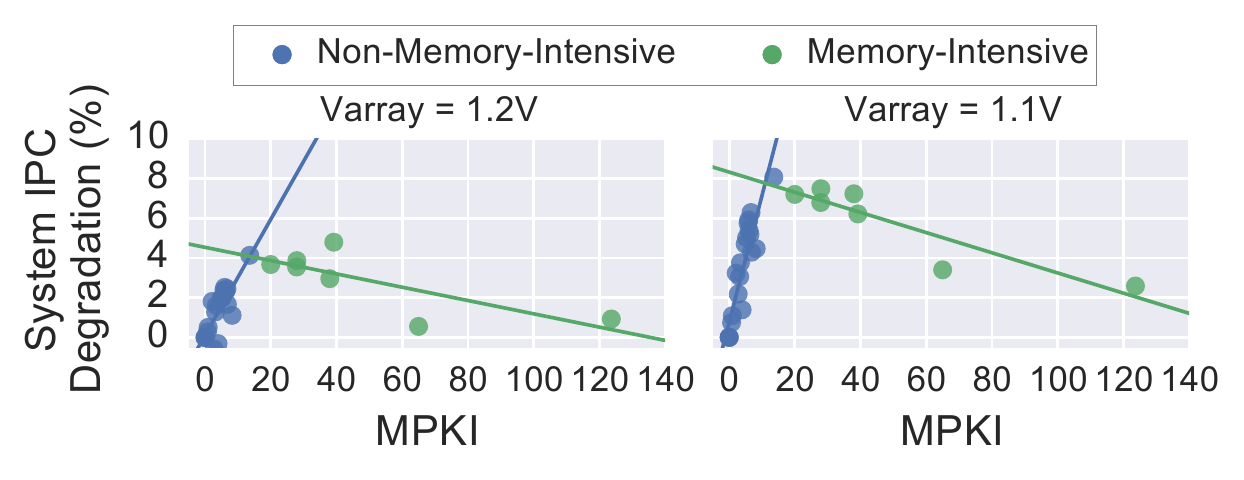}
    }

    \vspace{0.15in}
    \subcaptionbox{\fixV{Performance loss vs.\ memory stall time fraction.}\label{fig:memstall}}[\linewidth]
    {
        \includegraphics[width=\linewidth]{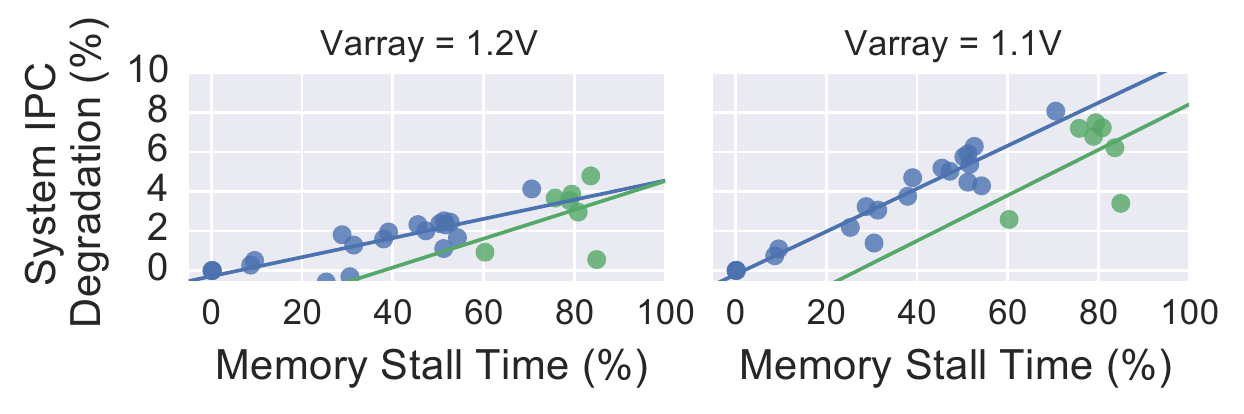}
    }
    \vspace{-0.15in}

    \caption{Relationship between performance loss (due to increased
    memory latency) and applications' characteristics: \fixIV{MPKI (a) and memory
    stall time fraction (b)}. Each data point
    represents a single application.}
    \label{fig:voltron_mot}
\end{figure}

\paratitle{Key Observation} We find that an application's performance loss due
to higher latency has a strong linear relationship with its memory \fix{demand
  \fixIII{(e.g., memory intensity)}}. \figref{voltron_mot} shows the
relationship between the performance loss of each application \fix{(due to
  reduced voltage)} and its memory \fix{demand} under two different
\fix{reduced-voltage} values (see \ssecref{meth} for our methodology). Each data
point represents a single application. \fixIV{\figref{mpki}} shows each
application's performance loss versus its \emph{memory intensity}, expressed
using the commonly-used metric MPKI (last-level cache misses per
kilo-instruction). \fixIV{\figref{memstall}} shows \fix{each application's
  performance loss versus its \emph{memory stall time}}, the fraction of
execution time for which memory requests stall the CPU's \fixIV{instruction
  window}  (i.e., reorder buffer). In \fixIV{\figref{mpki}}, we
see that the performance loss is a \fix{\emph{piecewise linear function}} based
on the MPKI. The observation that an application's \emph{sensitivity to memory
  latency} is correlated with MPKI has also been \fixV{made} and utilized by prior
\fixIV{works~\cite{kim-hpca2010, kim-micro2010, mutlu-isca2008, zheng-icpp2008,
    mutlu-micro2007, muralidhara-micro2011, das-hpca2013, das-micro2009,
    usui-taco2016, zhao-micro2014, das-isca2010}.}

When an application is \fix{\emph{not} memory-intensive} (i.e., \fixIV{has an} $MPKI < 15$), its performance
loss grows linearly with MPKI, becoming \emph{more sensitive} to memory latency.
\fix{ Latency-sensitive applications spend most of their time \fixIII{performing computation at the CPU cores} and
  issue memory requests infrequently. As a result, increasing the number of
  memory requests causes more stall cycles in the CPU. }

On the other hand, the performance of \fix{memory-intensive} applications (i.e.,
\fixIV{those with} $MPKI \ge 15$) is \emph{less sensitive} to memory latency as
the MPKI grows. This is because \fix{memory-intensive} applications experience
frequent cache misses and spend a large portion of their time waiting on pending
memory requests. As a result, their rate of progress is significantly affected
by the memory bandwidth, \fix{and therefore they are more} \emph{sensitive to
  memory throughput} instead of latency. \fix{With more \fixIII{outstanding}
  memory requests (i.e., higher MPKI), the memory \fixIII{system} is more likely
  to service them in parallel, leading to more \emph{memory-level
    parallelism}\fixIII{~\cite{kim-micro2010,mutlu-isca2008,glew-asploswac98,
      mutlu-hpca2003, mutlu-ieeemicro2006, lee-micro2009}}. Therefore, improved
  memory-level parallelism enables applications to tolerate \fixIII{higher
    latencies} more easily.}


\fixIV{\figref{memstall}} shows that an application's performance loss increases
with its \fixIV{instruction window (reorder buffer)} \emph{stall time
  \fixIV{fraction}} due to memory requests for both \fix{memory-intensive and
  non-memory-intensive} applications. \fix{A stalled \fixIV{instruction window}
  prevents the CPU from fetching or dispatching new
  instructions~\cite{mutlu-hpca2003}, thereby degrading the running
  application's performance.} This observation has also been \fixIII{made} and
utilized by prior
\fixV{works~\cite{ghose-isca2013,mutlu-ieeemicro2006,mutlu-isca2005,mutlu-hpca2003}}.

\paratitle{Performance Loss Predictor} Based on the observed linear
relationships \fixIV{between performance loss vs. MPKI and memory stall time
  fraction}, we use \emph{ordinary least squares (OLS)} regression to develop
a piecewise linear model for each application \fix{that can serve} as the performance loss
predictor for Voltron. Equation~\ref{eq:plm} shows the model, which takes the
following inputs: memory latency ($Latency=tRAS+tRP$), the application's MPKI, and
its memory stall time \fixIV{fraction}.

\vspace{-0.15in}
\begin{align}
  \footnotesize
    \mathrm{PredictedLoss}_{i} =
    \begin{cases}
        \label{eq:plm}
        \!\begin{aligned}
        &\alpha_{1} + \beta_{1}  \mathrm{Latency}_{i} + \beta_{2}
        \mathrm{App.MPKI}_{i} \\&+ \beta_{3}  \mathrm{App.StallTimeFraction}_{i} \\[1.5ex]
        \end{aligned} &\text{if } MPKI < 15 \\
        \!\begin{aligned}
        &\alpha_{2} + \beta_{4}  \mathrm{Latency}_{i} + \beta_{5}
        \mathrm{App.MPKI}_{i} \\&+ \beta_{6}  \mathrm{App.StallTimeFraction}_{i}
        \end{aligned} &\text{otherwise}
    \end{cases}
\end{align}
\vspace{-0.1in}
\begin{table}[h]
    \renewcommand{\arraystretch}{0.9}
    \centering
    \setlength{\tabcolsep}{.45em}
    \begin{tabular}{llll|llll}
        \toprule
        $\alpha_{1}$ & $\beta_{1}$ & $\beta_{2}$ & $\beta_{3}$  & $\alpha_{2}$ &
        $\beta_{4}$ & $\beta_{5}$ & $\beta_{6}$ \\
        \midrule
        -30.09 & 0.59 & 0.01 & 19.24 & -50.04 & 1.05 & -0.01 & 15.27 \\
        \bottomrule
    \end{tabular}
\end{table}

$\mathrm{PredictedLoss}_{i}$ is the predicted performance loss for the
application. The subscript $i$ refers to each data sample,
\fixIV{which describes a particular application's characteristics (MPKI and
  memory stall time fraction) and the memory latency associated with the
  selected voltage level.} \fixIV{To generate the data samples, we run a total
  of 27 workloads across 8 different voltage levels that range from 1.35V to
  \fixVI{0.90V}, at a 50mV step (see \ssecref{meth} for our methodology). In total, we
  generate 216 data samples for finding the coefficients (i.e., $\alpha$ and
  $\beta$ values) in our model. To avoid overfitting the model, we use
  \fixIII{the \emph{scikit-learn} machine learning toolkit}~\cite{scikit} to
  perform cross-validation, which randomly splits the data samples into a
  training set (151 samples) and a test set (65 samples)}. To assess the fit of
the model, we use a common metric, root-mean-square error (RMSE), which is 2.8
and 2.5 for the low-MPKI and high-MPKI pieces of the model, respectively.
Furthermore, we calculate the R\textsuperscript{2} value to be 0.75 and 0.90 for
the low-MPKI and high-MPKI models, respectively. \fixV{Therefore, the RMSE and
R\textsuperscript{2} metrics
  indicate that our model provides high accuracy for predicting the performance
  loss of applications under different \varr values.}

\paratitle{Array Voltage Selection} Using the performance loss predictor,
\voltron selects the minimum value of \varr that satisfies the given user target
for performance loss. \fix{Algorithm~\ref{voltron} depicts the array voltage
  selection component of Voltron.} The voltage selection \fixIII{algorithm is
  executed} at periodic intervals throughout the runtime of an application.
During each interval, the application's memory \fix{demand} is profiled. At the
end of an interval, \voltron uses the profile to iteratively compare the
\fix{performance loss} target to the predicted performance loss incurred by each
voltage level, starting from a minimum value of 0.90V. Then, \voltron selects
the minimum \varr that \fixIV{does not exceed} the performance loss target and
\fixIII{uses} this selected \varr\fixIII{\ as the DRAM supply voltage} in the
subsequent interval. In our evaluation, we provide \voltron with a total of 10
voltage levels (every 0.05V step from 0.90V to 1.35V) for selection.

\fix{

\begin{afloat}[h]

\algnewcommand\algorithmicto{\textbf{to}}
\algrenewtext{For}[3]{\algorithmicfor\ #1 $\gets$ #2 \algorithmicto\ #3 \algorithmicdo}

\algrenewcommand\algorithmicfunction{}
\algrenewcommand\algorithmicdo{}
\algrenewcommand\algorithmicindent{1.2em}
\algrenewcommand\alglinenumber[1]{\footnotesize\texttt{#1}}
\small

\begin{algorithmic}[1]
\Function{SelectArrayVoltage}{$target\_loss$}

\ForEach {$interval$} \Comment{Enter at the end of an interval}
  \State $profile$ = GetMemoryProfile()
  \State $Next$\varr = 1.35
  \For{\varr}{0.9}{1.3} \Comment{\fixV{Search for the smallest \varr that satisfies the
    performance loss target}}
    \State $predicted\_loss$ = Predict(Latency(\varr), $profile$.MPKI,
    $profile$.StallTime) \Comment{\fixV{Predict performance loss}}
    \If{$predicted\_loss \leq target\_loss$} \Comment{\fixV{Compare the
        predicted loss to the target}}
    \State $Next$\varr = \varr \Comment{\fixV{Use the current \varr for the next interval}}
        \State \fixIII{\textbf{break}}
    \EndIf
  \EndFor
  \State ApplyVoltage($Next$\varr) \Comment{\fixV{Apply the new \varr for the
      next interval}}
\EndFor

\EndFunction
\end{algorithmic}
    \caption{Array Voltage Selection}
\label{voltron}
\end{afloat}

}


\subsection{Implementation}

\voltron's two components require modest modifications \fix{to} different parts of the
system. In order to support array voltage scaling, \voltron requires minor
changes to the power delivery network of DIMMs, as commercially-available DIMMs
currently \fixIII{use} a single supply voltage for both the DRAM array and \fixIII{the} peripheral
circuitry. Note that this supply voltage goes through \emph{separate} power
pins: $V_{DD}$ and $V_{DDQ}$ for the DRAM array and peripheral circuitry,
respectively, on a modern DRAM chip~\cite{micronDDR3L_2Gb}. Therefore, to enable
independent voltage adjustment, we propose to partition the power delivery
network on the DIMM into two domains: one domain to supply only the DRAM array
($V_{DD}$) and the other domain to supply only the peripheral circuitry
($V_{DDQ}$). 

Performance-aware voltage control requires \myitem{i} performance monitoring
hardware that records the MPKI and memory stall time of each application; and
\myitem{ii} a control algorithm block, which predicts the performance loss at
different \varr values and accordingly selects the smallest acceptable \varr.
\voltron utilizes the performance counters that exist in most modern CPUs to
perform performance monitoring, thus requiring no additional hardware overhead.
\voltron reads these counter values and feeds them into the \fix{array voltage
  selection algorithm, which is} implemented in the system software layer.
\fix{Although reading the performance monitors has a small amount of software
  overhead, we believe the overhead is negligible because we do so only at the
  end of each interval \fixV{(i.e., every four million cycles in most of our
    evaluations; see sensitivity studies in \ssecref{interval})}.}

Voltron periodically executes this performance-aware voltage control
\fix{mechanism} during the runtime of the target application. During each time
interval, Voltron monitors the application's behavior through hardware counters.
At the end of an interval, the \fixIII{system software} \fix{executes the array
  voltage selection algorithm} to select the predicted \varr and
\fixIV{accordingly adjust} the timing parameters stored in the memory controller
for activation, restoration, and precharge. \fix{Note that there could be other
  (e.g., completely hardware-based) implementations of Voltron. We leave a
  detailed explanation of different implementations to future work.}

%
%
%


\section{System Evaluation}
\label{sec:eval}

In this section, we evaluate the system-level \fixIV{performance and energy}
impact of \voltron.  We present our evaluation methodology in \ssecref{meth}.
Next, we study the energy savings and performance loss when we use array voltage
scaling without any control (\ssecref{scaling_eval}).  We study how
performance-aware voltage control delivers overall system energy reduction with
only a modest \fix{amount of} performance loss
(\fixIII{Sections~\ref{ssec:pavc_eval} and \ref{ssec:energy_breakdown}}). We
then evaluate \fix{an enhanced version of} \voltron\fix{, which exploits}
spatial error locality (\ssecref{eval_el}). Finally,
Sections~\ref{ssec:eval_hetero} to \ref{ssec:interval} present sensitivity
studies of \voltron\fix{\ to various system and algorithm parameters}.

\subsection{Methodology}
\label{ssec:meth}

We evaluate \voltron using Ramulator~\cite{kim-cal2015}, a detailed and
\fixV{and cycle-accurate}
  \fix{open-source DRAM simulator~\cite{ram-github}}, integrated with a
  multi-core performance simulator. We model a low-power mobile system that
  consists of 4 ARM cores and DDR3L DRAM. \tabref{sys-config} shows our system
  parameters. Such a system resembles existing commodity devices, such as the
  Google Chromebook~\cite{chromebook} \fix{or the NVIDIA} SHIELD
  tablet~\cite{shield}. To model the energy consumption, we use
  McPAT~\cite{mcpat:micro} for the processor and DRAMPower~\cite{drampower} for
  the DRAM-based memory system. \fix{We open-source the code of Voltron
    \cite{volt-github}.}

\begin{table}[h]
  \renewcommand{\arraystretch}{0.9}
  \small
  \centering
    \setlength{\tabcolsep}{.6em}
    \begin{tabular}{ll}
        \toprule
        \multirow{2}{*}{Processor} & 4 ARM Cortex-A9 cores~\cite{arma9}, 2GHz, \\
        & 192-entry instruction window \\
        \midrule
        Cache & L1: 64KB/core, L2: 512KB/core, L3: 2MB shared \\
        \midrule
        Memory & \multirow{2}{*}{64/64-entry read/write request queue,
            FR-FCFS~\cite{rixner-isca2000,zuravleff-patent}} \\
        Controller & \\
        \midrule
        \multirow{2}{*}{DRAM} & DDR3L-1600~\cite{jedec-ddr3l}\\
        & 2 channels (1 rank and 8 banks per channel) \\
        \bottomrule
    \end{tabular}
  \caption{Evaluated system configuration.}
  \label{tab:sys-config}
\end{table}


\tabref{varr_latency} lists the latency values we evaluate for each DRAM array
voltage (\varr). The latency values are obtained from our SPICE model using data
from real devices (\ssecref{low_volt_latency}), \fix{which is available
  online~\cite{volt-github}}.\footnote{\fixIV{In this work, we do not have
  experimental data on the restoration latency (\tras) under reduced-voltage
  operation. This is because our reduced-voltage tests access cache lines
  sequentially from each DRAM row, and \tras overlaps with the latency of
  reading all of the cache lines from the row.  Instead of designing a separate
  test to measure \tras, we use our circuit simulation model
  (\ssecref{low_volt_latency}) to derive \tras values for reliable operation
  under different voltage levels. We leave the thorough experimental evaluation
  of \tras under reduced-voltage operation to future work.}} To account for
  manufacturing process variation,
we conservatively add in the same latency guardband (i.e., 38\%) used by
manufacturers at the nominal voltage level of 1.35V to each of our \fix{latency
  values}. We then round up each latency \fix{value} to the nearest clock cycle
time (i.e., 1.25ns).

\begin{table}[h]
  \small
  \centering
    \setlength{\tabcolsep}{.5em}
    \begin{tabular}{ll|ll}
        \toprule
        \varr & tRCD - tRP - tRAS (ns) & \varr & tRCD - tRP - tRAS (ns) \\
        \midrule

        1.35 & 13.75 - 13.75 - 36.25 &  1.10 & 15.00 - 16.25 - 40.00 \\
        1.30 & 13.75 - 13.75 - 36.25 &  1.05 & 16.25 - 17.50 - 41.25 \\
        1.25 & 13.75 - 15.00 - 36.25 &  1.00 & 17.50 - 18.75 - 45.00 \\
        1.20 & 13.75 - 15.00 - 37.50 &  0.95 & 18.75 - 21.25 - 48.75 \\
        1.15 & 15.00 - 15.00 - 37.50 &  0.90 & 21.25 - 26.25 - 52.50 \\
        \bottomrule
    \end{tabular}
  \caption{DRAM latency \fixIV{required for correct operation} for each evaluated \varr.}
  \label{tab:varr_latency}
\end{table}


\paratitle{Workloads} We evaluate 27~benchmarks from SPEC
CPU2006~\cite{spec2006} and YCSB~\cite{cooper-socc2010}, as shown in
\tabref{workload_list} \fix{along with each benchmark's \fixIII{L3 cache} MPKI,
  i.e., memory intensity}. \fix{We use the 27~benchmarks to form
  \emph{homogeneous} and \emph{heterogeneous} \fixIV{multiprogrammed}
  workloads.} \fix{For each \emph{homogeneous workload}, we replicate one of our
  benchmarks} by running one copy on each core to form a four-core
\fixIV{multiprogrammed} workload, as done in many past works \fixIII{that
  evaluate multi-core system performance}~\cite{lee-hpca2015,
  chang-sigmetrics2016, lee-hpca2013, nair-isca2013, nair-micro2014,
  shafiee-hpca2014, shevgoor-micro2013, chatterjee-micro2012}. \fixIV{Evaluating
  homogeneous workloads enables easier analysis and understanding of the
  system.} \fix{For each \emph{heterogeneous workload}, we combine four
  \emph{different} benchmarks \fix{to create a four-core workload}. We
  categorize the heterogeneous workloads by varying the fraction of
  \fix{memory-intensive} benchmarks in each workload (0\%, 25\%, 50\%, 75\%, and
  100\%). Each category consists of 10~workloads, resulting in a total of
  50~workloads across all categories.} Our simulation executes at least
500~million instructions on each core. \fix{We calculate system energy as} the
product of the average dissipated power (from both CPU and DRAM) and the
workload runtime. We measure system performance with the commonly-used
\emph{weighted speedup} (WS) metric~\cite{snavely-asplos2000}, which is a
measure of job throughput on a multi-core system~\cite{eyerman-ieeemicro2008}.


\begin{table}[h]
  \renewcommand{\arraystretch}{0.92}
  \centering
    \setlength{\tabcolsep}{.25em}
    \begin{tabular}{clrc|clrc}
        \toprule
        Number & Name & \fixV{L3} MPKI & & Number & Name & L3 MPKI \\
        \midrule
        0 & YCSB-a    & 6.66 & &  14 & h264ref     & 2.14   \\
        1 & YCSB-b    & 5.95 & &  15 & hmmer       & 6.33   \\
        2 & YCSB-c    & 5.74 & &  16 & libquantum  & 37.95  \\
        3 & YCSB-d    & 5.30 & &  17 & mcf         & 123.65 \\
        4 & YCSB-e    & 6.07 & &  18 & milc     & 27.91 \\
        5 & astar     & 3.43 & &  19 & namd     & 2.76 \\
        6 & bwaves    & 19.97& &  20 & omnetpp  & 27.87\\
        7 & bzip2     & 8.23 & &  21 & perlbench& 0.95 \\
        8 & cactusADM & 6.79 & &  22 & povray   & 0.01 \\
        9  & calculix    & 0.01  & & 23 & sjeng    & 0.73 \\
        10 & gamess      & 0.01  & & 24 & soplex   & 64.98\\
        11 & gcc         & 3.20  & & 25 & sphinx3  & 13.59\\
        12 & GemsFDTD    & 39.17 & & 26 & zeusmp   & 4.88 \\
        13 & gobmk       & 3.94  & & & & \\

        \bottomrule
    \end{tabular}

  \caption{Evaluated benchmarks with their \fixIII{respective} L3 MPKI values.}
  \label{tab:workload_list}
\end{table}




\ignore{
\subsection{Dynamic DRAM Energy Results}

Since higher DRAM access latency can potentially increase the energy
consumption, we perform a first order study to evaluate the \emph{dynamic}
energy consumption of the three fundamental access operations that we
demonstrate to be affected by \varr: \myitem{i} activation, \myitem{ii}
precharge, and \myitem{iii} restoration.

\figref{spice_energy_savings} shows the normalized dynamic energy consumption of
the three operations as voltage varies. As \varr lowers, the power consumption
of these operations reduces with high latency. We calculate their energy by
multiplying their power consumption with its latency at each \varr point. The
results show that even though latency increases, the power reduction still saves
\emph{dynamic} energy per operation. However, the dynamic energy savings have
diminished returns when \varr reduces below 1.1V since the latency is increasing
at an exponential rate (shown previously in Eq.~\ref{exp_decay},
\ssecref{spice}). For precharge, the dynamic energy of precharge starts to give
negative returns on energy savings. These results show that decreasing \varr
directly reduces the energy consumption of DRAM operations, but lower \varr does
not necessarily provide more energy savings. Now, we evaluate the system
performance and energy savings by taking into account other dynamic and static
energy consumption in both the DRAM and CPU.

\begin{figure}[!h]
    \centering
    \captionsetup[subfigure]{justification=centering}[\linewidth]
    {
        \includegraphics[width=\linewidth]{plots/spice_dram_op_energy_over_vdd}
        \vspace{-0.1in}
    }
    \caption{Dynamic energy of DRAM operations as \varr varies.}
    \label{fig:spice_energy_savings}
\end{figure}

}

\subsection{Impact of Array Voltage Scaling}
\label{ssec:scaling_eval}

In this section, we evaluate how array voltage scaling (\ssecref{avs}) affects
the system energy consumption and application performance \fix{of our homogeneous workloads} at different \varr
values. We split our discussion into two parts: the results for \fix{memory-intensive}
workloads (i.e., applications where MPKI \fix{$\geq$} 15 for each core), and the results
for \fix{non-memory-intensive} workloads.

\begin{figure}[ht]
    \centering
    \captionsetup[subfigure]{justification=centering}
    {
        \includegraphics[width=\linewidth]{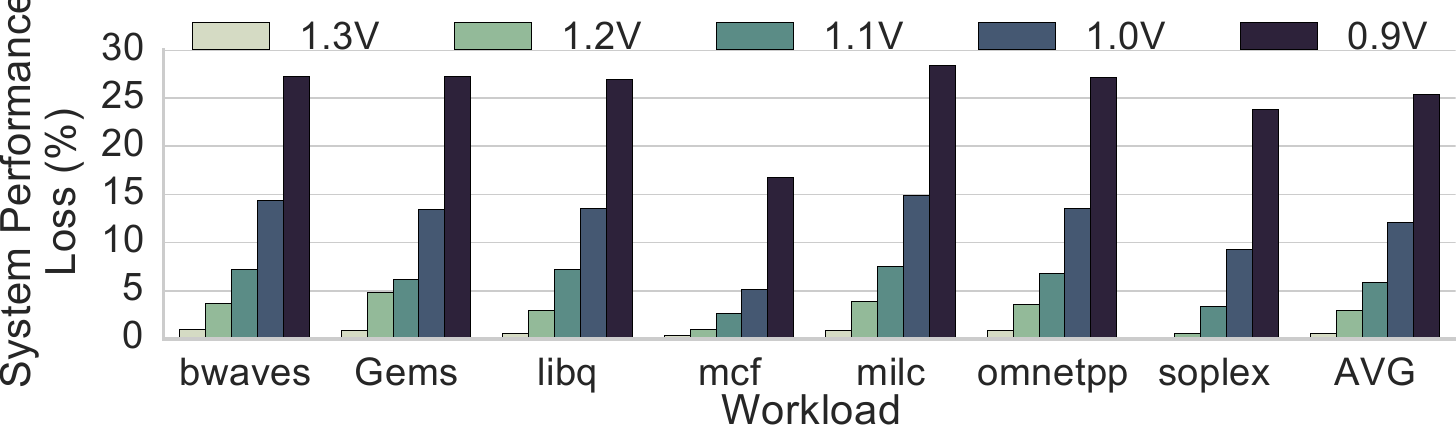}
        \vspace{0.1in}
    }

    {
        \includegraphics[width=\linewidth]{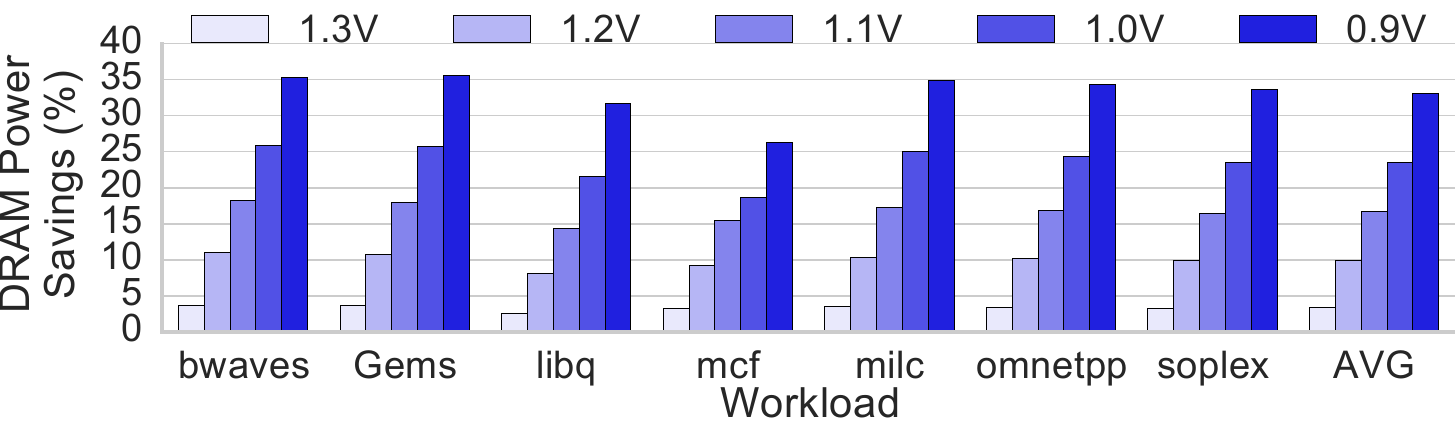}
        \vspace{0.1in}
    }

    {
        \includegraphics[width=\linewidth]{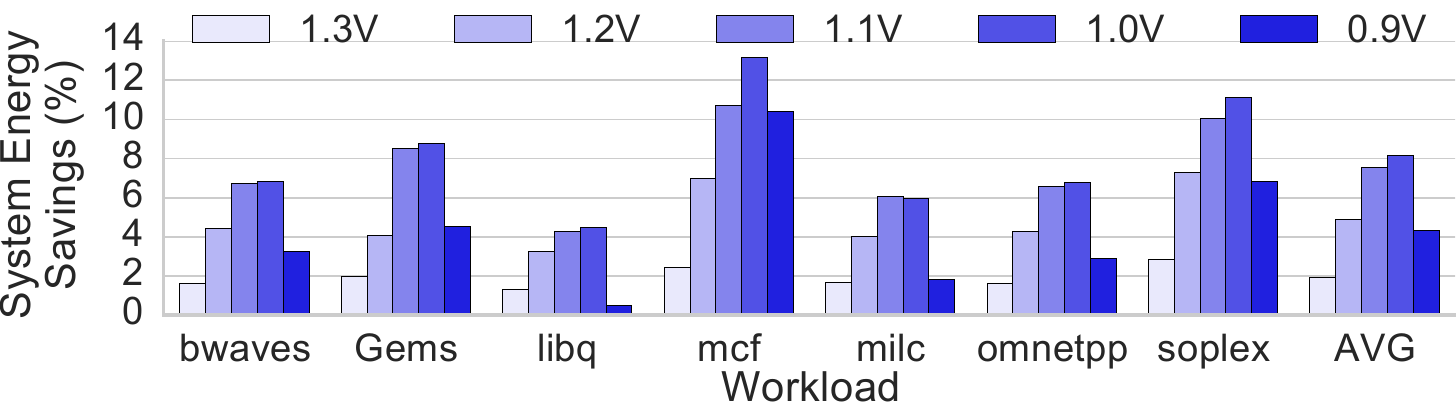}
        \vspace{-0.1in}
    }
    \caption{System performance loss and energy
    savings due to array voltage scaling for memory-intensive workloads.}
    \label{fig:vsweep_high_mpki}
\end{figure}

\paratitle{Memory-Intensive Workloads} \figref{vsweep_high_mpki} shows the
system performance (WS) loss, DRAM power reduction, and system energy reduction,
compared to a baseline DRAM with 1.35V, when we vary \varr from 1.30V to 0.90V.
We make three observations from these results.

First, system performance loss \fixIII{increases} as we lower \varr, due to the increased
DRAM access latency. However, different workloads \fix{experience} a different
rate of performance loss, as they tolerate memory latency differently. Among the
\fix{memory-intensive} workloads, \emph{mcf} exhibits the lowest performance
degradation since it has the highest memory intensity and \fix{high memory-level
  parallelism}, leading to high queuing delays in the memory controller. The
queuing delays \fix{and memory-level parallelism} hide the longer DRAM access
latency \fix{more than in other workloads}. Other workloads lose more
performance because they are \fixIII{less able to tolerate/hide the} increased latency. Therefore,
workloads with very high memory intensity \fix{and memory-level parallelism can
  be less sensitive} to the increased memory latency.


Second, DRAM power savings increase with lower \varr since reducing the DRAM
array voltage decreases \emph{both} the dynamic and static power components of
DRAM. \fix{However, \emph{system} energy savings does \fixIII{\emph{not}}
  monotonically increase with lower \varr.} We find that using $V_{array}$=0.9V
\fix{provides} lower system energy savings than \fixIV{using} $V_{array}$=1.0V,
as the processor takes \emph{much longer} to run the applications \fixIV{at
  $V_{array}$=0.9V}. In this case, the increase in static DRAM and CPU energy
\fix{outweighs} the dynamic DRAM energy savings.


Third, \fix{reducing \varr leads to a system energy reduction only when} the
reduction in DRAM energy outweighs the increase \fixVI{in} CPU energy (due to the
longer execution time). For \varr=1.1V, the system energy reduces by an average
of 7.6\%. Therefore, we conclude that array voltage scaling is an effective
technique that improves system energy consumption, with a small performance
loss, for \fix{memory-intensive} workloads.



\paratitle{\fix{Non-Memory-Intensive} Workloads} \label{ssec:comp}
\tabref{non_mem_intensive_results} summarizes the \fixIV{system performance
  loss} and energy savings of 20 \fix{non-memory-intensive} workloads as \varr
varies from 1.30V to 0.90V, over the performance and energy consumption under a
nominal \varr of 1.35V. Compared to the \fix{memory-intensive} workloads,
\fix{non-memory-intensive} workloads \fixIV{obtain} smaller system energy
savings, as the system energy is dominated by the processor. Although the
workloads are more compute-intensive, lowering \varr \emph{does} reduce their
system energy consumption, by decreasing the energy consumption of DRAM. For
example, at 1.2V, array voltage scaling achieves an overall system energy
savings of 2.5\% with a performance loss of only 1.4\%.


\begin{table}[h]
  \centering
    \setlength{\tabcolsep}{.45em}
    \begin{tabular}{crrrrr}
        \toprule
        \textbf{\varr} & 1.3V & 1.2V & 1.1V & 1.0V & 0.9V  \\
        \midrule

        \textbf{System Performance Loss} (\%) & 0.5 & 1.4 & 3.5 & 7.1 & 14.2 \\
        \textbf{DRAM Power Savings} (\%) & 3.4 & 10.4 & 16.5 & 22.7 & 29.0 \\
        \textbf{System Energy Savings} (\%) & 0.8 & 2.5 & 3.5 & 4.0 & 2.9 \\

        \bottomrule
    \end{tabular}

  \caption{System performance loss and energy savings due to array voltage
    scaling for \fix{non-memory-intensive} workloads.}
  \label{tab:non_mem_intensive_results}
\end{table}


\subsection{Effect of Performance-Aware Voltage Control}
\label{ssec:pavc_eval}

In this section, we evaluate the effectiveness of \fix{our complete proposal for Voltron,
which incorporates} \fixIII{our} \emph{performance-aware voltage control} mechanism
\fixIII{to drive the array voltage scaling component intelligently}. The \fixIII{performance-aware voltage control}
mechanism selects the lowest voltage level that satisfies the performance loss
bound (provided by the user or system designer) based on our performance model
(see \ssecref{pavc}). We evaluate \voltron with a target performance loss of
5\%. \voltron executes the performance-aware voltage control mechanism once
every four million cycles.\footnote{\fix{We evaluate the sensitivity to the \fixIII{frequency} at
  which we execute the mechanism (i.e., the interval \fix{length} of Voltron) in
  \ssecref{interval}.}} We quantitatively
compare Voltron \fixIV{to} \textit{\memdvfs}, a dynamic DRAM frequency and voltage
scaling mechanism proposed by prior work~\cite{david-icac2011}, which we
describe in \ssecref{dvfs}. Similar to the configuration used in the prior work,
we enable \memdvfs to switch dynamically between three frequency steps: 1600,
1333, and 1066 MT/s, which employ \fix{supply voltages of} 1.35V, 1.3V, and 1.25V, respectively.

\figref{memdvfs} shows the system performance (WS) loss, DRAM power savings, and
system energy savings due to \memdvfs and \voltron, compared to a baseline DRAM
with \fix{a supply voltage of} 1.35V. We show one graph per metric, where each graph uses boxplots to show
the distribution among all workloads. \fix{In each graph, we categorize the workloads as}
\fixV{either non-memory-intensive or memory-intensive}.
Each box illustrates the quartiles of the population, and the
whiskers indicate the minimum and maximum values. The red dot indicates the
mean. We make four major observations.

\begin{figure}[!h]
    \centering
    \subcaptionbox{\label{fig:volt_vs_mem_perf}}[0.49\linewidth][l]
    {
      \includegraphics[width=0.48\linewidth]{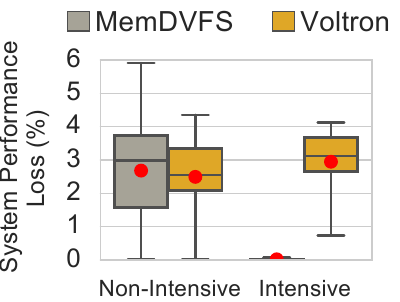}
    }
    \subcaptionbox{\label{fig:volt_vs_mem_dpow}}[0.49\linewidth][r]
    {
      \includegraphics[width=0.48\linewidth]{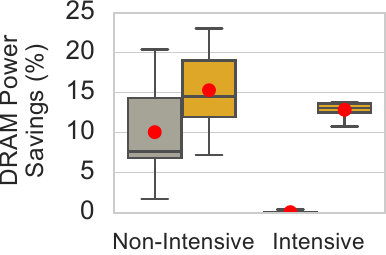}
    }

    \subcaptionbox{\label{fig:volt_vs_mem_syse}}[0.48\linewidth][c]
    {
      \includegraphics[width=0.48\linewidth]{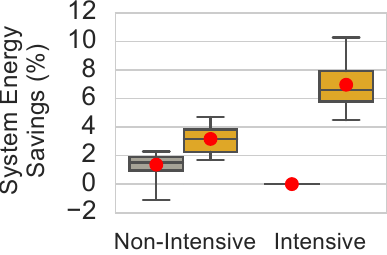}
    }
    \vspace{-0.15in}
    \caption{\fixIV{Performance and energy} comparison between \voltron and
      \memdvfs on \fixV{non-memory-intensive and memory-intensive workloads}.}
    \label{fig:memdvfs}
\end{figure}

First, \fix{as shown in \figref{volt_vs_mem_perf}}, \voltron consistently
selects a \varr value that satisfies the performance loss bound of 5\% across
all workloads. \voltron incurs \fix{ an average (maximum) performance loss of 2.5\% (4.4\%) and 2.9\% (4.1\%)}
for \fixV{non-memory-intensive and memory-intensive} workloads,
respectively. This demonstrates that our performance model enables \voltron to
select a low voltage value that saves energy while bounding performance loss
based on the user's requirement. We evaluate \voltron with a range of different
performance targets in \ssecref{vary_perf_target}.

Second, \memdvfs has almost zero effect on \fixIV{memory-intensive} workloads.
This is because \memdvfs avoids scaling DRAM frequency \fix{(and hence voltage)}
when an application's memory bandwidth \fixIV{utilization} is above a fixed
threshold. Reducing the frequency can result in a large performance loss since
the \fixIV{memory-intensive} workloads require high memory throughput. As
\fix{memory-intensive} applications have high memory bandwidth consumption that
easily exceeds the \fixIV{fixed} threshold \fixIV{used by \memdvfs}, \memdvfs
\emph{cannot} perform frequency and voltage scaling during most of the execution
time. These results are consistent with the results reported in
\memdvfs\cite{david-icac2011}. In contrast, \voltron reduces system energy
\fix{(shown in \figref{volt_vs_mem_syse})} by 7.0\% on average for
\fix{memory-intensive} workloads, at the cost of 2.9\% system performance loss,
which is well within the specified performance loss target of 5\% \fix{(shown in
  \figref{volt_vs_mem_perf})}.

Third, both \memdvfs and \voltron reduce the average system energy consumption
for \fix{non-memory-intensive} workloads. \memdvfs reduces system energy by
dynamically scaling the frequency and voltage of DRAM, which lowers the DRAM
power consumption \fix{(as shown in \figref{volt_vs_mem_dpow})}. By reducing the
DRAM array voltage to a lower value than \memdvfs, \voltron is able to provide a
slightly higher DRAM power and system energy reduction \fix{for
  \fix{non-memory-intensive} workloads} \fixIV{than \memdvfs}.

Fourth, although \voltron reduces the system energy with a small performance
loss, the average system energy efficiency, in terms of \emph{performance per
  watt} (not shown in the figure), still improves by 3.3\% and 7.4\% for
\fixIV{\mbox{non-memory-intensive} and memory-intensive} workloads, respectively, over
the baseline. \fix{Thus, we} demonstrate that \voltron is an effective mechanism
that improves \fixIV{system} energy efficiency not only on
\fix{non-memory-intensive} applications, but also \fixIV{(especially)} on
\fix{memory-intensive} workloads where prior work was unable to do so.

\fix{To summarize, across \fixIV{non-memory-intensive and memory-intensive}
  workloads, Voltron reduces the average system energy consumption by 3.2\% and
  7.0\% while limiting average system performance loss to only 2.5\% and 2.9\%,
  respectively.  Voltron ensures that no workload loses performance by more than
  the specified target of 5\%. We conclude that Voltron is an effective DRAM
  \fixIV{and system} energy reduction mechanism that significantly outperforms
  prior memory DVFS mechanisms.}

\fix{
\subsection{\fixIII{System Energy Breakdown}}
\label{ssec:energy_breakdown}

To demonstrate the source of energy savings from Voltron, \figref{breakdown}
compares the system energy breakdown of Voltron to the baseline\fix{, which}
operates at the nominal voltage level \fix{of 1.35V}. The breakdown shows the
average CPU and DRAM energy consumption across workloads\fix{, which} are
categorized into \fixIV{non-memory-intensive and memory-intensive workloads}.
We make \fix{two observations from the figure}.

\figputHWL{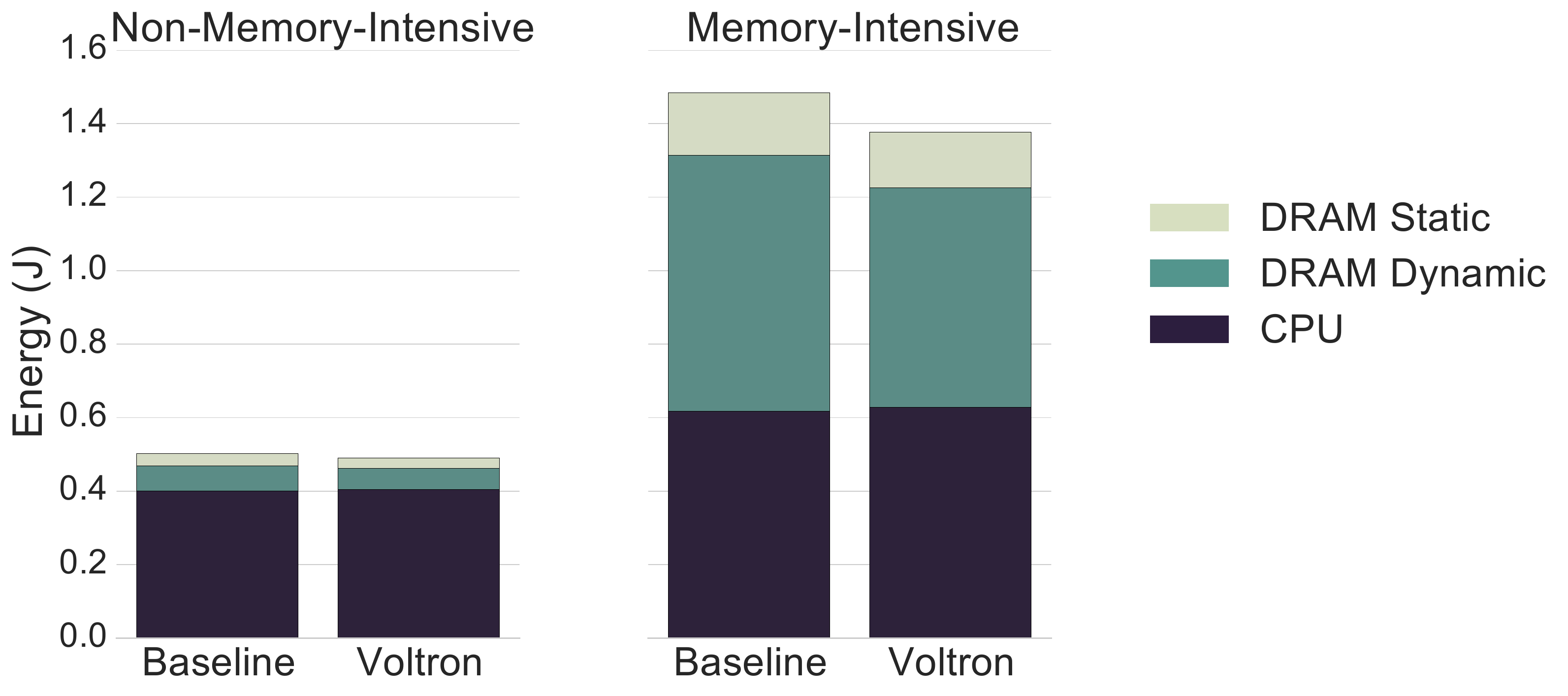}{Breakdown of system energy consumption (lower is better).}{breakdown}

First, in the \fixIV{non-memory-intensive} workloads, \fix{the} CPU consumes an
average of 80\% of the total system energy when \fix{the} DRAM uses the nominal
voltage level. As a result, Voltron has less potential \fix{to reduce} the
overall system energy \fix{as} it \fix{reduces \emph{only}} the DRAM energy,
which \fix{makes up only} 20\% of the \fix{total} system energy. Second, DRAM
consumes an average of 53\% of the total system energy in the
\fix{memory-intensive} workloads.  \fixIII{As a result, Voltron has a larger
  room for potential improvement for memory-intensive workloads than for}
\fix{non-memory-intensive} workloads. Across the \fix{memory-intensive}
workloads, Voltron reduces the average dynamic and static DRAM energy by 14\%
and 11\%, respectively. However, Voltron increases the CPU energy consumption by
1.7\%, because \fix{the application} incurs a small system performance
degradation (due to \fix{the} increased \fix{memory} access latency), which is
within \fix{our 5\%} performance loss target (as shown in \ssecref{pavc_eval}).
\fixIV{We conclude that Voltron is effective in reducing DRAM energy, and it is
  an effective system energy reduction mechanism, especially when DRAM is a
  major consumer of energy in the system.} }

\ignore{
\subsection{Comparison to Memory DVFS}

In this section, we quantitatively compare Voltron against \textit{\memdvfs}, a
dynamic DRAM frequency and voltage scaling mechanism proposed by a prior
work~\cite{david-icac2011}, which we already described in \ssecref{dvfs}.
Similar to the configuration used in the prior work, we enable \memdvfs to
dynamically switch between three frequency steps: 1600, 1333, and 1066 MT/s,
which employ 1.35V, 1.3V, and 1.25V, respectively.

\figref{memdvfs} shows the system performance loss and system energy savings due
to \memdvfs and \voltron. Each metric (y-axis) is shown in a single plot that
uses boxplots to show the distribution among multiple workloads, which are
categorized into high and low memory intensity (x-axis). Each box demonstrates
the quartiles of the population and the whiskers indicate the minimum and
maximum value. The red dot on each box indicates the mean. We make several major
observations.

\begin{figure}[!h]
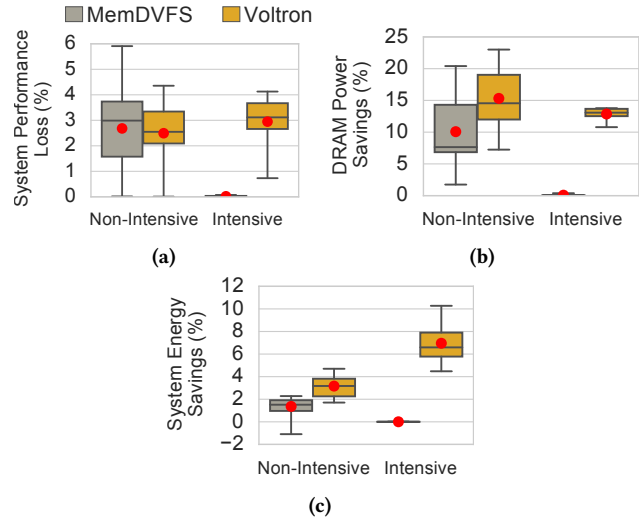

    \centering
    \subcaptionbox{\label{fig:volt_vs_mem_perf}}[0.49\linewidth][l]
    {
      \includegraphics[width=0.48\linewidth]{sim_results/paper_plots/homo_wklds/voltron_homo_wklds_mpki_box_perfloss}
    }
    \subcaptionbox{\label{fig:volt_vs_mem_dpow}}[0.49\linewidth][r]
    {
      \includegraphics[width=0.48\linewidth]{sim_results/paper_plots/homo_wklds/voltron_homo_wklds_mpki_box_drampowersavings}
    }

    \subcaptionbox{\label{fig:volt_vs_mem_syse}}[0.48\linewidth][c]
    {
      \includegraphics[width=0.48\linewidth]{sim_results/paper_plots/homo_wklds/voltron_homo_wklds_mpki_box_sysenergysavings}
    }
    \vspace{-0.15in}
    \caption{\fixIV{Performance and energy} comparison between \voltron and
      \memdvfs on \fixV{non-memory-intensive and memory-intensive workloads}.}
    \label{fig:memdvfs}
\end{figure}

First, for the \fix{memory-intensive} workloads, \memdvfs has almost zero effect on
system energy and performance. The reason is that \memdvfs avoids scaling DRAM
frequency when an application's memory bandwidth is above a fixed threshold
since reducing the frequency can result in a large performance loss due to lower
data throughput. As \fix{memory-intensive} applications have high memory bandwidth
consumption that is easily above the threshold, \memdvfs cannot perform
frequency and voltage scaling during most of the applications' runtime. These
results are consistent with that of reported in \memdvfs\cite{david-icac2011}. In
contrast, \voltron reduces system energy by 7.0\% on average for the \fix{memory-intensive}
workloads at the cost of 2.9\% system performance loss, which meets the
specified target of 5\%.

Second, both \memdvfs and \voltron reduce system energy consumption for low
memory-intensity workloads. By reducing the DRAM array voltage to a lower value
than \memdvfs, \voltron is able to provide slightly higher reduction. Third,
although \voltron reduces the system energy at a small performance loss, the
system energy efficiency, in terms of \emph{performance per watt}, still
improves by 7.4\% and 3.3\% for \fix{memory-intensive and non-memory-intensive} workloads,
respectively, on average over the baseline. We demonstrate that \voltron is a new
optimization technique that enables energy efficiency improvement not only on
non-memory-intensive applications, but also on memory-intensive workloads that
prior work cannot achieve.
}

\response{
\subsection{Effect of Exploiting \fixIV{Spatial} Locality \fixV{of Errors}}
\label{ssec:eval_el}

In \ssecref{spatial}, our experimental results show that errors due to reduced
voltage concentrate in certain regions, specifically in select DRAM banks for
some vendors' DIMMs. This implies that when we lower the voltage, only the banks
with errors require a higher access latency to read or write data correctly,
whereas error-free banks can be accessed reliably with the standard latency.
Therefore, in this section, we enhance our \voltron mechanism by exploiting the
spatial locality of \fixIV{errors caused by reduced-voltage operations}. The key
idea is to dynamically change the access latency \fixIII{on a per-bank basis
  (i.e., based on the DRAM banks being accessed)} to account for the reliability
of each bank.
\fixIII{In other words, we would like to increase the latency only for banks
  that would otherwise experience \emph{errors}, and do so just enough such that
  these banks operate reliably.}

\fix{For our evaluation, we model the behavior based on a subset
  \fixIII{(three)} of Vendor C's DIMMs, which show that the number of banks with
  errors increases as we reduce the \fixIV{supply} voltage
  \fixIII{(\ssecref{spatial})}. We observe that these DIMMs start experiencing
  errors at 1.1V using the standard latency values. However, only one bank
  observes errors when we reduce the voltage level from 1.15V to 1.1V
  \fixIII{(i.e., 50mV reduction)}. We evaluate a \fixIV{conservative} model that
  increases the number of banks that need higher latency by one for every 50mV
  reduction from the nominal voltage of 1.35V. Note that this model is
  conservative, because we start increasing the latency when the voltage is
  reduced to 1.3V, which is much higher than the lowest voltage level (1.15V)
  for which we observe that DIMMs operate reliably without \fixIV{requiring} a
  latency increase. Based on this conservative model, we choose the banks
  \fixIV{whose latencies should} increase \emph{sequentially} starting from the
  first bank, while the remaining banks operate at the standard latency. For
  example, at 1.25V (100mV lower than the nominal voltage of 1.35V), \voltron
  needs to increase the latency for the first two out of the eight banks to
  ensure reliable \fixIII{operation}. }


\fix{\figref{voltron_bankloc} compares the system performance and energy
efficiency of our bank-error locality aware version of \voltron (denoted as
\emph{Voltron+BL}) to the previously-evaluated \voltron mechanism, which is not
aware of such locality.} \fix{By increasing the memory latency for only} a
subset of banks at each voltage step, Voltron+BL reduces the average performance
loss from 2.9\% to 1.8\% and \fix{increases} the average system energy savings
from 7.0\% to 7.3\% for \fix{memory-intensive} workloads, with similar
improvements for \fix{non-memory-intensive} workloads. We show that enhancing
\voltron by adding awareness of the spatial locality of errors can further
mitigate the latency penalty due to reduced voltage, \fixIV{even with \fixV{the}
  conservative bank error locality model we assume and evaluate in this
  example.} \fix{We believe that a mechanism that exploits \fixIV{spatial} error
  locality at a finer granularity could lead to even higher performance and
  energy savings, but we leave \fixIV{such an evaluation} to future work.}

\begin{figure}[!h]
    \centering
    \subcaptionbox*{}[0.49\linewidth]
    {
      \includegraphics[width=0.49\linewidth]{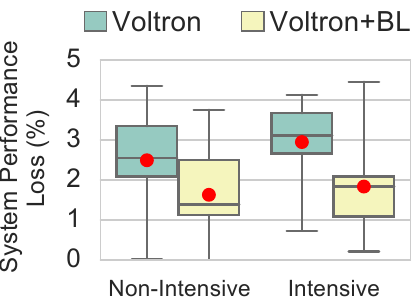}
    }
    \subcaptionbox*{}[0.49\linewidth]
    {
      \includegraphics[width=0.49\linewidth]{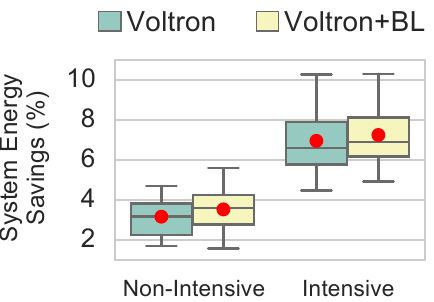}
    }
    \vspace{-0.25in}
    \caption{\fixIV{Performance and energy benefits} of exploiting bank-error locality in Voltron (denoted as
      Voltron+BL) on \fixIV{non-memory-intensive and memory-intensive workloads}.}
    \label{fig:voltron_bankloc}
\end{figure}


\subsection{Effect on Heterogeneous Workloads}
\label{ssec:eval_hetero}

\begin{figure*}[!h]
    \centering
    \subcaptionbox*{}[0.49\linewidth]
    {
      \includegraphics[scale=1.0]{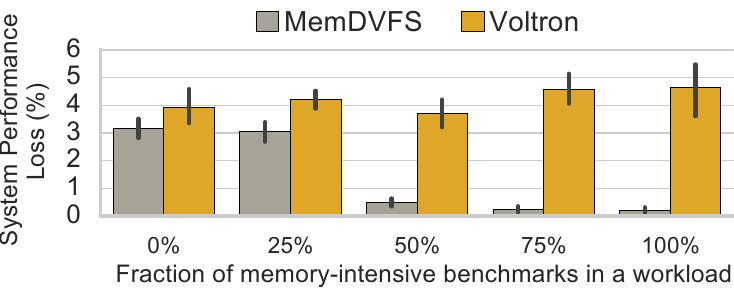}
    }
    \subcaptionbox*{}[0.49\linewidth]
    {
      \includegraphics[scale=1.0]{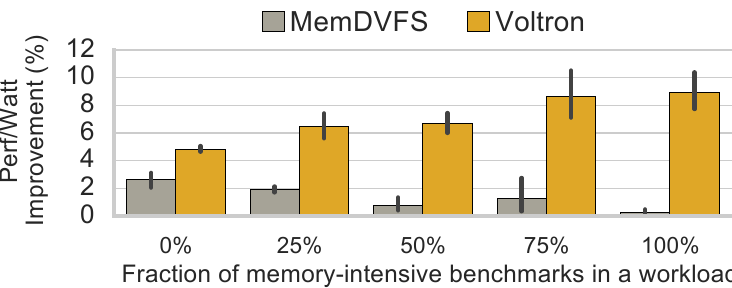}
    }
    \vspace{-0.25in}

    \caption{\response{System performance loss and energy efficiency improvement of Voltron
        and MemDVFS across 50 different heterogeneous workload mixes.}}
    \label{fig:voltron_hetero}
\end{figure*}

So far, we have evaluated \voltron on \fixIII{\emph{homogeneous} multi-core} workloads, where
each workload consists of the same benchmark running on all cores. In this
section, we evaluate the effect of \voltron on \textit{heterogeneous} workloads,
where each workload consists of \emph{different} benchmarks running on each
core. We categorize the workloads based on the fraction of memory-intensive
benchmarks in the workload (0\%, 25\%, 50\%, 75\%, and 100\%). Each category
consists of 10~workloads, resulting in a total of 50~workloads across all
categories.


\fix{\figref{voltron_hetero} shows the system performance loss and energy
efficiency improvement (in terms of performance per watt) with \voltron and with
\memdvfs for heterogeneous workloads. The error bars indicate the 95\%
confidence intervals \fixIII{across all workloads in the category}. We make two
observations from the figure.}
\fix{First, for each category of the heterogeneous workloads, \voltron is able
  to meet the 5\% performance loss target on average.
However, since \voltron is not designed to provide a \emph{hard} performance
guarantee for every single workload, \voltron exceeds the performance loss
target for 10 out of the 50~workloads, though it exceeds the target by only
0.76\% on average. Second,} the energy efficiency improvement due to
\voltron\fixIII{\ becomes larger} as the memory intensity of the workload
increases. This is because the fraction of system energy coming from memory
grows with higher memory intensity, \fixIII{due to the higher amount of} memory
traffic. Therefore, the memory energy reduction from \voltron has \fix{a
  greater} impact at the system level with more \fix{memory-intensive}
workloads. On the other hand, \memdvfs becomes \emph{less} effective with higher
memory intensity, as the memory bandwidth \fixIV{utilization more frequently}
exceeds the fixed threshold \fixIII{employed by \memdvfs}. Thus, \memdvfs has
\fix{a} smaller opportunity to scale the frequency and voltage. \fix{We conclude
  that \voltron is an effective mechanism that can adapt to different
  applications' characteristics to improve system energy efficiency.}


\subsection{Effect of Varying the Performance Target}
\label{ssec:vary_perf_target}

\figref{voltron_sweep} shows the performance loss and energy efficiency
\fixV{improvement due to} \voltron as we vary the system performance loss target for
\fixIV{both homogeneous and} heterogeneous workloads. For each target, we use a
boxplot to show the distribution across all workloads. In total, we evaluate
\voltron on 1001 combinations of workloads and performance targets: \fixIV{27
  homogeneous workloads $\times$ 13 targets + 50 heterogeneous workloads
  $\times$ 13 targets}. The first major observation is that \voltron's
performance-aware voltage control mechanism adapts to different performance
targets by dynamically selecting different voltage values at runtime. Across all
1001~runs, \voltron\fixIII{\ keeps performance within the performance loss}
target for 84.5\% of them. Even though \voltron cannot enforce a
\fixIII{\emph{hard}} performance guarantee for all workloads, \fixIII{it}
exceeds the target by only 0.68\% on average for those workloads \fixIII{where
  it does} not strictly meet the target.

%

\begin{figure}[!h]
    \centering
    \subcaptionbox*{}[\linewidth][c]
    {
      \includegraphics[width=\linewidth]{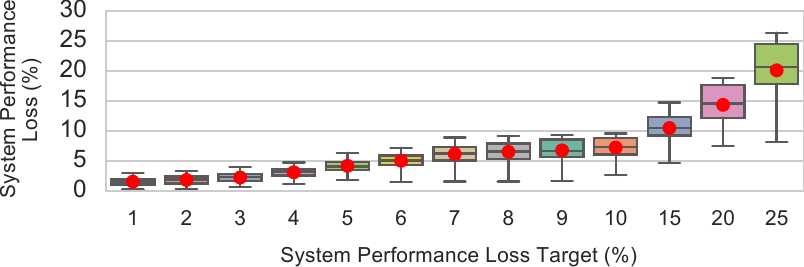}
      \vspace{-0.2in}
    }
    \subcaptionbox*{}[\linewidth][c]
    {
      \includegraphics[width=\linewidth]{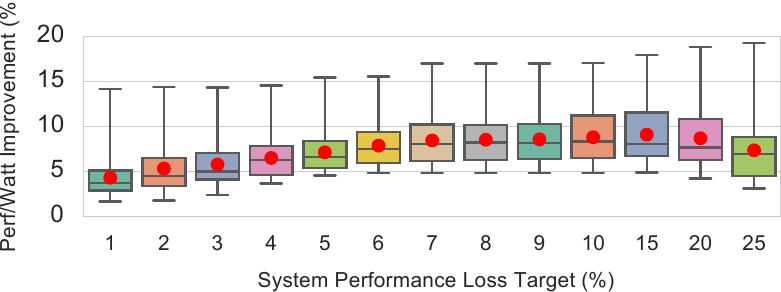}
    }
    \vspace{-0.2in}
    \caption{System performance loss and energy efficiency improvement of Voltron as the system performance loss target varies.}
    \label{fig:voltron_sweep}
\end{figure}

Second, system energy efficiency increases with higher performance loss targets,
but the gains plateau at around a target of 10\%. Beyond the 10\% target,
\voltron starts selecting smaller \varr values (e.g., 0.9V) that result in much
higher memory latency, which in turn increases both the CPU runtime and system
energy. \fix{In \ssecref{scaling_eval}, we observed that employing a \varr value
  less than 1.0V can result in smaller system energy savings than using
  \varr=1.0V.}

\fixIV{We conclude that,}
compared to prior work on memory DVFS, \voltron is a more
flexible mechanism, as it allows the \fixIV{users or system} designers to select
a performance and energy trade-off that best suits their target system or
applications.

\subsection{Sensitivity to the Profile Interval Length}
\label{ssec:interval}

\figref{interval} shows the average \fixV{system} energy efficiency improvement
due to \voltron with different profile interval lengths \fixIV{measured across
  27 homogeneous workloads}. As the \fixIV{profile} interval length increases
beyond two million cycles, we observe that the energy efficiency \fixIII{benefit
  of \voltron} \fixV{starts reducing}. This is because longer intervals prevent
\voltron from making \fixV{faster \varr} adjustments based on the
\fixV{collected new profile} information. Nonetheless, \voltron consistently
improves system energy efficiency \fixIV{for all evaluated profile interval
  lengths}.

\begin{figure}[h]
\begin{minipage}{\linewidth}
\begin{center}
\includegraphics[width=\linewidth]{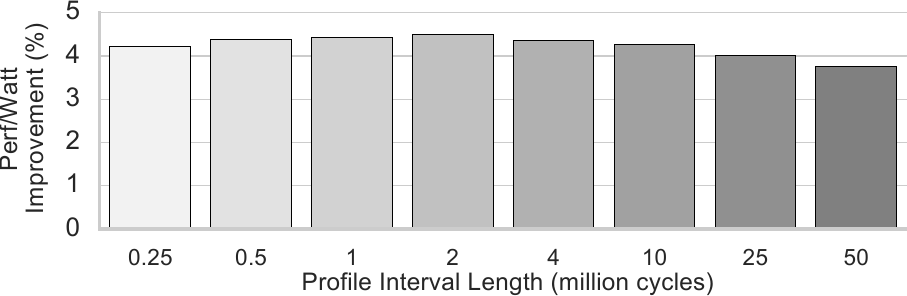}
\end{center}
\vspace{-0.1in}
\caption{Sensitivity of Voltron's \fixIV{system energy efficiency improvement} to profile interval length.\label{fig:interval}}
\end{minipage}
\end{figure}

} 


\section{Related Work}

\response{To our knowledge, this is the \fixIII{first work}} to \myitem{i}
experimentally \changes{characterize the reliability and performance of modern
  low-power DRAM chips under different supply voltages}, and \myitem{ii}
\fixIV{introduce} a new mechanism that reduces DRAM energy while retaining high
memory \fixIV{data} throughput by \fixIV{adjusting} the DRAM array voltage. We
briefly discuss other prior work in DRAM energy reduction.

\ignore{\paratitle{DRAM Frequency and Voltage Scaling} Many prior works have proposed to
reduce DRAM energy by adjusting the memory channel frequency and/or the DRAM
supply voltage dynamically. Deng et al.~\cite{deng-asplos2011} propose MemScale,
which scales the frequency of DRAM at runtime based on a performance predictor
of an in-order processor. Other work focuses on developing management policies
to improve system energy efficiency by coordinating DRAM \emph{DFS} with DVFS on
the CPU~\cite{deng-micro2012, deng-islped2012, begum-iiswc2015} or
GPU~\cite{paul-isca2015}. In addition to frequency scaling, David et
al.~\cite{david-icac2011} propose to scale the DRAM supply voltage along with
the memory channel frequency, based on the memory bandwidth utilization of
applications.}

\paratitle{DRAM Frequency and Voltage Scaling} \changes{We have already
discussed the most relevant works on DRAM frequency and voltage scaling, and
provided qualitative comparisons in \ssecref{dvfs}.} \changes{ None of these
works provide a comprehensive experimental analysis \fixIII{of} the impact of
using a wide range of supply voltages on DRAM reliability, latency, and refresh.
In contrast to all these works, our work focuses on a detailed experimental
characterization of real DRAM chips as the supply voltage varies. Our study
provides fundamental observations for potential mechanisms that can mitigate
DRAM and system energy consumption. Furthermore, frequency scaling hurts memory
throughput, and thus significantly degrades \fixIV{the} system performance of
\fixIV{especially} memory-intensive workloads (see \ssecref{avs} for our
quantitative comparison). We demonstrate the importance \fix{and benefits} of
\fix{exploiting} our \fix{experimental} observations by proposing \voltron, one
\fixIV{new example} mechanism that uses our observations to reduce DRAM and system
energy without sacrificing memory throughput. }




\ignore{vs. our work. They focus on benchmarks that don't utilize BW. There are
BW sensitive apps, such as GPU, but is good at hiding latency. We focus on mem
intensive workloads. In addition, we conduct experimental study on the DRAM
behavior which prior work does not have. Our approach can be combine with these
works to enable a comprehensive technique to improve energy efficiency for a
wide range of workloads. Contradicting comments...bc we show that freq scale
doesn't quite work for compute-intensive workloads. Unclear what the breakdown
between CPU and DRAM in these prior works. }

\paratitle{Low-Power Modes for DRAM} Modern DRAM chips support various low-power
standby modes. \fix{Entering and exiting these modes incurs some amount of
  latency, which delays memory requests that must be serviced.}
To increase the opportunities to exploit these low-power modes, several prior
works propose mechanisms that increase periods of memory idleness through data
placement (e.g.,~\cite{lebeck-asplos2000, fan-islped2001}) and memory traffic
reshaping (e.g.,~\cite{aggarwal-hpca2008, bi-hpca2010, amin-islped2010,
  lyuh-dac2004, diniz-isca2007}). Exploiting low-power modes is orthogonal to
our work on studying the impact of \fixIV{reduced-voltage operation in DRAM}.
Furthermore, low-power modes have a smaller effect on memory-intensive
workloads, which exhibit little idleness in memory accesses, \fixIV{whereas, as
  we showed in \ssecref{pavc_eval}, our mechanism is especially effective on
  memory-intensive workloads.}

\paratitle{Low-Power DDR DRAM Chips} Low-power DDR (LPDDR)~\cite{jedec-lpddr3,
jedec-lpddr4} is a specific type of DRAM that is optimized for low-power systems
like mobile devices. To reduce power consumption, LPDDRx (currently in its 4th
generation) employs a few major design changes that differ from conventional
DDRx chips. First, LPDDRx uses a low-voltage swing I/O interface that consumes
40\% less I/O power than DDR4 DRAM~\cite{choi-memcon2013}. Second, it supports
additional low-power modes with a lower supply voltage. Since the LPDDRx array
design remains the same as DDRx, our observations on the correlation between
access latency and array voltage are applicable to LPDDRx DRAM \fixIV{as well}.
Our proposed \voltron approach can provide significant benefits in LPDDRx,
\fixIII{since array} energy consumption is significantly \emph{higher} than the
energy consumption of \fixIII{peripheral circuitry in LPDDRx
  chips}~\cite{choi-memcon2013}. \changes{We leave the detailed evaluation of
  LPDDRx chips for future work since our current experimental platform is not
  capable of evaluating them.}

\paratitle{Low-Power DRAM Architectures} \fixIV{Prior} works
(e.g.,~\cite{udipi-isca2010, zhang-isca2014, cooper-balis-ieeemicro2010,
  chatterjee-hpca2017}) propose to modify the DRAM chip architecture to reduce
the \act power by activating only a fraction of a row instead of the entire row.
Another common technique, called sub-ranking, reduces dynamic DRAM power by
accessing data from a subset of chips from a DRAM module~\cite{zheng-micro2008,
  yoon-isca2011, ware-iccd2006}. \fix{A couple of} prior
works~\cite{malladi-isca2012, yoon-isca2012} propose DRAM module architectures
that integrate many low-frequency LPDDR chips to enable DRAM power reduction.
These proposed changes to DRAM chips or DIMMs are orthogonal to our work.

\fix{ \paratitle{Reducing Refresh Power} In modern DRAM chips, although
different DRAM cells have widely different retention times~\cite{liu-isca2013,
  kim-edl2009}, memory controllers conservatively refresh \emph{all} of the
cells based on the retention time of a small fraction of weak cells, which have
the longest retention time out of all of the cells. To reduce DRAM refresh
power, many prior works (e.g., \fixV{\cite{liu-isca2012, agrawal-hpca2014,
    qureshi-dsn2015, liu-isca2013,
    venkatesan-hpca2006,bhati-isca2015,lin-iccd2012,ohsawa-islped1998,
    patel-isca2017, khan-sigmetrics2014, khan-dsn2016, khan2016case}}) propose
mechanisms to reduce unnecessary refresh operations, and, thus, refresh power,
by characterizing the retention time profile (i.e., the \fixIV{data} retention
behavior of each cell) within the DRAM chips. However, these techniques do not
reduce the power of \emph{other} DRAM operations, and these prior works do
\emph{not} provide an experimental characterization of the effect of reduced
voltage level\fixIV{s} on \fixIV{data} retention time.


\paratitle{Improving DRAM Energy Efficiency by Reducing Latency or
Improving Parallelism} \fixIV{Various} prior works
(e.g.,\fixIV{~\cite{hassan-hpca2016, lee-hpca2015, chang-hpca2014,
lee-thesis2016, lee-hpca2013,
lee-sigmetrics2017,lee-arxiv2016,kim-isca2012, lee-thesis2016,
lee-taco2016, seshadri-micro2013, chang2016low, lee-pact2015}}) improve
DRAM energy efficiency by reducing the execution time through techniques
that reduce the DRAM access latency or improve parallelism between memory
requests.  These mechanisms are orthogonal to ours, because they do
\fixIV{not reduce} the voltage level of DRAM. }

\fixIII{\paratitle{Experimental Studies of DRAM Chips} Recent works
\fixIV{experimentally investigate various} reliability, \fixIV{data} retention,
and latency characteristics of modern DRAM chips~\cite{liu-isca2012,
  liu-isca2013, kim-isca2014, chang-sigmetrics2016, lee-hpca2015,
  lee-sigmetrics2017, chandrasekar-date2014, khan-sigmetrics2014,
  jung-memsys2016, jung-patmos2016,
  hassan-hpca2017,khan-dsn2016,patel-isca2017,lee-arxiv2016,lee-thesis2016,kim-thesis2015,
  meza2015revisiting, schroeder-sigmetrics2009, sridharan-asplos2015,
  sridharan2012study}. None of these works study these characteristics under
reduced-voltage operation, which we do in this paper.}


\fixIII{\paratitle{Reduced-Voltage Operation in SRAM Caches} Prior works propose
different techniques to enable SRAM caches to operate under reduced voltage
levels
(e.g.,~\cite{alameldeen-isca2011,alameldeen-tc2011,wilkerson-ieeemicro2009,chishti-micro2009,roberts-dsd2007,wilkerson-isca2008}).
These works are orthogonal to our experimental study because we focus on
\fixIV{understanding and enabling reduced-voltage operation in} DRAM, which is a
\fixIV{significantly} different memory technology than SRAM.}


\section{Conclusion}

\changes{ This paper provides the first experimental study that comprehensively
characterizes and analyzes the behavior of DRAM chips when the supply voltage is
reduced below its nominal value. We demonstrate, \fixIV{using 124 DDR3L DRAM
  chips}, that the \fix{DRAM} supply voltage can be reliably reduced to a
certain level, beyond which errors arise within the data. We then experimentally
demonstrate the relationship between the supply voltage and the latency of the
fundamental DRAM operations (activation, restoration, and precharge). \fixIV{We
  show that bit errors caused by reduced-voltage operation can be eliminated by
  increasing the latency of the three fundamental DRAM operations.} By
\fix{changing} the memory controller configuration to allow for the longer
latency of these operations, we can thus \emph{further} lower the supply voltage
without inducing errors in the data. We also \fixIV{experimentally} characterize
the relationship between reduced supply voltage and error locations, stored data
patterns, temperature, and data retention.


Based on these observations, we propose and evaluate \voltron, a low-cost energy
reduction mechanism that reduces DRAM energy \emph{without} \fix{affecting}
memory data throughput. \voltron reduces the supply voltage for \emph{only} the
DRAM array, while maintaining the nominal voltage for the peripheral circuitry
to continue operating the memory channel at a high frequency. \voltron uses a
new piecewise linear performance model to find the array supply voltage that
maximizes the system energy reduction within a given performance loss target.
\fix{Our experimental evaluations across a wide variety of workloads}
demonstrate that \voltron significantly reduces system energy \fix{consumption}
with only \fix{very} modest performance loss.

We conclude that it is very promising to understand and exploit reduced-voltage
operation in modern DRAM chips. We hope that the experimental characterization,
analysis, and optimization techniques presented in this paper will enable the
development of other new mechanisms that can effectively \fixIII{exploit the
trade-offs between voltage, reliability, and latency in DRAM to improve system
performance, efficiency, and/or reliability.}}

\begin{acks}

We thank our shepherd Benny Van Houdt, the anonymous reviewers of SIGMETRICS
2017, and the SAFARI group members for feedback. We acknowledge the support of
Google, Intel, \fix{NVIDIA}, Samsung, VMware, and the United States Department
of Energy. This research was supported
in part by the ISTC-CC, SRC, and NSF (grants 1212962 and 1320531). Kevin Chang
is supported in part by an SRCEA/Intel Fellowship.

\end{acks}

\bibliographystyle{IEEEtranS}
\bibliography{paper}

\newpage

\appendix
\label{sec:appendix}

\section*{Appendix}

\section{FPGA Schematic of DRAM Power Pins}
\label{sec:pin_layout}

\fix{\figref{pin_layout} shows a schematic of the DRAM pins that our FPGA
board~\cite{ml605_schematic} connects to (see \secref{fpga} for our experimental
methodology). Since there are a large number of pins that are used for different
purposes (e.g., data address), we zoom in on the right side of the figure to
focus on the power pins that we adjust for our experiments in this paper.}
\fixIII{Power pin numbering information can be found on the datasheets provided
  by all major vendors (e.g., \cite{micronDDR3L_2Gb, hynix-ddr3l,
    samsungddr3l_2Gb}).} In particular, we tune the VCC1V5 pin \fix{on the
  FPGA}, which is directly connected to all of the $V_{DD}$ and $V_{DDQ}$ pins
on the DIMM. The reference voltage VTTVREF is automatically adjusted by the DRAM
to half of VCC1V5.

\figputHW{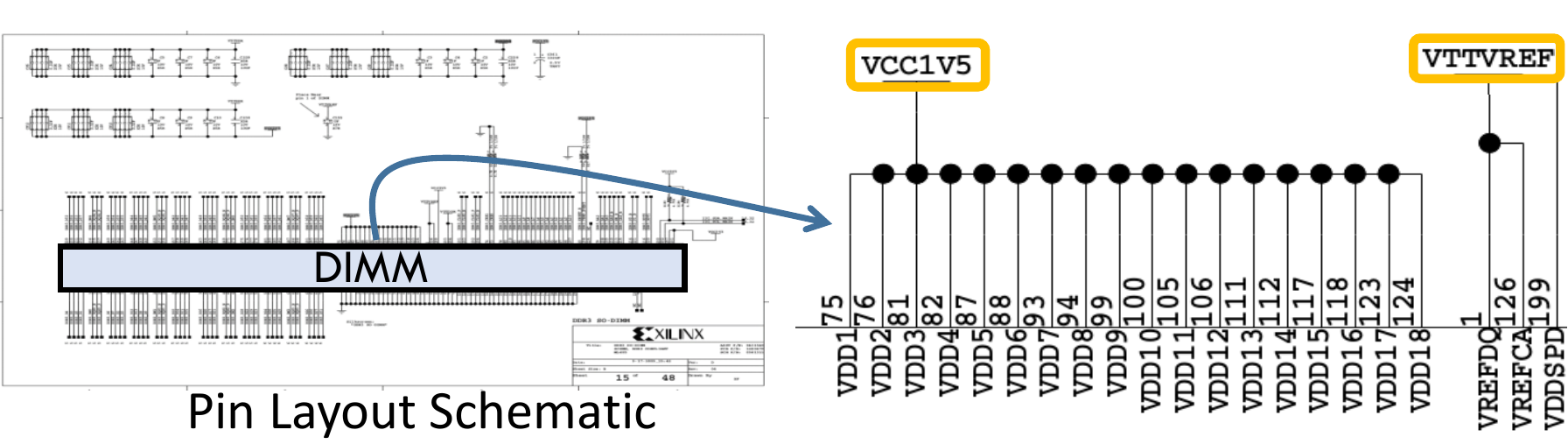}{DRAM power pins \fix{controlled by the ML605 FPGA board}.}

\section{Effect of Data Pattern on Error Rate}
\label{sec:datapatt}

\fix{As
discussed in \ssecref{volt_sensitivity}, we do \emph{not} observe a significant
effect \fix{of} different stored data patterns \fix{on the DRAM error rate when we
  reduce the supply voltage.} \figref{datapatt} shows the average bit error rate
(BER) of three different data patterns (\patt{aa}, \patt{cc}, and \patt{ff})
across different supply voltage levels for each vendor. Each data pattern
represents the byte value (shown in hex) that we fill into the DRAM. The error
bars indicate the 95\% confidence interval.  We make two observations from the
figure.



\figputHWL{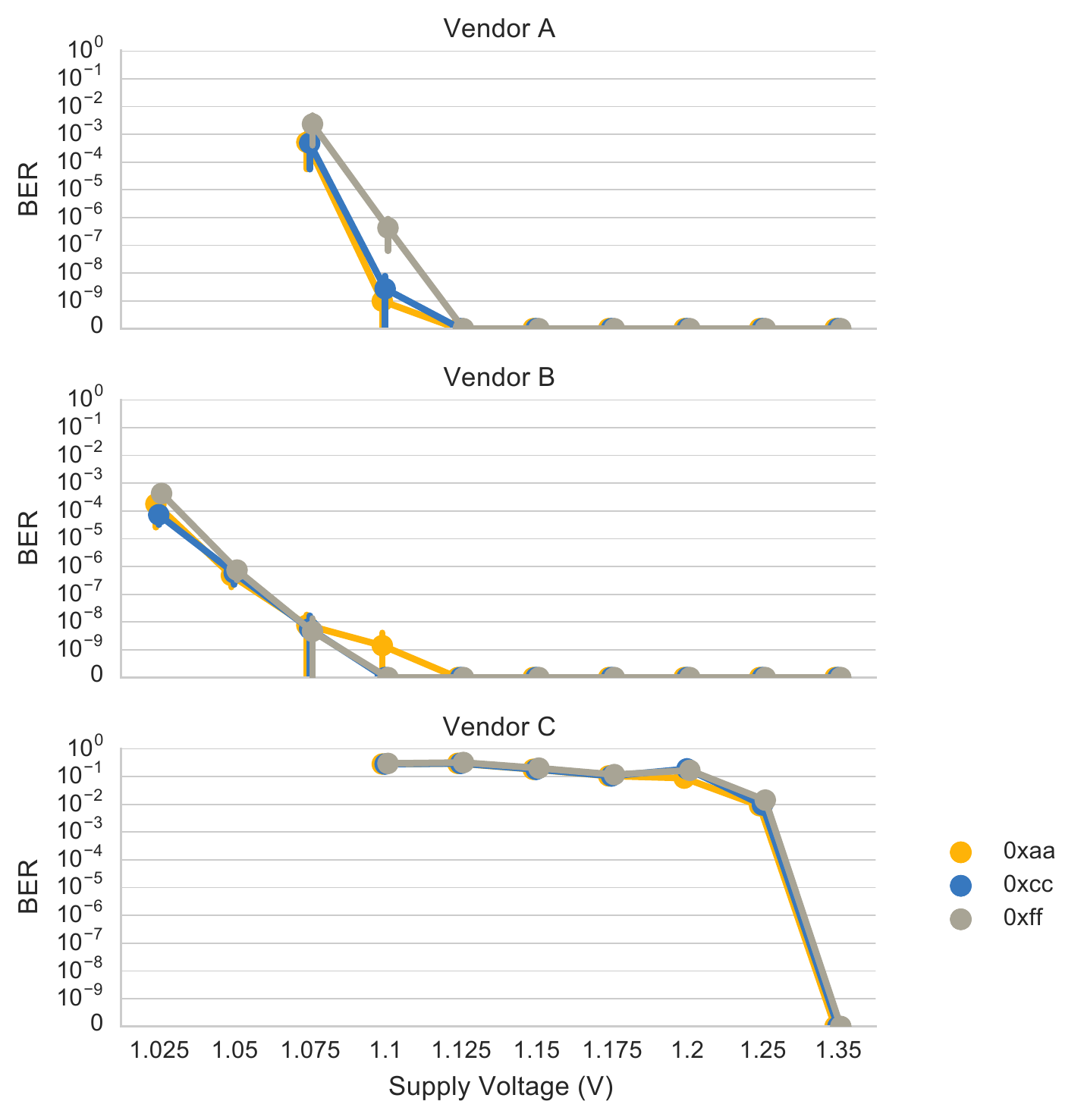}{Effect of \fixIII{stored data pattern} on \fix{bit error rate (BER)} across different supply voltage levels.}{datapatt}

First, the BER increases as we reduce the supply
voltage for all three data patterns. We made a similar observation in
\ssecref{volt_sensitivity}, which shows that the fraction of errors increases as the
supply voltage drops. We explained our hypothesis on the cause of the errors, and
used both experiments and simulations to test the
hypothesis, in \ssecref{low_volt_latency}.

Second, we do \emph{not} observe a significant difference across the BER values from
the three different data patterns. We attempt to answer the following question:
\fix{Do} different data patterns induce BER values that are statistically different
from each other at each voltage level? To answer this, we conduct a one-way
ANOVA (analysis of variance) test across the measured BERs from all three data patterns at each supply
voltage level to calculate a \emph{p-value}. If the p-value is below 0.05, we
can claim that these three data patterns induce a statistically-significant
difference on the error rate. \tabref{pval} shows the calculated p-value at each
supply voltage level. At certain supply voltage levels, we do not have a p-value
listed (shown as --- or $\triangle$ in the table), \fixIII{either} because there
are no errors \fixIV{(indicated as ---)} or we cannot
reliably access data \fixIV{from the DIMMs} even if the access latency is
higher than the standard value \fixIV{(indicated as $\triangle$)}.


\begin{table}[h]
  \small
  \centering
    \begin{tabular}{crrr}
        \toprule
        Supply & \multicolumn{3}{c}{\bf Vendor} \\

         \cmidrule(lr){2-4}
        Voltage & \multicolumn{1}{c}{A} & \multicolumn{1}{c}{B} & \multicolumn{1}{c}{C} \\
        \midrule
        1.305  & --- & --- & --- \\
        1.250  & --- & --- & \textbf{0.000000} \\
        1.200   & --- & --- & 0.029947 \\
        1.175 & --- & --- & 0.856793 \\
        1.150  & --- & --- & 0.872205 \\
        1.125 & --- & 0.375906 & 0.897489 \\
        1.100 & \textbf{0.028592} & 0.375906 & \textbf{0.000000} \\
        1.075 & 0.103073 & 0.907960 & $\triangle$ \\
        1.050 & $\triangle$  & 0.651482 & $\triangle$ \\
        1.025 & $\triangle$  & \textbf{0.025167} & $\triangle$ \\
        \bottomrule
    \end{tabular}

  \caption{Calculated p-values from the BERs across three data patterns at each
    supply voltage level. A p-value less \fixIII{than} 0.05 indicates that the BER is
    statistically different across the three data patterns (indicated in bold). \fix{--- indicates
    that the BER is zero. \fixIV{$\triangle$ indicates that we cannot reliably access data from the DIMM}.}\vspace{-15pt}}
  \label{tab:pval}
\end{table}


Using the one-way ANOVA test, we find that using different data patterns does
\emph{not} have a statistically significant (i.e., p-value $\geq$ 0.05) effect
on the error rate at \emph{all} supply voltage levels. Significant effects
(i.e., p-value < 0.05) occur at 1.100V for Vendor~A, at 1.025V for Vendor~B, and
at both 1.250V and 1.100V for Vendor~C. As a result, our study does \emph{not}
\fixIV{provide} enough evidence to conclude that using any of the three data
patterns (\patt{aa}, \patt{cc}, and \patt{ff}) induces higher or lower error
rates than the other two patterns at reduced voltage levels. }

%
%
%
%

\fix{
\section{SPICE Simulation Model}
\label{spice_model}

\figputWS{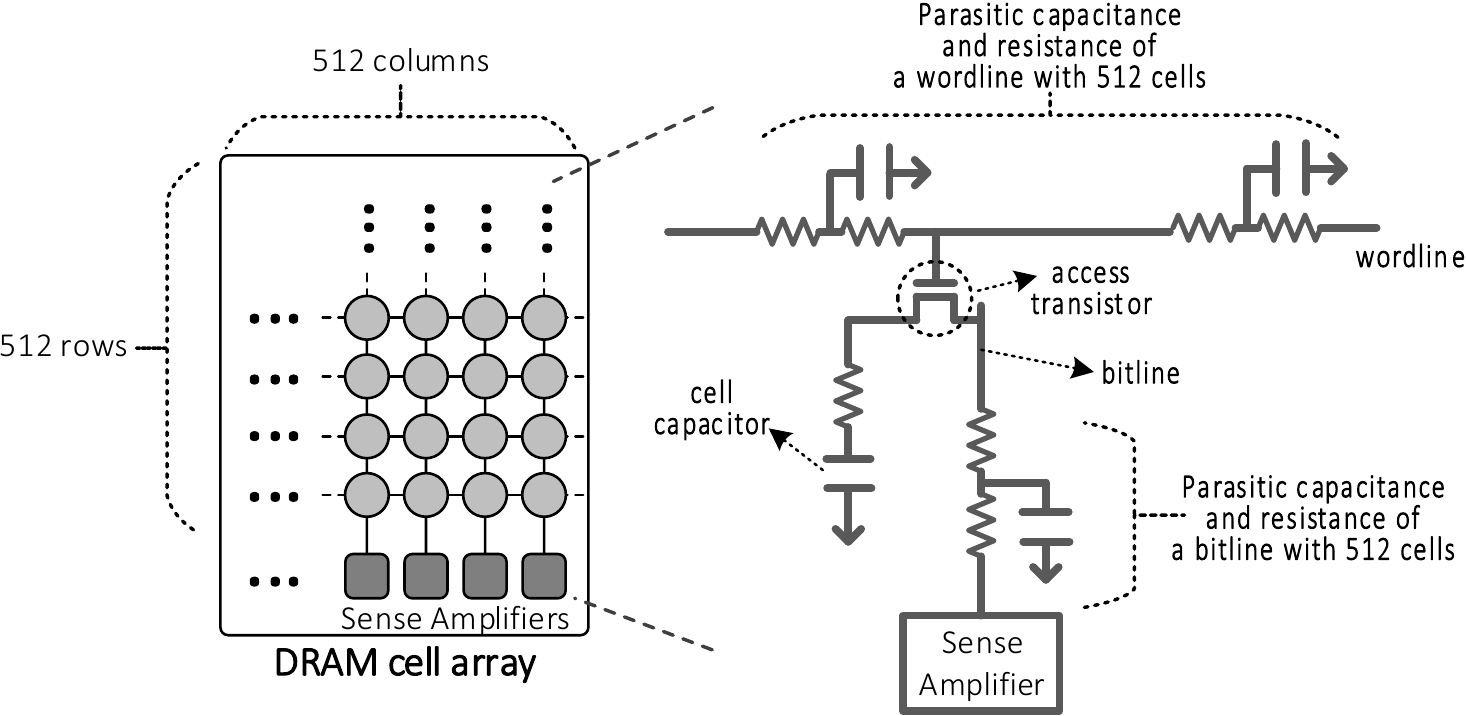}{0.8}{Our SPICE model \fixIV{schematic} of a DRAM cell array.}

We perform circuit-level SPICE simulations to understand in detail how the DRAM cell
arrays operate at low supply voltage. We model a DRAM cell array in SPICE,
and simulate its behavior for different supply voltages. We have released our
SPICE model online~\cite{volt-github}.

\paratitle{DRAM Cell Array Model} We build a detailed cell array model, as shown
in \figref{spice_model}. In the cell array, the DRAM cells are organized as $512
x 512$ array, which is a common organization in modern DRAM
chips~\cite{vogelsang-micro2010}. Each column is vertical, and corresponds to
512~cells sharing a bitline that connects to a sense amplifier. Due to the
bitline wire and the cells that are connected to the bitline, there is parasitic
resistance and capacitance on each bitline. Each row consists of 512~cells
sharing the same wordline, which also has parasitic resistance and capacitance.
The amount of parasitic resistance and capacitance on the bitlines and wordlines
is a major factor that affects the latency of DRAM operations accessing a cell
array~\cite{lee-hpca2013,lee-sigmetrics2017}.



\paratitle{Simulation Methodology} We use the LTspice~\cite{ltspice} SPICE simulator
to perform our simulations.
To find the access latency of the DRAM
operations under different supply voltages, we build a DRAM cell array using
technology parameters that we derive from a \SI{55}{\nano\meter} DRAM
model~\cite{vogelsang-micro2010} and from a \SI{45}{\nano\meter} process technology model~\cite{ptm,
  zhao-isqed2006}. By default, we assume that the cell capacitance is \SI{24}{\femto\farad} and
the bitline capacitance is \SI{144}{\femto\farad}~\cite{vogelsang-micro2010}. The nominal
\varr is 1.35V, and we perform simulations to obtain the latency of DRAM operations
at every 25mV step from 1.35V down to 0.9V.  The results of our SPICE simulations
are discussed in Section~\ref{ssec:volt_sensitivity} and \ref{ssec:low_volt_latency}.
}


\newpage

\section{Spatial Distribution of Errors}
\label{spatial}

\begin{figure}[!h]
    \centering
    \subcaptionbox{Supply voltage=1.075V.}[\linewidth][l]
    {
        \includegraphics[width=\linewidth]{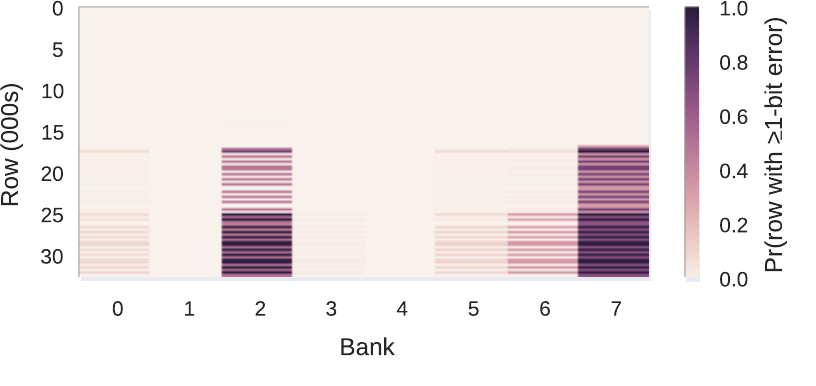}
    }

    \subcaptionbox{Supply voltage=1.1V.}[\linewidth][r]
    {
        \includegraphics[width=\linewidth]{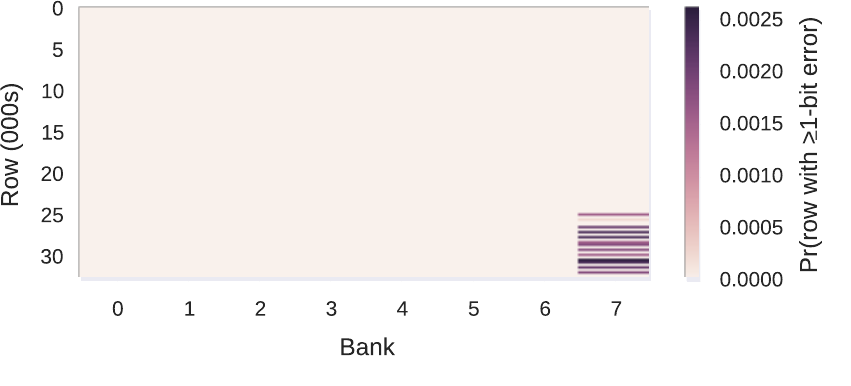}
    }

    \caption{Probability of error occurrence due to \fixIV{reduced-voltage operation} in a DIMM from \fix{Vendor~A}.}
    \label{fig:full_loc_A}
\end{figure}

\begin{figure}[!h]
    \centering
    \subcaptionbox{Supply voltage=1.025V.}[\linewidth][l]
    {
        \includegraphics[width=\linewidth]{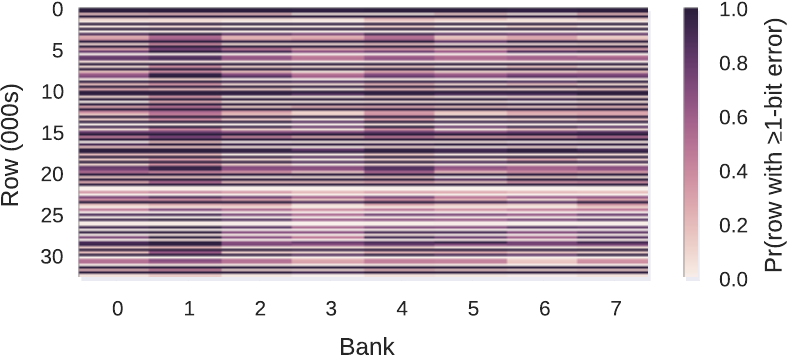}
    }

    \subcaptionbox{Supply voltage=1.05V.}[\linewidth][r]
    {
        \includegraphics[width=\linewidth]{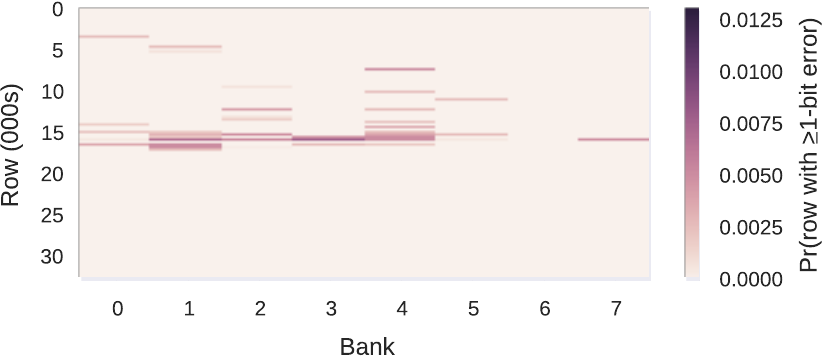}
    }

    \subcaptionbox{Supply voltage=1.1V.}[\linewidth][c]
    {
        \includegraphics[width=\linewidth]{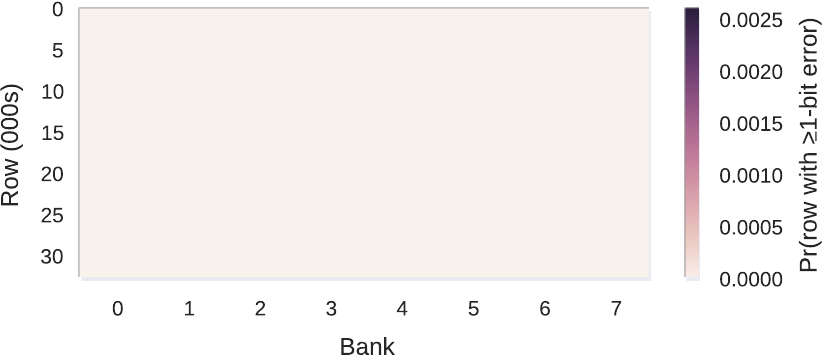}
    }
    \caption{Probability of error occurrence due to \fixIV{reduced-voltage operation} in a DIMM from \fix{Vendor~B}.}
    \label{fig:full_loc_B}
\end{figure}

\begin{figure}[!h]
    \centering
    \subcaptionbox{Supply voltage=1.1V.}[\linewidth][l]
    {
        \includegraphics[width=\linewidth]{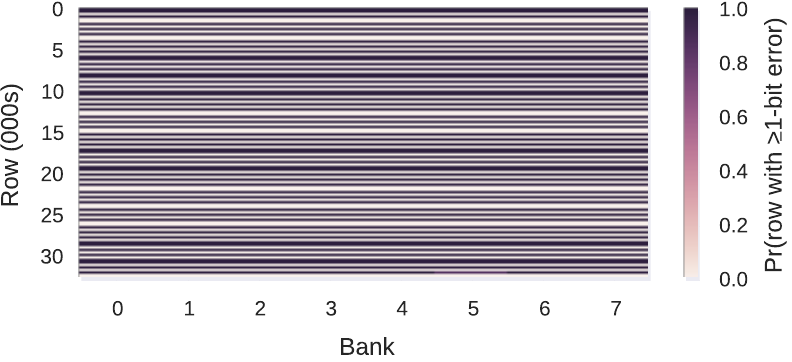}
    }
    \subcaptionbox{Supply voltage=1.15V.}[\linewidth][l]
    {
        \includegraphics[width=\linewidth]{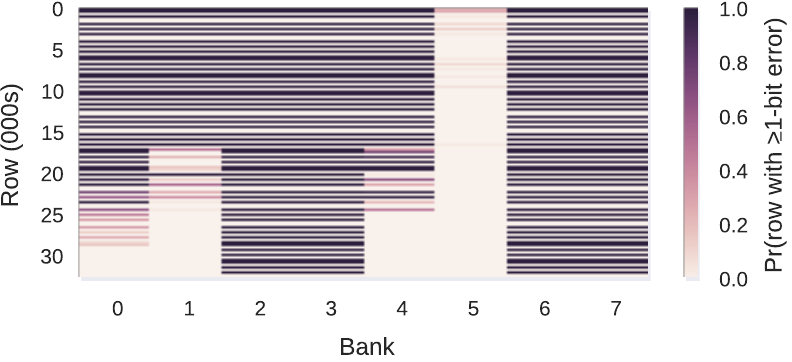}
    }
    %
    \subcaptionbox{Supply voltage=1.2V.}[\linewidth][l]
    {
        \includegraphics[width=\linewidth]{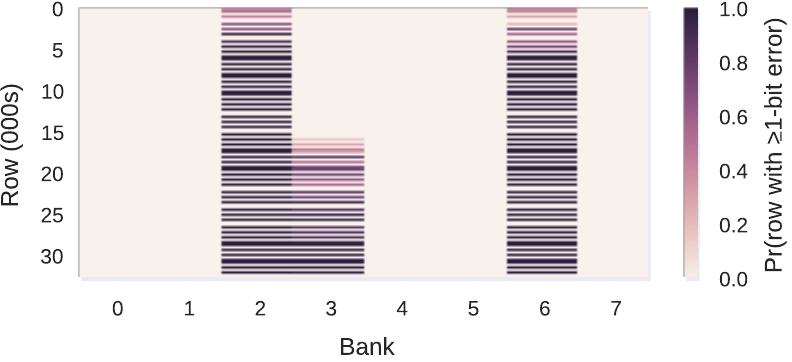}
    }
    \caption{Probability of error occurrence due to \fixIV{reduced-voltage operation} in a DIMM from \fix{Vendor~C}.}
    \label{fig:full_loc_C}
\end{figure}

In this section, we expand upon the spatial locality data presented in
\ssecref{spatial}. \fix{Figures~\ref{fig:full_loc_A}, \ref{fig:full_loc_B}, and
\ref{fig:full_loc_C} show the physical locations of errors that occur when the
supply voltage is reduced for a representative DIMM from Vendors~A, B, and C,
respectively.} At higher voltage levels, even if errors occur, they
tend to cluster in certain regions of a DIMM. However, as we reduce the supply
voltage further, the number of errors increases, and the errors start to spread
across the \fix{entire} DIMM.

\newpage
\
\newpage

\section{Full Information of Every Tested DIMM}
\label{sec:dimm_info}

\tabref{modules} lists the parameters of every DRAM module that we evaluate,
\fixIII{along with the \vmin we discovered for each module based on our
  experimental characterization (\ssecref{volt_sensitivity}).} \fix{We provide
  all results for all DIMMs in our GitHub repository~\cite{volt-github}.}
\definecolor{lightgray}{gray}{0.95}

\begin{table*}[t]

\setlength{\tabcolsep}{4pt}
\centering

\begin{tabular}{ccccccccccccc}

\toprule

\multirow{2}{*}[-2pt]{\em \small \centering Vendor} &
\multirow{2}{*}[-2pt]{\em \small \centering Module} &
                     {\em \small \centering Date$^{\ast}$} &
\multicolumn{4}{c}{\em \small \centering Timing$^{\dagger}$} &
\multicolumn{2}{c}{\em \small \centering Organization} &
\multicolumn{4}{c}{\em \small \centering Chip}
\\

\cmidrule(lr){3-3} \cmidrule(lr){4-7} \cmidrule(lr){8-9} \cmidrule(lr){10-13}

 &
 & {\em \small (yy-ww)}
 & {\em \small Freq~(MT/s)}
 & {\em \small tRCD (ns)}
 & {\em \small tRP (ns)}
 & {\em \small tRAS (ns)}
 & {\em \small Size (GB)$^{\ddagger}$}
 & {\em \small Chips$^{\star}$}
 & {\em \small Size (Gb)}
 & {\em \small Pins}
 & {\em \small Die Version$^{\S}$}
 & {\em \small \fix{V\textsubscript{min}} (V)$^{\circ}$}
 \\

\midrule


 & \module{A}{ 1}{} & 15-46 & 1600 & 13.75 & 13.75 & 35 & 2
& 4 & 4 & $\times$16  & $\mathcal{B}$ & 1.100 \\
 & \module{A}{ 2}{} & 15-47 & 1600 & 13.75 & 13.75 & 35 & 2
& 4 & 4 & $\times$16  & $\mathcal{B}$ & 1.125 \\
 & \module{A}{ 3}{} & 15-44 & 1600 & 13.75 & 13.75 & 35 & 2
& 4 & 4 & $\times$16  & $\mathcal{F}$ & 1.125 \\
 & \module{A}{ 4}{} & 16-01 & 1600 & 13.75 & 13.75 & 35 & 2
& 4 & 4 & $\times$16  & $\mathcal{F}$ & 1.125 \\
 & \module{A}{ 5}{} & 16-01 & 1600 & 13.75 & 13.75 & 35 & 2
& 4 & 4 & $\times$16  & $\mathcal{F}$ & 1.125 \\
 & \module{A}{ 6}{} & 16-10 & 1600 & 13.75 & 13.75 & 35 & 2
& 4 & 4 & $\times$16  & $\mathcal{F}$ & 1.125 \\
 & \module{A}{ 7}{} & 16-12 & 1600 & 13.75 & 13.75 & 35 & 2
& 4 & 4 & $\times$16  & $\mathcal{F}$ & 1.125 \\
 & \module{A}{ 8}{} & 16-09 & 1600 & 13.75 & 13.75 & 35 & 2
& 4 & 4 & $\times$16  & $\mathcal{F}$ & 1.125 \\
 & \module{A}{ 9}{} & 16-11 & 1600 & 13.75 & 13.75 & 35 & 2
& 4 & 4 & $\times$16  & $\mathcal{F}$ & 1.100 \\
\multirow{-10}{5em}{\centering {\large\em A} \\ {\tiny \quad \\} \centering
  Total of \\ \centering 10 DIMMs}
 & \module{A}{10}{} & 16-10 & 1600 & 13.75 & 13.75 & 35 & 2
& 4 & 4 & $\times$16  & $\mathcal{F}$ & 1.125 \\

\midrule

 & \module{B}{ 1}{} & 14-34 & 1600 & 13.75 & 13.75 & 35 & 2 &
4 & 4 & $\times$16  & $\mathcal{Q}$ & 1.100 \\
 & \module{B}{ 2}{} & 14-34 & 1600 & 13.75 & 13.75 & 35 & 2 &
4 & 4 & $\times$16  & $\mathcal{Q}$ & 1.150 \\
 & \module{B}{ 3}{} & 14-26 & 1600 & 13.75 & 13.75 & 35 & 2 &
4 & 4 & $\times$16  & $\mathcal{Q}$ & 1.100 \\
 & \module{B}{ 4}{} & 14-30 & 1600 & 13.75 & 13.75 & 35 & 2 &
4 & 4 & $\times$16  & $\mathcal{Q}$ & 1.100 \\
 & \module{B}{ 5}{} & 14-34 & 1600 & 13.75 & 13.75 & 35 & 2 &
4 & 4 & $\times$16  & $\mathcal{Q}$ & 1.125 \\
 & \module{B}{ 6}{} & 14-32 & 1600 & 13.75 & 13.75 & 35 & 2 &
4 & 4 & $\times$16  & $\mathcal{Q}$ & 1.125 \\
 & \module{B}{ 7}{} & 14-34 & 1600 & 13.75 & 13.75 & 35 & 2 &
4 & 4 & $\times$16  & $\mathcal{Q}$ & 1.100 \\
 & \module{B}{ 8}{} & 14-30 & 1600 & 13.75 & 13.75 & 35 & 2 &
4 & 4 & $\times$16  & $\mathcal{Q}$ & 1.125 \\
 & \module{B}{ 9}{} & 14-23 & 1600 & 13.75 & 13.75 & 35 & 2 &
4 & 4 & $\times$16  & $\mathcal{Q}$ & 1.125 \\
 & \module{B}{10}{} & 14-21 & 1600 & 13.75 & 13.75 & 35 & 2 &
4 & 4 & $\times$16  & $\mathcal{Q}$ & 1.125 \\
 & \module{B}{11}{} & 14-31 & 1600 & 13.75 & 13.75 & 35 & 2 &
4 & 4 & $\times$16  & $\mathcal{Q}$ & 1.100 \\
\multirow{-12}{5em}{\centering {\large\em B} \\ {\tiny \quad \\} \centering
  Total of \\ \centering 12 DIMMs}
 & \module{B}{12}{} & 15-08 & 1600 & 13.75 & 13.75 & 35  & 2 & 4 & 4 &
$\times$16  & $\mathcal{Q}$ & 1.100 \\

\midrule

 & \module{C}{ 1}{} & 15-33 & 1600 & 13.75 & 13.75 & 35  & 2 & 4 & 4 & $\times$16
& $\mathcal{A}$ & 1.300 \\
 & \module{C}{ 2}{} & 15-33 & 1600 & 13.75 & 13.75 & 35  & 2
& 4 & 4 & $\times$16 & $\mathcal{A}$ & 1.250 \\
 & \module{C}{ 3}{} & 15-33 & 1600 & 13.75 & 13.75 & 35  & 2
& 4 & 4 & $\times$16  & $\mathcal{A}$ & 1.150 \\
 & \module{C}{ 4}{} & 15-33 & 1600 & 13.75 & 13.75 & 35  & 2
& 4 & 4 & $\times$16  & $\mathcal{A}$ & 1.150 \\
 & \module{C}{ 5}{} & 15-33 & 1600 & 13.75 & 13.75 & 35  & 2
& 4 & 4 & $\times$16  & $\mathcal{C}$ & 1.300 \\
 & \module{C}{ 6}{} & 15-33 & 1600 & 13.75 & 13.75 & 35  & 2
& 4 & 4 & $\times$16  & $\mathcal{C}$ & 1.300 \\
 & \module{C}{ 7}{} & 15-33 & 1600 & 13.75 & 13.75 & 35  & 2
& 4 & 4 & $\times$16  & $\mathcal{C}$ & 1.300 \\
 & \module{C}{ 8}{} & 15-33 & 1600 & 13.75 & 13.75 & 35  & 2
& 4 & 4 & $\times$16  & $\mathcal{C}$ & 1.250 \\
\multirow{-9}{5em}{\centering {\large\em C} \\ {\tiny \quad \\} \centering
  Total of \\ \centering 9 DIMMs}
 & \module{C}{9}{} & 15-33 & 1600 & 13.75 & 13.75 & 35  & 2
& 4 & 4 & $\times$16  & $\mathcal{C}$ & 1.300 \\

\bottomrule
\end{tabular}

\begin{tabular}{l}
\small
{$\,$} \\

{$\ast$ The manufacturing date in the format of year-week
(yy-ww). For example, 15-01 indicates that the DIMM was manufactured during}\\
{the first week of 2015.} \\

{$\,$} \vspace{-7pt}\\

$\dagger$ The timing factors associated with each DIMM:\\
\hspace{0.1in}$Freq$: the channel frequency\\
\hspace{0.1in}$tRCD$: the minimum required latency for an \act to complete \\
\hspace{0.1in}$tRP$: the minimum required latency for a \pre to complete \\
\hspace{0.1in}$tRAS$: the minimum required latency for to restore the charge in an activated row of cells \\

{$\,$} \vspace{-7pt}\\

{$\ddagger$ The maximum DRAM module size supported by our testing platform is 2GB.} \\

{$\,$} \vspace{-7pt}\\

{$\star$ The number of DRAM chips mounted on each DRAM module. }\\

{$\,$} \vspace{-7pt}\\

{$\S$ The DRAM die versions that are marked on the chip package. } \\

{$\,$} \vspace{-7pt}\\

${\circ}$ The \emph{minimum voltage level} that allows error-free \changes{operation}, as
described in \ssecref{volt_sensitivity}.

{$\,$} \\

\end{tabular}

\caption{Characteristics of the evaluated DDR3L DIMMs.}
\vspace{-35pt}
\label{tab:modules}
\end{table*}

\end{document}